# Li-iontronics in single-crystalline $T$-$Nb_2O_5$ thin films with vertical ionic transport channels

Hyeon Han[1,7*], Quentin Jacquet[2,7,8], Zhen Jiang[3,7], Farheen N. Sayed[2], Jae-Chun Jeon[1], Arpit Sharma[1], Aaron M. Schankler[3], Arvin Kakekhani[3], Holger L. Meyerheim[1], Jucheol Park[4], Sang Yeol Nam[4], Kent J. Griffith[5], Laura Simonelli[6], Andrew M. Rappe[3*], Clare P. Grey[2*], Stuart S. P. Parkin[1*]

[1]Max Planck Institute of Microstructure Physics, 06120 Halle (Saale), Germany.

[2]Yusuf Hamied Department of Chemistry, University of Cambridge, Lensfield Road, Cambridge CB2 1EW, UK.

[3]Department of Chemistry, University of Pennsylvania, Philadelphia, Pennsylvania 19104-6323, USA.

[4]Test Analysis Research Center, Gumi Electronics and Information Technology Research Institute, Gumi, 39171, Republic of Korea.

[5]Department of Chemistry, Northwestern University, Evanston, IL, 60208, USA.

[6]ALBA Synchrotron Light Source, Cerdanyola del Vallès, Barcelona, Spain.

[7]These authors contributed equally.

[8]Present address: Univ. Grenoble Alpes, CEA, CNRS, IRIG-SyMMES, 38000 Grenoble, France.

*e-mail: hyeonhan21@gmail.com, rappe@sas.upenn.edu, cpg27@cam.ac.uk, stuart.parkin@mpi-halle.mpg.de

**The niobium oxide polymorph *T*-$Nb_2O_5$ has been extensively investigated in its bulk form especially for applications in fast-charging batteries and electrochemical (pseudo)capacitors. Its crystal structure that has two-dimensional (2D) layers with very low steric hindrance allows for fast Li-ion migration. However, since its discovery in 1941, the growth of single-crystalline thin films and its electronic applications have not yet been realized, likely due to its large orthorhombic unit cell along with the existence of many polymorphs. Here we demonstrate the epitaxial growth of single-crystalline *T*-$Nb_2O_5$ thin films, critically with the ionic transport channels oriented perpendicular to the film's surface. These vertical 2D channels enable fast Li-ion migration which we show gives rise to a colossal insulator-metal transition where the resistivity drops by eleven orders of magnitude due to the population of the initially empty Nb $4d^0$ states by electrons. Moreover, we reveal multiple unexplored phase transitions with distinct crystal and electronic structures over a wide range of Li-ion concentrations by comprehensive *in situ* experiments and theoretical calculations, that allow for the reversible and repeatable manipulation of these phases and their distinct electronic properties. This work paves the way to the exploration of novel thin films with ionic channels and their potential applications.**

The control of the electronic properties of materials via voltage biasing forms the cornerstone of many electronic applications. A recent trend in the field has been to leverage ionic liquid gating (ILG) of oxides, enabling the electric-field control of insulator-to-metal transitions (IMT). These electronic transitions via ILG are typically induced by the electrochemical intercalation of $O^{2-}$ or $H^+$ ions[1-13], which can, however, be sluggish, poorly controlled or involve degradation of the electrolyte[6]. Alternatively, intercalation of $Li^+$ ions into oxides can provide fast ion diffusion and, thereby, is fundamental to diverse applications, ranging from Li-ion batteries[14-19] and electrochromics[20-22] to electronic devices[23-30]. $WO_3$ is known as one of the best performing electrochemical materials for Li-ionic gating, owing to the large resistance change (up to ≈7 orders of magnitude), fast response, and good endurance[9,10,29]. However, this material and its thin film form can show a limited voltage operation range due to the conversion reaction that occurs with Li-ions[31]. Thus, finding new thin film systems that exhibit rapid and large changes in properties with high stability via Li-ion intercalation can boost various applications. Moreover, understanding the correlation between structure, electrochemical, and electronic properties during Li-ion insertion is needed to realize repeatable manipulation of these properties.

*T*-$Nb_2O_5$ is a promising material, which is known as an anode material for applications in batteries and electrochemical capacitors/pseudocapacitors[32-38]. Fast Li-ion diffusion in *T*-$Nb_2O_5$ is enabled by its crystal structure, consisting of two-dimensional (2D) layers at 4g Wyckoff positions of the space group *Pbam*, with very low steric hindrance for intra-layer Li-ion diffusion[37,39] (Fig. 1a and Supplementary Fig. 1). Moreover, *T*-$Nb_2O_5$ is a $d^0$ insulator, and Li-ion intercalation is expected to increase the electronic conductivity by filling the Nb 4*d* levels (left inset of Fig. 1b), making *T*-$Nb_2O_5$ a promising candidate for switchable electronic applications. However, since the discovery of *T*-$Nb_2O_5$ in 1941[37], the single-crystalline thin film growth and its electronic properties have not been demonstrated. This difficulty is likely due to its large orthorhombic unit cell ($a$ = 6.175 Å, $b$ = 29.175 Å, $c$ = 3.930 Å)[39], its metastability[32], and the presence of many other $Nb_2O_5$ polymorphs that are close in energy[38].

Here, we demonstrate epitaxial growth of high-quality single-crystalline *T*-$Nb_2O_5$ thin films oriented with their (180) plane parallel to the surface such that 2D open channels in the crystal structure are ideally oriented perpendicular the film surface. Li-containing ionic liquid gating (Li-ILG) of *T*-$Nb_2O_5$ thin films results in a fast and colossal insulator-metal transition with a ≈11 order of magnitude increase in the resistivity of the material during the early stages of Li insertion. Multiple unexplored phase transitions, including an orthorhombic metal and a monoclinic metal are found in thin film forms, while the bulk material further reveals a tetragonal insulating phase at high Li concentrations. These new phases were observed by comprehensive *in situ* experiments and rationalized with the help of density-functional theory (DFT) calculations. This understanding enables repeatable and durable control of electronic properties by operating within the reversible phase transition range of the crystalline thin films. Furthermore, a tunable metallization voltage is demonstrated by altering the chemical potential of the gate electrode via the use of lithiated counter electrodes.

## Growth of epitaxial thin films having vertical ionic channels

Pulsed laser deposition (PLD) was used to grow *T*-$Nb_2O_5$ thin films on (001)- and (110)-oriented substrates of $LaAlO_3$ (LAO) and $(La_{0.18}Sr_{0.82})(Al_{0.59}Ta_{0.41})O_3$ (LSAT). Various polymorphs including the *TT*-, *T*-, *B*-, and *H*-$Nb_2O_5$ phases were identified by X-ray diffraction (XRD) that strongly depend on the growth temperature (Supplementary Fig. 2a). Single phase *T*-$Nb_2O_5$ (180) films were obtained for growth temperatures between 600 and 650 °C. Cross-sectional scanning transmission electron microscopy (STEM) was used to obtain high angle annular dark field (HAADF) images and fast Fourier transformation (FFT) patterns (Fig. 1b and Supplementary Fig. 3). When using conventional (001)-oriented substrates, multi-domains with the 4-fold symmetry, rotated in-plane by 90° with respect to each other, were formed, as previously observed for growth on $SrTiO_3$ (001)[40]. However, the growth of 2-fold symmetry *T*-$Nb_2O_5$ (180) thin films was realized by using (110)-oriented substrates. Reflection high-energy electron diffraction (RHEED) patterns and XRD phi-scans further confirmed this finding (see Supplementary Section 1). The anisotropic in-plane geometry of the (110)-oriented substrates likely prohibits the formation of multi-domains[41]. Both thin films, exhibit vertically oriented two-dimensional 4g layers (the green dashed lines in Fig. 1b and Supplementary Fig. 3) that are ideal for Li-ion transport.

## Structural and electronic phase diagram by Li insertion

The evolution of the crystal structure correlated with corresponding changes of the electronic properties in *T*-$Nb_2O_5$ via Li-ion intercalation was investigated by various *in situ* and *ex situ* methods. XRD patterns and resistance were measured during Li-ionic liquid gating (Li-ILG) of a single-crystalline $T$-$Nb_2O_5$/LSAT (110) thin film device using Au/Ru as the gate electrode (Fig. 2a and Supplementary Fig. 7). Between a gate voltage ($V_g$) of 0 and 3 V, there is no noticeable change in the (180) reflection position, while the resistance starts to decrease at $V_g$ ≈2 V, suggesting the onset of an insulator to metal transition without a significant structural modification. Beginning with $V_g$ = 3.5 to 4 V, the (180) diffraction peak starts to shift toward lower angles without any change in resistance; this shift was assigned to a monoclinic phase transition by using reciprocal space maps (RSMs) (Supplementary Fig. 8). As shown in Supplementary Fig. 9, in the range between $V_g$ = 4 to –1 V, the (180) peak and resistance return to their original values, indicating the reversibility of the structural and electronic changes. At $V_g$ = 6 V, concomitantly with the irreversible increase of resistance suggesting a metal to insulator transition, an amorphization of the film was observed, evidenced by a decrease of the (180) peak intensity together with TEM images (Supplementary Fig. 10), indicating a degradation of the thin film at high Li concentration.

*In situ* XRD experiments were performed on polycrystalline $T$-$Nb_2O_5$ powder in a typical battery configuration[42] (Figs 2b-2f and Supplementary Section 2). During Li-ion insertion, a first structural phase transition is revealed at $x$ = 0.8 in $Li_xNb_2O_5$ with a large shift in the (180) peak position along with the disappearance of the (181) peak (Fig. 2b and Supplementary Fig. 11). In addition, at a higher Li-ion content, $x \approx 1.8$, a second structural transition is observed. To unambiguously identify these unexplored structures, we performed synchrotron-based XRD (SXRD) for the $Li_{0.8}Nb_2O_5$ and $Li_3Nb_2O_5$ phases. Rietveld refinement of *ex situ* SXRD pattern of $Li_{0.8}Nb_2O_5$ shows that the transition corresponds to a monoclinic ($m$-) distortion ($\gamma \approx 94°$) of the initial orthorhombic ($o$-) $T$-$Nb_2O_5$ structure without a significant change in the Nb framework (Fig. 2c, Supplementary Fig. 12 and Table 1). Interestingly, ≈25 % of the crystal volume remains in the orthorhombic phase, while both phases are active for Li intercalation (Supplementary Fig. 15). This suggests that the driving force for the transition to the monoclinic phase of the powder is weak and, hence, in competition with other effects preventing the transition, such as strain or compositional heterogeneities. The second structural transition of the powder involves the formation of a new tetragonal ($t$-) Li-rich layered rock-salt derivative with an approximate composition of $Li_3Nb_2O_5$ (Fig. 2d, Supplementary Fig. 13 and Table 2). A recent report mentions the formation of a cubic phase, which can be viewed as a parent phase of this newly discovered tetragonal rock-salt phase but which is characterized by a disordered Li/Nb site occupancy. Formed by lithiation of an amorphous $Nb_2O_5$ powder, it possesses excellent capacity at higher current densities and with a conductivity which is larger by ≈4 orders of magnitude as compared to the pristine[43], but we note that this phase still belongs to the insulating regime and is in an irreversible transition range for the initially crystalline thin film form. Having solved all phases in the $Li_xNb_2O_5$, we report the evolution of the cell parameters as a function of Li-ion content together with the voltage (Fig. 2e).

The correlation between the *in situ* powder XRD/SXRD and *in situ* thin film XRD measurements allows us to study the structural evolution as a function of Li concentration and $V_g$ (Fig. 2f). The single-crystalline thin film morphology enables monitoring of the change in electrical resistance as a function of Li-ion concentration and carrier mobility via Hall effect measurements (will be described in Fig. 4h later). These measurements therefore allow the correlation of the film`s electronic properties with its structure. Finally, the Li-ion concentration-dependent electronic and structural phase diagrams of the $T$-$Nb_2O_5$ thin film are obtained (Fig. 3a), revealing multi-step phase transitions between the initial orthorhombic insulator ($x < 0.3$), an orthorhombic metal ($0.3 \leq x < 0.8$), a monoclinic metal ($0.8 \leq x < 1.8$), and an irreversible transition to an amorphous insulator ($x \geq 1.8$). The tetragonal phase observed by *in situ* SXRD of the powder sample was not seen in thin films, possibly because of a clamping effect via the substrate, which hinders this phase change and thus drives the amorphization. In particular, the early lithiation stage in $T$-$Nb_2O_5$ is promising for fast and reversible electronic applications due to the sharp conductivity change without the significant structural change.

## DFT calculations on *T*-$Nb_2O_5$ via lithiation

We investigated the evolution of the structural and electronic properties of $T$-$Nb_2O_5$ during lithiation through *ab initio* study. Here, we designed a new structural simulation model (Fig. 3b) for pristine $T$-$Nb_2O_5$ ($Nb_{16.8}O_{42}$) by eliminating the fractional occupancy of Nb and the resulting charge imbalance (see Supplementary Section 3.2), resulting in a DFT-derived band gap of ≈2.3 eV (Fig. 3c). Next, we studied the energetic and kinetic performance of Li interstitials in our model unit cell of [$T$-$Nb_2O_5$]. We found that Li interstitials are located within the 4g layer atop two-coordinated oxygen atoms with a marginally stronger binding energy ($\Delta E_b$ = -2.06 eV, defined in Methods) than that of other sites (see Supplementary Section 3.3). Then, we investigated Li diffusion along the minimum energy path from this site to sites atop neighboring O atoms (Fig. 3g) and found an activation barrier of $E_a$ = 0.26 eV. Since the diffusion barriers and energy differences for Li at different sites are both small, there is a high probability that the incorporated Li-ions are not trapped in fixed locations.

To avoid spurious interactions between periodic images along the *c* direction, which has a small lattice periodicity, we model low Li concentrations by using supercells. With one extra Li inserted into an ($a \times b \times 3c$) super cell, the band gap is found to be greatly reduced (Fig. 3d) from 2.3 eV (in [$T$-$Nb_2O_5$]) to 0.3 eV (in $Li_{x=0.02}$-[$T$-$Nb_2O_5$] ($x$ = Li/Nb)), because a new filled defect band just below the Fermi energy ($E_F$) is induced by this Li interstitial (see Supplementary Fig. 17b). After that, the material becomes metallic once a second extra Li is intercalated into the ($a \times b \times 3c$) super cell, as shown in Fig. 3e. This is because the newly induced band crosses the $E_F$ (see Supplementary Fig. 17c). In addition, we note that the size of the simulation supercell (especially that of the *c*-axis) affects the Li concentration required to induce metallization (see Supplementary Section 3.4). Therefore, *ab initio* calculations can provide a qualitative explanation for the rapid onset of metallization at low Li concentration.

Finally, we investigated the differential binding energy ($\Delta E_b$) as a function of Li interstitial concentration in our [$T$-$Nb_2O_5$] model. For low Li concentrations ($x$ = Li/Nb < 0.2), the obtained $\Delta E_b$ fluctuates around -2 eV (-1.95 eV – -2.24 eV) (see Fig. 3h). With increasing Li concentration, the magnitude of $\Delta E_b$ gradually decreases but it remains smaller than -1.5 eV until at least Li/Nb = 0.5. Moreover, we found that at higher Li concentration (Li/Nb ≈ 0.5), a phase transformation from orthorhombic to monoclinic is energetically favorable. Figure 3i shows the relative change of the total energy of the $Li_{x=0.54}$-[$T$-$Nb_2O_5$] structure as a function of the monoclinic tilt angle varying between 90° and 92.5°. The monoclinic phase with a tilt angle of about 91.4° is more stable than the orthorhombic one by 0.44 meV per atom, and it is consistent for other metastable configurations of $Li_{x=0.54}$-[$T$-$Nb_2O_5$] (see Supplementary Figs. 20 and 21). Entropy corrections at the finite temperature (300 K) originating from phonon contributions also stabilize the monoclinic phase by additional 0.58 meV/atom (for a total free energy difference of 1.02 meV/atom) compared with the orthorhombic phase. For comparison, we also considered the stability of the monoclinic phase for the case of the pristine the un-lithiated [$T$-$Nb_2O_5$] unit cell (Fig. 3i) and found that in this case the orthorhombic phase is more stable. Therefore, the calculations predict that $T$-$Nb_2O_5$ will undergo a phase transition from orthorhombic to monoclinic at Li/Nb ≈ 0.5, in agreement with our experimental findings.

## Electrochemical and electronic properties via Li intercalation

Epitaxial *T*-$Nb_2O_5$ single-crystalline thin films deposited on LSAT (110) substrates were characterized electrochemically in a typical Li-ion battery configuration allowing the quantification of inserted Li depending on the reaction speed. Li metal and carbonate based electrolyte with $LiPF_6$ salt were used as anode/reference electrode and electrolyte, respectively (see methods and Supplementary Fig. 22). Galvanostatic cycling at current densities ranging from 1.43 - 14.3 A/g delivered the expected voltage versus capacity profile with reversible capacities ranging from 130 – 80 mAh/g (Fig. 2a and 2b). This indicates another figure of merit representing the excellent intercalation kinetics for the epitaxial film in accord with previous reports on nanoparticles[34]. The electrochemical properties were also demonstrated by cyclic voltammetry for 16, 80, and 160 nm thick films (Fig. 2c and Supplementary Fig. 23).

To explore the electronic property changes of epitaxial *T*-$Nb_2O_5$ thin films upon lithiation, Hall devices for Li-ionic liquid gating (Li-ILG) were fabricated with the Au/Ru gate electrode (Fig. 4d and Supplementary Fig. 24a). Before placing 0.3 M Li-TFSI in EMIM-TFSI (Li-IL) onto the device, we first measured temperature-dependent resistivity (RT) curves of pristine *T*-$Nb_2O_5$ films using a high resistance meter. The room temperature resistivity of the *T*-$Nb_2O_5$ film is $2.78 \times 10^8$ Ω cm (Supplementary Fig. 25), which is comparable with previous reports ranging from $10^7$ to $10^9$ Ω cm[43,44]. After placing IL onto the device, the resistivity drops to $\approx 5 \times 10^{-1}$ Ω cm due to the lower resistance of IL ($\approx 10^{14}$ Ω for *T*-$Nb_2O_5$ while $\approx 10^6$ Ω for IL when a source-drain current of 1 µA is applied, see Supplementary Fig. 26). Therefore, before Li intercalation, the resistance of the device is determined by the resistance of IL. We emphasize that the high limit of the resistivity accessible to the *T*-$Nb_2O_5$ material is given by the *ex-situ* measured value, while at the low limit, the resistivity is governed by the gated *T*-$Nb_2O_5$ film. The source-drain current ($I_{sd}$) or voltage ($V_{sd}$) dependent resistance changes of IL alone show that the resistance of IL decreases as increasing $I_{sd}/V_{sd}$ (Supplementary Figs. 26), which is likely due to the leakage from the IL decomposition at high voltages. Therefore, by adjusting the operating parameters of our device (for example using a lower $I_{sd}/V_{sd}$), we can measure a larger resistivity change during gating. Note that, for simplicity of the resistivity calculations of the device, all resistivities were calculated using the film thickness and the channel dimensions of $65 \times 30$ µm$^2$. The RT curves at different gate voltages are shown in Fig. 2f. Notably, the resistivity after metallization is $\approx 2 \times 10^{-3}$ Ω cm, which is ≈11 orders of magnitude smaller than that of pristine $T$-$Nb_2O_5$. Such a colossal insulator-metal transition is remarkable compared to the previously reported metallization by ILG[1-13] (See Supplementary Fig. 28 and Table 3).

To investigate the origin of metallization, the devices were gated between $V_g$ =3 and -2 V using conventional ionic liquids (ILs) including DEME-TFSI, EMIM-TFSI, and 0.3 M Li-TFSI in EMIM-TFSI (Li-IL) (Fig. 2g). While Li-ILG induces metallization, no noticeable change in resistivity was observed for pure IL. This contrasts with ILG of other oxide films such as $VO_2$, $WO_3$, and $SrCoO_{2.5}$, all of which show metallization in the absence of Li-ions due to oxygen or hydrogen ion migration[2-10]. The effect of the substrate orientation and film thickness on ILG is summarized in Supplementary Fig. 29. All films grown on (001)- and (110)-oriented substrates show good metallization behaviour owing to vertical ionic transport channels. For thicknesses ranging from 16 to 160 nm and gating with sufficiently slow voltage sweep rates, the resistivity at $V_g$ =3 V is close to $10^{-3}$ Ω cm, demonstrating that the metallization happens in the bulk of the material. This is further confirmed by hard X-ray absorption near-edge structure (XANES) measurements (Supplementary Fig. 30), where Nb K- absorption edge shifts to lower energy upon gating to positive potentials. This indicates the reduction of the Nb oxidation state in the bulk of the film, in agreement with prior work on a polycrystalline sample[45]. When we performed high voltage gating (Supplementary Figs. 31 and 32), the resistivity increased as the gating voltage was increased above 4.5 V, indicating the onset of a metal to insulator transition. The voltage-dependent carrier concentration, mobility, and resistivity changes are summarized in Fig. 2h. The electronic transitions correlate well with the *in situ* XRD data in Figs. 2f and 3a.

## Tunable and low voltage operation by chemical potential control

The working voltage of a Li-ion battery cell depends on the difference in chemical potential between the cathode and the anode[18]. To exploit this, the Au gate electrode was replaced by conventional Li-containing electrode materials, namely $LiFePO_4$ (LFP), $LiCoO_2$ (LCO), and $Li_xNb_2O_5$ (LNbO). The Li intercalation potentials in these materials are ≈ 3.5, 4.0 and 1.8 V *vs*. $Li^+$/Li for LFP, LCO, and LNbO, respectively, and are all within the electrochemical stability range of the IL used in this study. The fabrication process is described in Supplementary Section 6 and Fig. 3e, and the $V_g$-dependent resistivity (RV) curves with different gate electrodes are shown in Supplementary Fig. 35. The critical metallization voltage ($V_c$) is defined from the peak of the normalized $d\rho/dV$ curves. The chemical potential (V *vs*. Li/$Li^+$) dependent $V_c$ plot (Supplementary Fig. 35c) shows that $V_c$ tends to decrease as the potential difference between the gate electrode and *T*-$Nb_2O_5$ decreases. Notably, the LNbO gate electrode leads to coupled resistivity changes between the LNbO gate electrode and $Nb_2O_5$ channel due to the ion exchange with each other, and it reveals a significant decrease in the $V_c$ value of ≈0.54 V compared to the Au electrode (Fig. 2i) which has been extensively used in typical ILG devices[2-5]. In particular, such potential control of $V_c$ is not possible with conventional ILG based on proton insertion or oxygen ion loss, because the $V_c$ will be determined by the more poorly determined electrochemical process that occurs at the counter (gate) electrode (including double layer formation, electrolyte degradation and oxygen evolution).

## Pulsed voltage gating of *T*-$Nb_2O_5$ and $WO_3$ thin film devices

We performed pulsed voltage gating on *T*-$Nb_2O_5$ and $WO_3$ thin film devices to compare the kinetics of Li insertion and metallization (Supplementary Section 7). To explore the effects of the crystal orientation and the presence of crystallinity, devices were fabricated using an Au gate electrode, as depicted in Fig. 2d, for a 30 nm polycrystalline *T*-$Nb_2O_5$ thin film grown on an YSZ (001) substrate (See Supplementary Fig. 36) and compared to the single-crystalline thin films deposited on LSAT (110) having the same thickness (Supplementary Fig. 39). The single-crystalline *T*-$Nb_2O_5$ device shows metallization after a single pulse of 4.3 V with a pulse width (*W*) of 0.8 s (Supplementary Fig. 39a) while the polycrystalline *T*-$Nb_2O_5$ becomes metallic after 45 pulses, indicating that the vertical orientation of ionic channels plays a crucial role for the fast ionic gating. The critical resistivity for metallization was defined as $2 \times 10^{-3}$ Ω cm, which shows a decrease in resistivity with decreasing temperature (Fig. 4f and supplementary Fig. 38b). The pulse voltage (H)-dependent resistance changes of the single-crystalline *T*-$Nb_2O_5$ device are shown in Supplementary Fig. 37. Moreover, a 30 nm epitaxial $WO_3$ film (Supplementary Fig. 37) is gated with Li-IL at 3.5 V and pure IL (DEME-TFSI) at 4.3 V and shows metallization after 8 and ≈1500 pulses, respectively, showing that $Li^+$ ion gating is much faster than the $O^{2-}/H^+$ ion gating. Note that pulsed voltage gating above 4 V on $WO_3$ with Li-IL leads to increased resistance compared to 3.5 V (Supplementary Fig. 38c), which is likely due to a conversion reaction[31], limiting higher voltage operation. Single-crystalline *T*-$Nb_2O_5$ features a resistance change by ≈11 orders from initially insulating film to the gated metallic film by a single pulse while for $WO_3$ it is ≈6 orders of magnitude (Supplementary Fig. 39b). The pulsed width (W)-dependent resistance changes (Supplementary Fig. 40) further reveal the larger resistance changes of the single-crystalline $T$-$Nb_2O_5$ thin film compared to the $WO_3$ thin film in the electrochemical reaction region.

The single-crystalline $T$-$Nb_2O_5$ device with an Au gate electrode reveals 1 order of magnitude resistance change over ≈$3.5\times10^5$ pulses (3.8 V/-2 V with a pulse width of 50 msec), illustrating good reversibility (Supplementary Fig. 41). In particular, by replacing the Au gate electrode with Li-$Nb_2O_5$ (Fig. 4e), coupled responses between twin $T$-$Nb_2O_5$ devices are realized (Fig. 4j). The resistances of the two devices oscillate out of phase for 3 orders of magnitude for more than $10^3$ pulses, when applying 3 V/-3 V and a pulse width of 50 ms.

In summary, we have realized the growth of single-crystalline epitaxial $T$-$Nb_2O_5$ thin films with vertically formed 2D channels that provide paths for fast ionic migration. This morphology allows us to study the evolution of the electronic and structural properties during Li-ILG by comprehensive *in situ* and *ex situ* experiments. They reveal unexplored sequential phase transformations, including an orthorhombic insulator, an orthorhombic metal, a monoclinic metal, and a degraded insulating phase as Li concentration is increased. DFT calculations further support that the monoclinic phase (approximately $Li_1Nb_2O_5$) is energetically favourable and metallic. Defect electronic states near the Fermi energy from Li-ion migration lead to abrupt changes in resistivity. Thus, the $T$-$Nb_2O_5$ films with vertical ionic transport channels undergo significant electrical change in an early stage of Li insertion into the initially insulating $d^0$-films, leading to a colossal insulator-metal transition with a eleven orders of magnitude change in resistance. The $T$-$Nb_2O_5$ film shows an even larger and faster resistance change and a wider voltage operation range via Li interaction, as compared with $WO_3$ thin films which are one of the best electrochemical materials. Moreover, coupled electronic responses between twin $T$-$Nb_2O_5$ devices are demonstrated through ionic exchange between each other. This work showcases a synergistic experiment-theory approach to develop new ionically channelled devices for diverse applications, including thin film batteries, electrochromic devices, neuromorphic devices, and electrochemical random-access memory (ECRAM).

**Acknowledgments.** This project has received funding from the European Union's Horizon 2020 research and innovation program under grant agreement No 737109. Funding has been provided by the Alexander von Humboldt Foundation in the framework of the Alexander von Humboldt Professorship to S.S.P.P. endowed by the Federal Ministry of Education and Research. The electrochemical theory of Z.J. and A.K. was supported by the U.S. Department of Energy, Office of Science, Basic Energy Sciences, under Award # DE-SC0019281. F.N.S acknowledges funding from the Faraday Institution CATMAT project (FIRG016). The oxide structure and phase transition theory of A.M.S. and A.M.R. was supported by the Office of Naval Research, under grant N00014-20-1-2701. The authors acknowledge computational support from the National Energy Research Scientific Computing Center (NERSC) of the DOE and the High-Performance Computing Modernization Office (HPCMO) of the U.S. Department of Defense (DOD). We thank Sang Ho Oh`s group at KENTECH for the discussion on the film structure. We thank Charles Guillemard at ALBA synchrotron and Jeong Hyo Jin at GERI for their assistance with gating of XANES and TEM samples, respectively.

**Author contributions.** H.H., Q.J., C.P.G, and S.S.P.P. conceived the project. H.H. grew and characterized thin films. H.H. and A.S. fabricated ILG devices. H.H. performed transport measurements. H.H. and J.-C.J. carried out the pulsed voltage gating and high resistance measurements. Q.J. and F.N.S. contributed to electrochemical measurements. Q.J., F.N.S., and K.J.G. analysed powder structures. H.H. and H.L.M. analysed thin film structures. S.Y.N. and J.P. performed TEM. L.S. carried out XANES. Z.J., A.M.S., A.K. performed DFT calculations. H.H. was the lead researcher. C.P.G, A.M.R., and S.S.P.P. supervised the project. H.H., Q.J, Z.J., C.P.G, A.M.R., and S.S.P.P. wrote the manuscript with contributions from all authors.

**Competing interests.** The authors declare no competing interests.

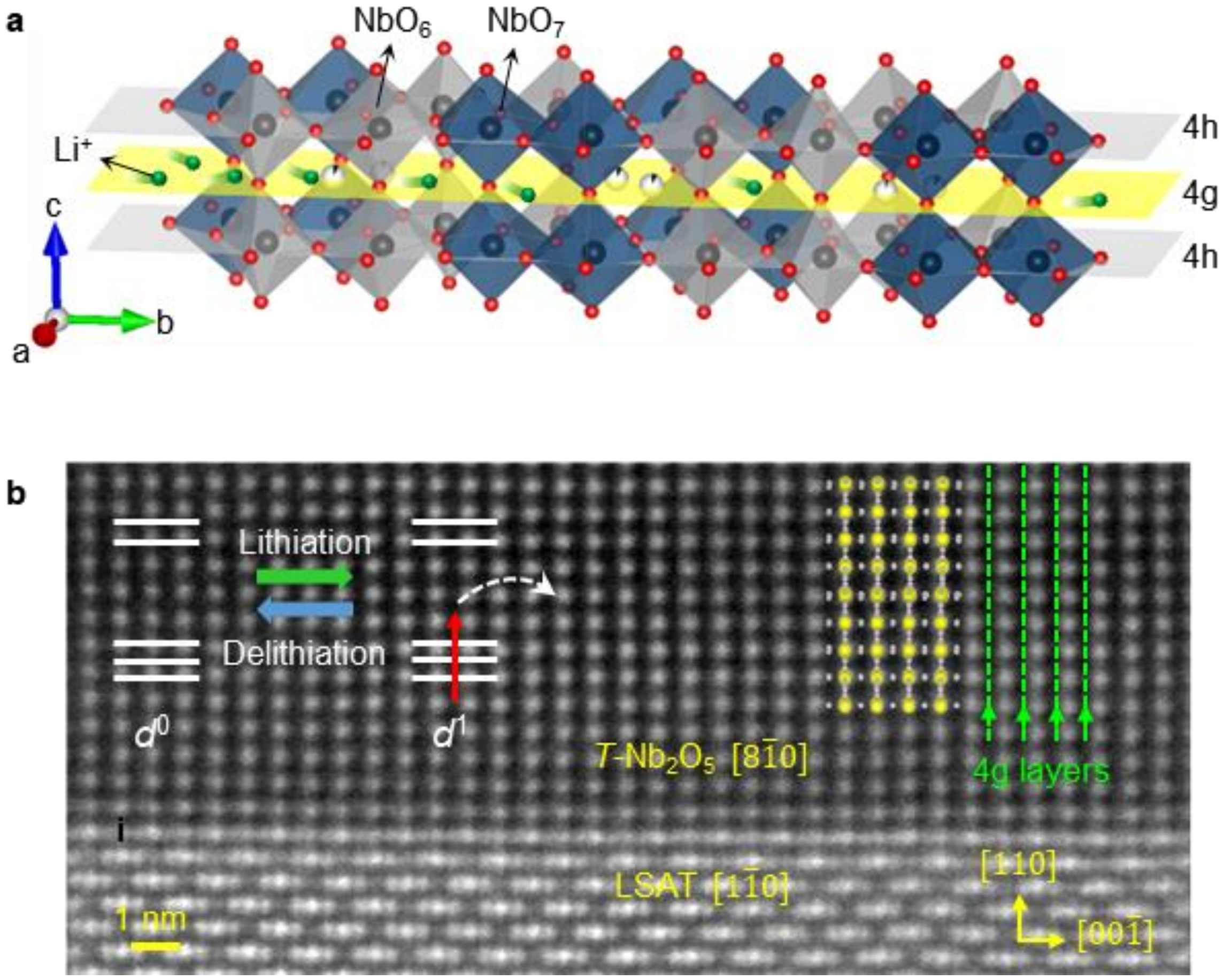

**Fig. 1. Structure of epitaxial *T*-$Nb_2O_5$ thin films. a**, Schematic showing Li-ion migration into *T*-$Nb_2O_5$. The black, red, and green spheres denote Nb, O, and Li ions, respectively. The grey and navy polyhedra denote distorted octahedra ($NbO_6$) and pentagonal bipyramids ($NbO_7$), respectively. The grey and yellow planes represent the 4h and 4g layers, respectively. The loosely packed 2D 4g layer provides for fast Li-ion transport. **b**, Cross-sectional HAADF-STEM image of a single-crystalline *T*-$Nb_2O_5$ thin film grown on a LSAT (110) substrate viewed along the LSAT [1$\bar{1}$0] direction. Left inset: Nb 4*d* orbital state changes from $d^0$ to $d^1$ due to electron doping via Li intercalation. Overlaid yellow and grey spheres represent Nb and O atoms, respectively. Green dashed lines represent the vertical ionic transport channels (4g layers) viewed from *T*-$Nb_2O_5$ [8$\bar{1}$0].

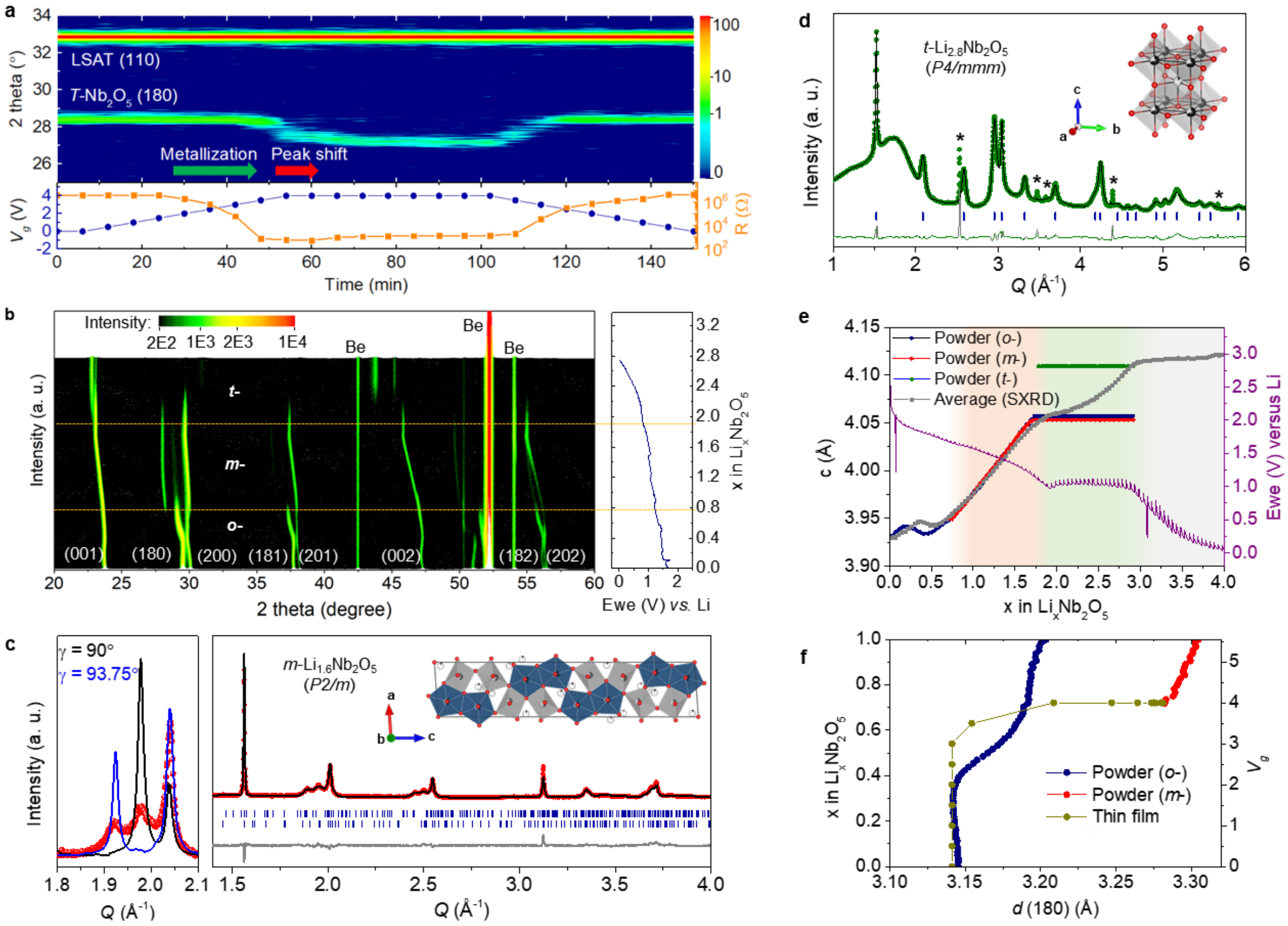

**Fig. 2. Sequential structural phase transitions of *T*-$Nb_2O_5$ via Li-ion intercalation. a**, *In situ* XRD and resistance measurements during Li-ILG of a 50 nm *T*-$Nb_2O_5$/LSAT(110). **b**, *In situ* XRD pattern for powder *T*-$Nb_2O_5$ until the deep discharge potential of 0.005 V *vs*. Li with C/34 scan rate for electrochemical cycling. "Be" indicates peaks from the Be window. Horizontal yellow dashed lines indicate the boundary of the transition. **c**. left panel: *Ex-situ* SXRD pattern (red) of $Li_{1.6}Nb_2O_5$ and simulated patterns of the *Pbam* (black) and *P*2/*m* (blue). The (180) reflection is split by monoclinic tilting. Right panel: SXRD pattern and Rietveld refinements of *m*-$Li_{1.6}Nb_2O_5$. Red circles, black line, and grey line represent the observed, calculated, and difference patterns, respectively. The black spheres, red spheres, grey polyhedra, and navy polyhedra in the unit cell represent Nb ions, O ions, octahedra ($NbO_6$), and pentagonal bipyramids ($NbO_7$), respectively. **d**. Rietveld refinement of an *in situ* SXRD pattern measured at 5 mV using the *t*-phase. The *in situ* cell produces peaks marked as *. Green circles, black line, and grey line represent the observed, calculated, and difference patterns, respectively. Blue vertical bars represent the Bragg position. The black spheres, red spheres, and grey polyhedra in the unit cell denote Nb ions, O ions, and octahedra ($NbO_6$), respectively. Li positions are not considered in all refined structures due to the low scattering of Li ions. **e**. *c*-lattice parameter as a function of *x* in $Li_xNb_2O_5$. Grey dots represent the average *c*-parameter obtained from the SXRD data. Blue, red, and green dots denote the *c* parameter of the *o*-, *m*-, and *t*- phases, respectively, extracted from the *in situ* XRD data. The purple line is the voltage curve obtained during the Galvanostatic discharge-charge measurement. Color zones highlight four regions: i) pristine *o*-phase from $Li_0Nb_2O_5$ to $Li_{0.8}Nb_2O_5$ (white), ii) a 25/75 % mixture of the *o*- and *m*-phases from $Li_{0.8}Nb_2O_5$ to $Li_{1.8}Nb_2O_5$ (pink), iii) progressive formation of the *t*-phase from $Li_{1.8}Nb_2O_5$ to $Li_3Nb_2O_5$ (green), iv) no change of the XRD pattern from $Li_3Nb_2O_5$ to $Li_4Nb_2O_5$ (grey). **f**. Comparison of the d(180)-spacing for the powder and thin film XRD.

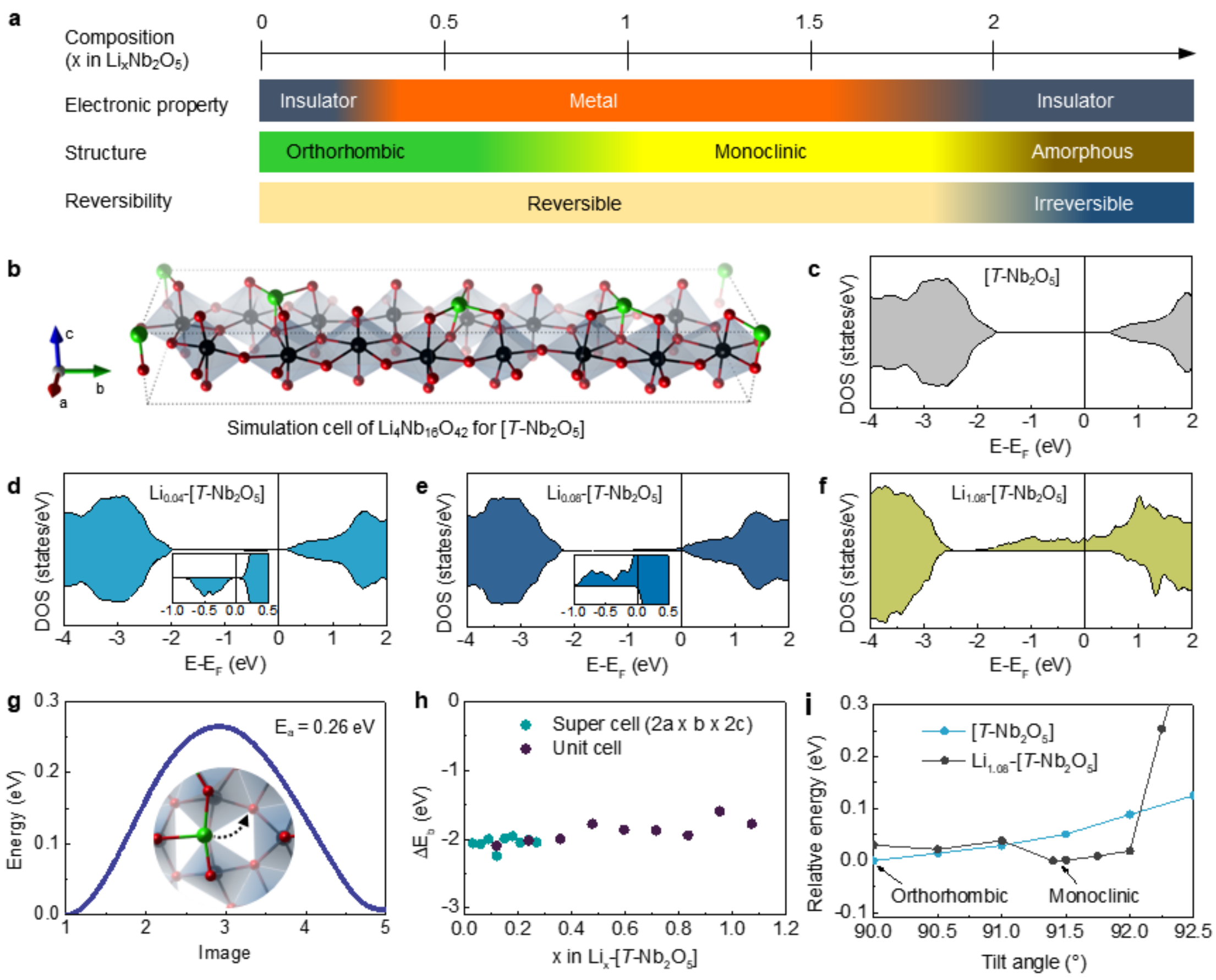

**Fig. 3. Electronic and structural phases of *T*-$Nb_2O_5$ via Li intercalation. a**. Electronic and structural phase diagram of *T*-$Nb_2O_5$ film versus Li intercalation. The phase transitions and their reversibility are presented schematically. **b**. Structural model of the orthorhombic model system ($Li_4Nb_{16}O_{42}$) which was employed to simulate the conventional (pristine) unit cell of *T*-$Nb_2O_5$ ($Nb_{16.8}O_{42}$, shown in Supplementary Fig. 16), in which 4 Li are added per unit cell to replace the charge of the fractionally distributed 0.8 Nb atom/cell within the 4g layer. This model for *T*-$Nb_2O_5$ characterized by a $Li_4Nb_{16}O_{42}$ stoichiometry is considered as unlithiated and referred to as [*T*-$Nb_2O_5$] throughout. **c**. Total density of states (DOS) of the [*T*-$Nb_2O_5$] model unit cell. **d**. and **e**. Total DOS with one and two extra Li intercalated into the ($a \times b \times 3c$) super cell of our simulation model, (i.e., $Li_{0.04}$-[*T*-$Nb_2O_5$] and $Li_{0.08}$-[*T*-$Nb_2O_5$]) **f,** Total density of states (DOS) of the monoclinic $Li_{1.08}$-[*T*-$Nb_2O_5$] structure. **g,** Diffusion performance of the one extra Li in the primitive $Li_{0.12}$-[*T*-$Nb_2O_5$] unit cell including free-energy profile and diffusion barrier ($E_a$). **h,** Evolution of the differential binding energy ($\Delta E_b$) for low Li concentration within the supercell ($2a \times b \times 2c$) and at high-concentration of Li within the primitive [*T*-$Nb_2O_5$] unit cell of **i,** The relative energy evolution of two lithiation states ([*T*-$Nb_2O_5$] and $Li_{1.08}$-[*T*-$Nb_2O_5$]) as a function of the monoclinic angle in the range between 90° and 92.5°.

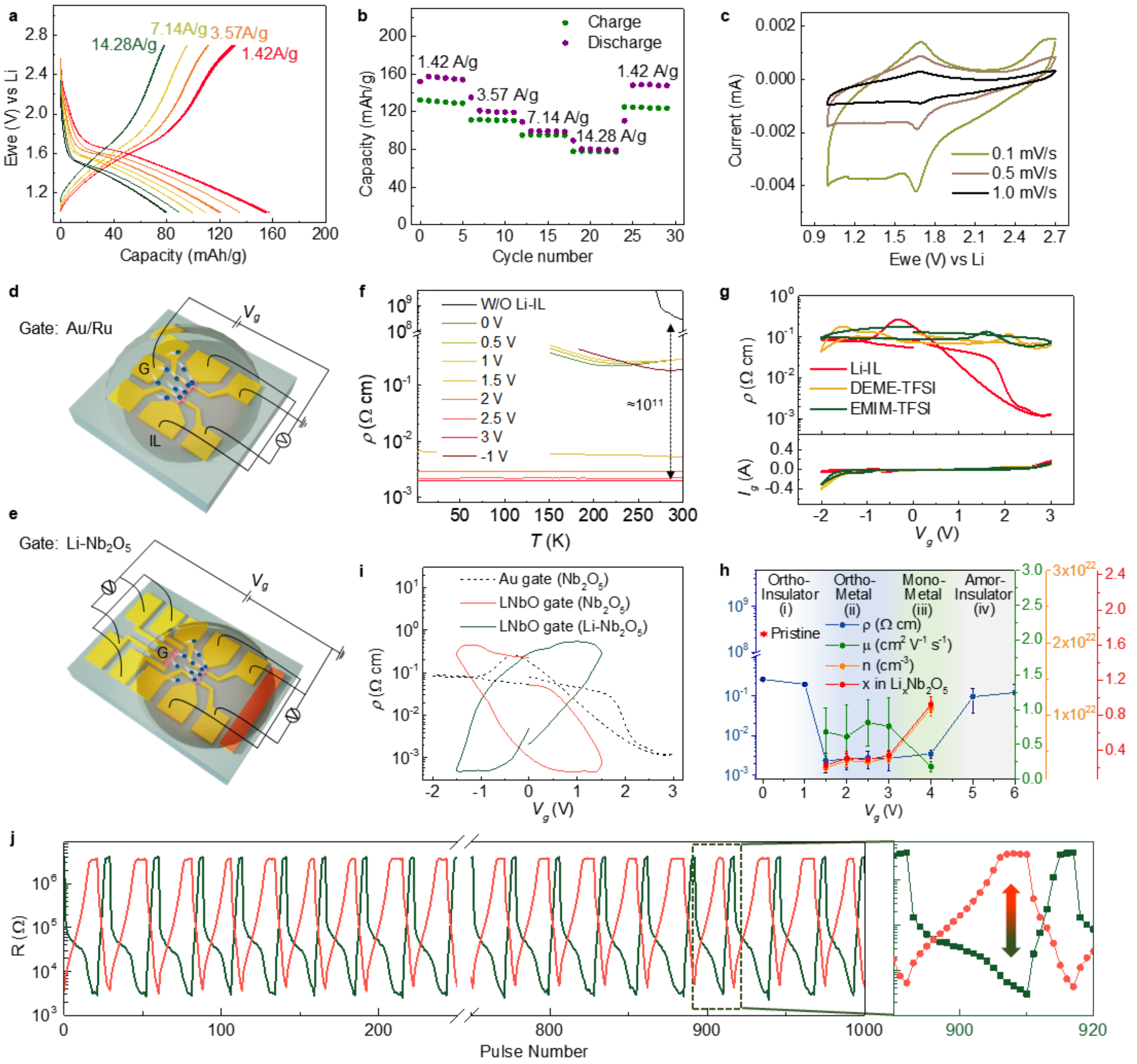

**Fig. 4. Electrochemical and electronic properties of epitaxial *T*-$Nb_2O_5$ thin films. a,** Galvanostatic discharge-charge curves and **b**, Specific capacity obtained at each cycle at different current rates for a 40 nm *T*-$Nb_2O_5$/LSAT (110). An irreversible capacity is observed at the slowest current densities, which is attributed to corrosion of parts of the device. **c**, Cyclic voltammogram recorded at different scan rates. The second cycle of each cycling rate is shown. Schematics of the ILG device **d**, with the Au gate electrode and **e**, using the LNbO gate electrode. G represents gate electrode. Blue spheres denote Li-ion migrations from gating. **f**, Temperature-dependent resistivity curves at different gate voltages. **g,** $V_g$-dependent resistivity and leakage current curves for several ILs. The sweeping rate was 16 mV/s. **h**, $V_g$ dependent carrier concentration (n), Li composition (x), mobility (μ), and resistivity (ρ) curves from Hall measurements at 200 K for $Li_xNb_2O_5$. The mean and standard deviation are represented with the error bars after three measurements. The resistivity of the pristine film is indicated by the red pentacle. Multi-step transitions are shown with increasing $V_g$. *i.e.*, (i) orthorhombic insulator (white regime) (ii) orthorhombic metal (blue), (iii) monoclinic metal (red), and (iv) insulating state (grey). The carrier concentration suddenly increases, and the mobility starts to decrease at ≈4 V, indicating carrier scattering at high Li concentrations. The Li concentrations are obtained from Hall measurements, assuming that each Li atom creates one charge carrier. **i,** $V_g$-dependent ρ curves for the 16 nm *T*-$Nb_2O_5$/LSAT(110) devices using the LNbO gate electrode (blue and light blue curves) and the Au electrode (the dotted black curve). The sweeping rate was 16 mV/s. **j**, Pulse voltage gating of twin *T*-$Nb_2O_5$ devices. The device structure is depicted in Fig. 2e. Pulse voltages of 3 V/-3 V were applied with a pulse with of 50 msec. The channel resistances were measured at 1 µA ($I_{sd}$). The film thickness and channel size were 30 nm and 60×30 µm$^2$, respectively.

## Methods

**Film growth.** A reflection high energy electron diffraction (RHEED)-assisted pulsed laser deposition (PLD) system using a 248 nm KrF excimer laser was employed to optimize the growth conditions of $Nb_2O_5$ thin films by varying the growth temperature from 500 to 900 ˚C. The laser fluence, oxygen partial pressure ($pO_2$), and repetition rate were 1 J $cm^{-2}$, 10 mT, and 6 Hz, respectively. The optimized growth temperature of $T$-$Nb_2O_5$ thin films was ≈620 ˚C. The heating and cooling rate were 30 and 10 ˚C/min, respectively. Substrates of LSAT (001), LAO (001), LSAT (110), and LAO (110) were used to study substrate orientation dependent domain structures. A YSZ (001) substrate was used to grow a polycrystalline $T$-$Nb_2O_5$ thin film. For gate electrode potential control, LFP and LCO thin films were grown on LSAT (110) substrates by varying $pO_2$ at room temperature. The laser fluence and repetition rate were 1 J $cm^{-2}$ and 6 Hz, respectively. The optimal $pO_2$ was 1 mT and 10 mT for the LFP and LCO thin films, respectively, as determined by XPS characterization. A $WO_3$ thin film was grown on a LAO (001) substrate. The growth temperature, laser fluence, oxygen partial pressure ($pO_2$), and repetition rate were 600 ˚C, 1 J $cm^{-2}$, 200 mT, and 6 Hz, respectively. The film thickness was characterized by X-ray reflectivity (XRR) measurements or fringes of theta-2theta scans, and the $T$-$Nb_2O_5$ thickness was further confirmed by TEM measurements.

**Ionic liquid gating device fabrication.** Standard photo-lithographic techniques were used to fabricate ionic liquid gating devices. Substrates with a size of 5×5 $mm^2$ were used. A $T$-$Nb_2O_5$ channel with a size of 65 × 30 $\mu m^2$ was etched and then Ru (5 nm) and Au (70 nm) layers were successively deposited using ion beam sputtering (SCIA coat 200) for both the gate electrode and channel contacts. The IL covered both the $T$-$Nb_2O_5$ and the gate electrode. For the gate electrode potential control, Li-ion containing oxides were deposited on top of the Au/Ru gate electrode.

For the LNbO gate electrode device fabrication, both channel and gate were etched, and then Au (70 nm)/Ru (5 nm) layers were deposited to make channel and gate contacts. Then, the reference electrode (LFP) was deposited using pulsed laser deposition. After the device fabrication, Li-ions are moved from the LFP to the gate electrode by ILG to make LNbO gate electrode. Then, the gating was applied to the $Nb_2O_5$ channel using the LNbO gate electrode.

For the *in situ* XRD measurements of the thin films, a $T$-$Nb_2O_5$ channel with a dimension of 2 × 2 $mm^2$ was etched, then Au (70 nm)/Ru(5 nm) layers were deposited for the gate electrode and channel contacts. The device was then attached using double-sided tape to a specially designed sample holder. The IL was placed on the device surface, and then Kapton film was attached to reduce the thickness of the IL. Resistance and θ−2θ scans were measured during *in situ* ILG.

**Thin film characterization.** θ−2θ scans, phi scans, and *in situ* XRD on the thin films were carried out using a Bruker d8 Discovery X-ray diffractometer with Cu-Kα radiation. RSM measurements were performed using a Ga jet X-ray source (λ=1.34 Å) and a six-circle diffractometer equipped with a Pilatus 100 K pixel detector. HAADF-STEM imaging was performed using a JEOL ARM200F with a spherical aberration corrector (CEOS GmbH) operated at 200 kV. X-ray photoelectron spectroscopy (XPS) (K-Alpha, Thermo Scientific) was conducted for Li-ion containing oxides. The film surface was gently cleaned by cluster ion etching prior to the measurement.

Diethylmethyl(2-methoxyethyl)ammonium bis(trifluoromethylsulfonyl)imide (DEME-TFSI), 1-Ethyl-3-methylimidazolium bis(trifluoromethylsulfonyl)imide (EMIM-TFSI), and Li-IL were used for the ILG. For the Li-IL, a lithium bis(trifluoromethanesulfonyl)imide (Li-TFSI) powder was dissolved in the EMIM-IL at 50 °C for 2 h to achieve a solution of molality 0.3 mol $kg^{-1}$. Each IL was dried in a vacuum chamber ($10^{-6}$ mbar) at 105 °C for at least 10 h prior to use.

Transport measurements of the ILG devices were carried out in a physical property measurement system (PPMS, Quantum Design). Gate voltages were applied using a Keithley 2450A source meter. For the resistance measurements, a constant current of 1 μA was applied using a Keithley 6221 current source, and the voltage was measured by a Keithley 2182A nano-voltmeter. The gate voltage is applied between the gate electrode and the *T*-$Nb_2O_5$ channel, while monitoring the resistance of the channel. The substrate is insulating, thus the voltage is applied through IL, resulting in Li-ion migration from gating. The resistance of the pristine *T*-$Nb_2O_5$ and $WO_3$ thin films was measured by a high resistance meter (B2985A, KEYSIGHT). 1000 data points were averaged at each temperature in a probe station. Li-ILG of thin films for the *ex situ* XRD, STEM, and XANES measurements was carried out using a polytetrafluoroethylene (PTFE) boat with an Au plate (the gate electrode) covered with the Li-IL.

The pulsed voltage gating was performed in a probe station (CRX; Lake Shore). Gate voltages were applied using a Keithley 2636B source meter. A constant current of 1 μA was applied using a Keithley 6221 current source, and the voltage was measured by a Keithley 2182A nano-voltmeter to measure the channel resistance. The 6221/2182A system is advantageous for the fast readouts after the pulsed gating and provides electrically floating between each device.

The Nb K-edge X-ray absorption spectra (XAS) were acquired at the CLÆSS beamline at the ALBA synchrotron[46]. The synchrotron radiation emitted by a wiggler source was monochromatized using a double crystal Si (311) monochromator, while higher harmonics were rejected by proper choice of angles and coatings of the collimating and focusing mirrors. The samples were mounted in a liquid nitrogen cryostat, and the spectra were recorded in fluorescence mode at 80 K by means of a multichannel silicon drift detector. The sample normal and the fluorescence detector were at 60 and 90 degrees with respect to the incoming beam, respectively. The fluorescence detector dead time was kept around 4.5 % at 19,600 eV for both samples for a better comparison.

***In situ* powder XRD.** Free-standing electrodes for *in situ* XRD measurements with coupled electrochemistry was prepared by mixing 90 wt. % *T*-$Nb_2O_5$ powder (Sigma-Aldrich), 5 wt. % Polytetrafluoroethylene (PTFE) binder, 5 wt. % carbon black (Timcal SuperP). The mixed powder was pressed and rolled onto a flat surface to give a homogenous film. The film was formed into a disc with a diameter of 13 mm and dried in a Büchi oven at 100 °C under dynamic vacuum ($10^{-2}$ mbar) for 12 h before transferring into an argon filled glovebox. A customized electrochemical cell equipped with a Be window was used to prepare cells for *in situ* XRD. The 70 μL of LP57 electrolyte (1 M $LiPF_6$, ethylene carbonate (EC)/ethyl methyl carbonate (EMC) 3/7, SoulBrain MI) was added to the film followed by 16 mm glass fiber separator. Battery cycling was conducted using a Land cycler at room temperature between open circuit voltage and 0.005 V at C/34 rate (34 hours for a full charge/discharge). *In situ* XRD data were collected at 300 K on a Panalytical Empyrean diffractometer equipped with a Ni filter using Cu–Kα radiation (λ = 1.5406 Å) in Bragg-Bentano geometry.

**Electrochemical characterization.** For electrochemical cycling of thin films in a pouch cell, the samples were deposited on non-conducting LSAT substrates giving highly oriented single-crystalline films. For the current collector, Au was deposited/patterned on the surface of the thin film. The Cu tab was connected on the gold pattern (Supplementary Fig. 7). The other components of cell, anode, electrolyte, and separator were the same as those used for the *in situ* powder XRD experiments. The C-rate was defined based on a 201.7 mAh/g, i.e., 1 electron reduction per $Nb_2O_5$. For cyclic voltammetry experiments, three different scan rates of 0.1, 0.5, and 1.0 mV/s, were used. Galvanostatic charge discharge was also performed with current densities of 14.28, 7.14, 3.57, and 1.43 A/g. The thin film capacity was determined from the dimensions (surface × thickness) and the specific capacity of $T$-$Nb_2O_5$ (175 mAh/g). The film thickness was determined by X-ray reflectivity measurements (Supplementary Fig. 2b) and further confirmed by TEM measurements. The electroactive surface is considered to be delimited by the Au pattern (0.12 $mm^2$).

**DFT calculations.** The unit cell of un-lithiated [$T$-$Nb_2O_5$] ($Li_4Nb_{16}O_{42}$ in our model) is described in Supplementary Section 3.2. We modeled the diffusion of one Li-ion in the unit cell of $T$-$Nb_2O_5$ by the nudged elastic band method with climbing image (CI-NEB). The free-energy profile of the most favorable diffusion pathway has been shown in Fig. 3g.

The differential binding energy ($\Delta E_b$) of each Li interstitial reaction in the [$T$-$Nb_2O_5$] model (as shown in Fig. 4g) was calculated as:

$$\Delta E_b = E(\mathrm{Li}_x\text{-}[T\text{-}\mathrm{Nb_2O_5}]) - E(\mathrm{Li}_{x'}\text{-}[T\text{-}\mathrm{Nb_2O_5}]) - 0.5(x\text{-}x') \times n_{\mathrm{Nb}}\, E(\mathrm{Li}) \quad (n_{\mathrm{Nb}}= 16.8 \text{ or } 67.2)$$

where $n_{\mathrm{Nb}}$ indicates the number of Nb atoms in the cell, and $0.5(x\text{-}x') \times n_{\mathrm{Nb}}$ is the number of extra Li atoms intercalated into the material. $E$ indicates the DFT energies of the compositions calculated from *ab initio* studies. The reference energy of $E$(Li) is calculated from bulk Li metal. The energy of [$T$-$Nb_2O_5$] is calculated from our unit cell model of $Li_4Nb_{16}O_{42}$ and super cell of $Li_{16}Nb_{64}O_{168}$, respectively. All possible interstitial sites for each Li atom were considered and the site providing the most negative $\Delta E_b$ is the most energetically favorable location for that atom (these sites are shown in Supplementary Fig. 20). Such arrangement of interstitial Li is iteratively used as a starting configuration for the higher concentration simulations. More detailed computational methods are shown in Supplementary Section 3.1.

**Data availability.** The main data that support the results of this study are available within this Article and Supplementary Information.

**References_Methods**

# Supplementary Information

## Table of Contents

## 1. Structural analysis of pristine *T*-$Nb_2O_5$ thin films

Niobium pentoxide ($Nb_2O_5$) exhibits many polymorphs such as *TT*-, *T*-, *M*-, *H*-, *B*-, *N*-, *R*-$Nb_2O_5$[1]. Among them, *T*-$Nb_2O_5$ (*T*: Tief in German) is one of the fastest $Li^+$-ion conductors found in oxides, featuring two-dimensional 4g atomic layers which provide exceptionally fast Li ion transport paths with very low steric hindrance (Supplementary Fig. 1). The single *T*-$Nb_2O_5$ (180) thin film is obtained at the growth temperatures between 600 and 650 ˚C by using PLD (Supplementary Fig. 2). Phi scans of the films grown on (001)-oriented substrates show multi-domains with 4-fold symmetry that the domains are in-plane rotated 90˚ each other (Supplementary Fig. 4b), whereas those of films grown on (110)-oriented substrates reveal 2-fold symmetry (Supplementary Fig. 4d). The substrate-orientation dependent domain structures are consistent with the reflection high-energy electron diffraction (RHEED) patterns (Supplementary Fig. 2b and 2c) and STEM-HADDF images (Supplementary Fig. 3). Namely, the film grown on LSAT (110) shows anisotropic in-plane diffraction patterns of *T*-$Nb_2O_5$ $[8\bar{1}0]$ and $[001]$ which are rotated 90˚ each other, while the film grown on LAO (001) exhibits the mixed diffraction patterns of *T*-$Nb_2O_5$ $[8\bar{1}0]$ and $[001]$.

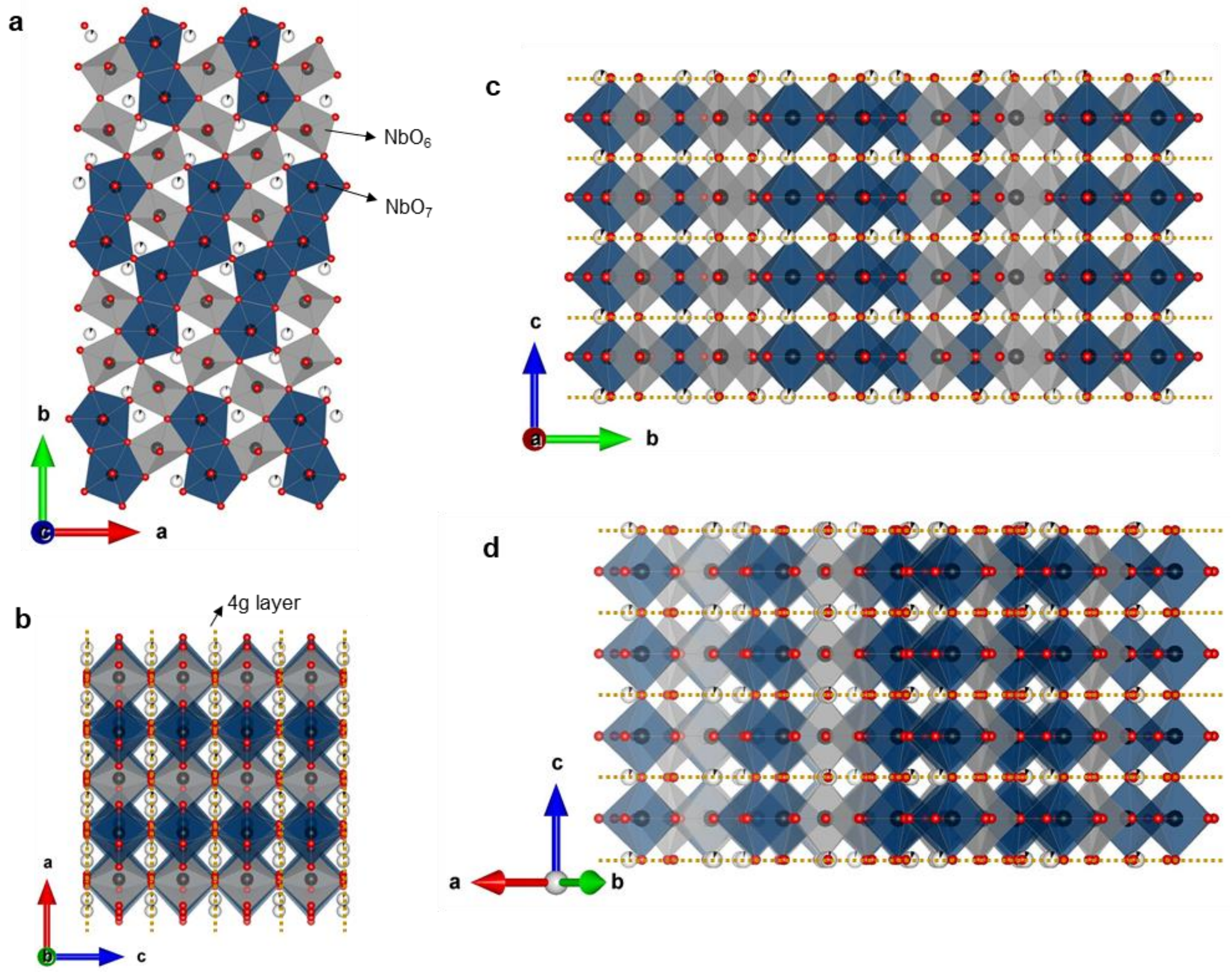

**Supplementary Fig. 1. Crystal structure of *T*-$Nb_2O_5$.** Schematic structures viewed along **a**, [001], **b**, [010], **c**, [100], and **d**, [180]. The (partially filled) navy and red spheres represent (partially occupied) Nb and O atoms, respectively. The grey and navy polyhedra denote distorted octahedra ($NbO_6$) and pentagonal bipyramids ($NbO_7$), respectively. The dashed dark yellow lines represent loosely packed 4g atomic layers.

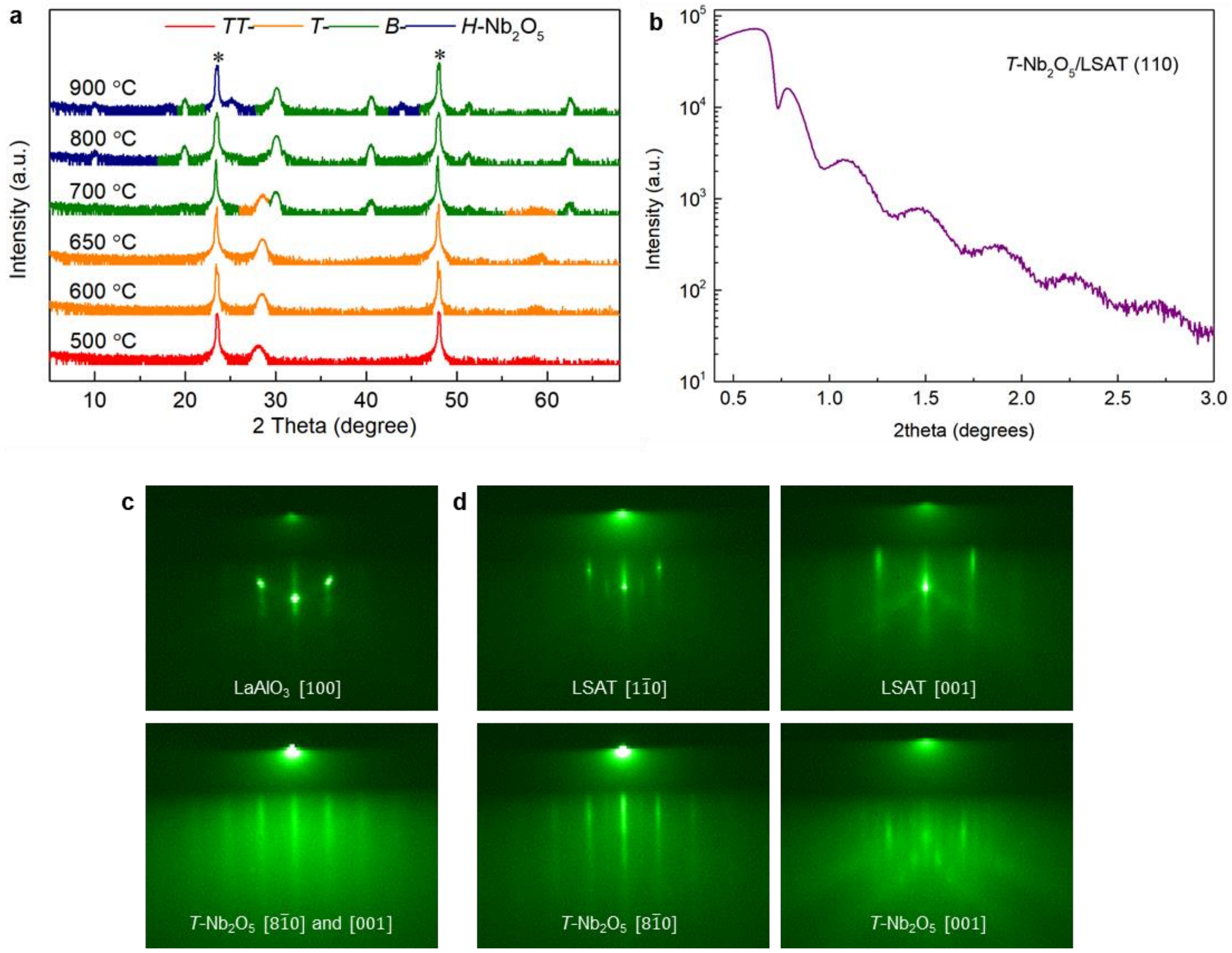

**Supplementary Fig. 2. Optimization of epitaxial *T*-$Nb_2O_5$ thin film growth and RHEED patterns. a**, Theta-2theta XRD scans with various growth temperatures grown on LAO (001) substrates. The results reveals many $Nb_2O_5$ polymorph formations such as *TT*-, *T*-, *B*-, and *H*-$Nb_2O_5$. Nonetheless, the growth temperatures between 600 and 650 °C reveal single *T*-$Nb_2O_5$ phase formation. **b**, X-ray reflectivity (XRR) of a *T*-$Nb_2O_5$ thin film grown on a LSAT (110) substrate at 620 °C. The film thickness of ≈16 nm was obtained from the periodicity of intensity oscillations. RHEED patterns of *T*-$Nb_2O_5$ thin films grown on **c**, LAO (001) and **d**, LSAT (110) substrates at the growth temperature of 650 °C. The top and bottom images represent before and after deposition, respectively. The film grown on the LSAT (110) substrate shows anisotropic in-plane diffractions of *T*-$Nb_2O_5$ $[8\bar{1}0]$ and $[001]$. While, the film grown on the LAO (001) substrate exhibits a mixture of *T*-$Nb_2O_5$ $[8\bar{1}0]$ and $[001]$ diffractions, indicating existence of multi-domains with 90° rotation each other.

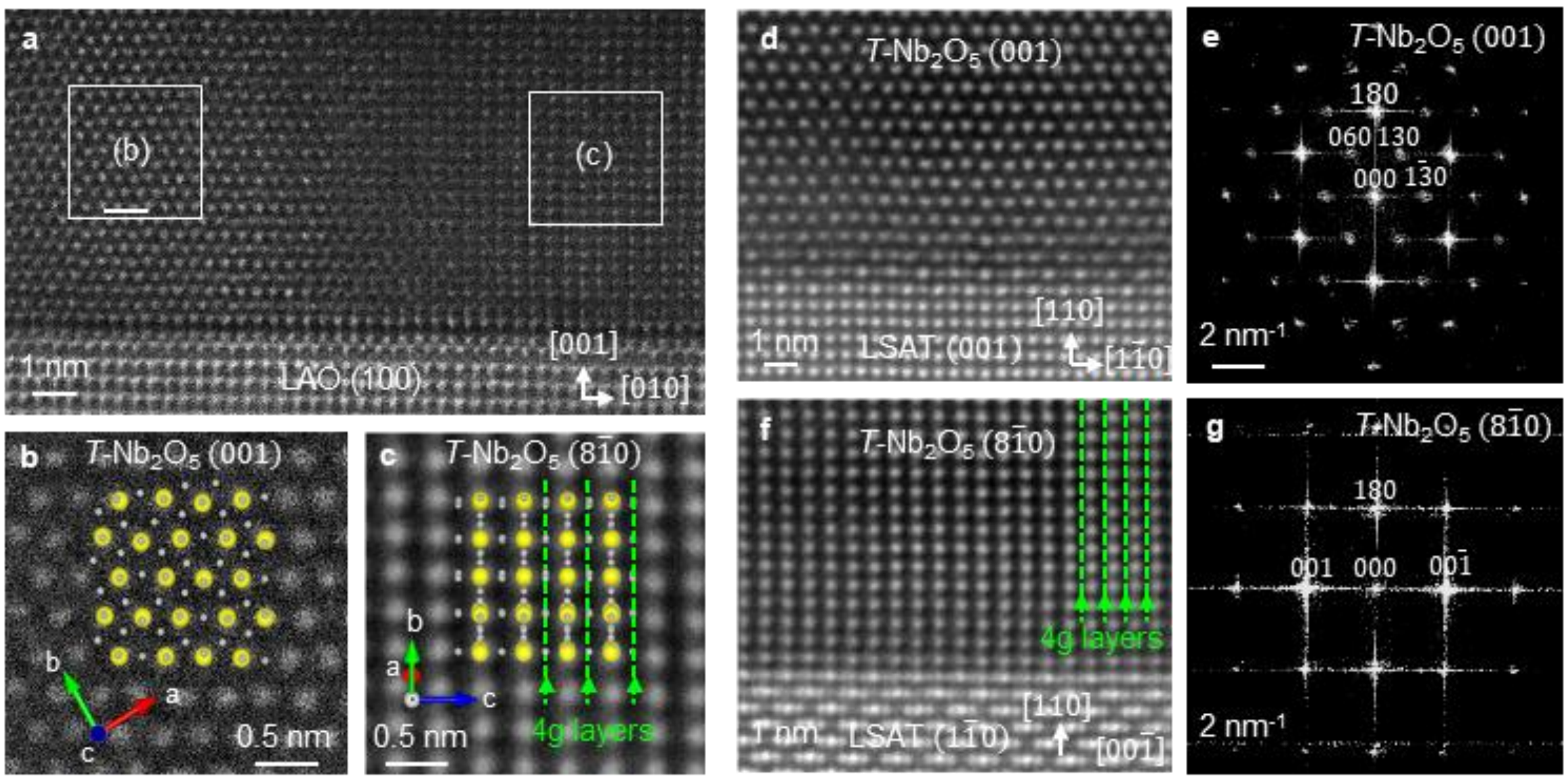

**Supplementary Fig. 3. Cross-sectional HAADF-STEM images of *T*-$Nb_2O_5$ thin films grown on LAO (001) and LSAT (110) substrates.** **a**, Cross-sectional HAADF-STEM image of a *T*-$Nb_2O_5$ thin film grown on a LAO (001) substrate viewed along LAO [100], revealing the presence of multi-domains. **b-c,** Magnified STEM images of each domain labelled in (**a**), showing domains of *T*-$Nb_2O_5$ [001] and [8$\bar{1}$0], respectively. Overlaid yellow and grey spheres represent Nb and O atoms, respectively. Green dashed lines represent the vertical ionic transport channels (4g layers) viewed from *T*-$Nb_2O_5$ [8$\bar{1}$0]. **d**, a STEM image and **e**, FFT pattern of *T*-$Nb_2O_5$ thin film grown on a LSAT (110) substrate viewed along the LSAT [001]. **f**, a STEM image and **g**, FFT pattern of *T*-$Nb_2O_5$ thin film grown on a LSAT (110) substrate viewed along the LSAT [1$\bar{1}$0].

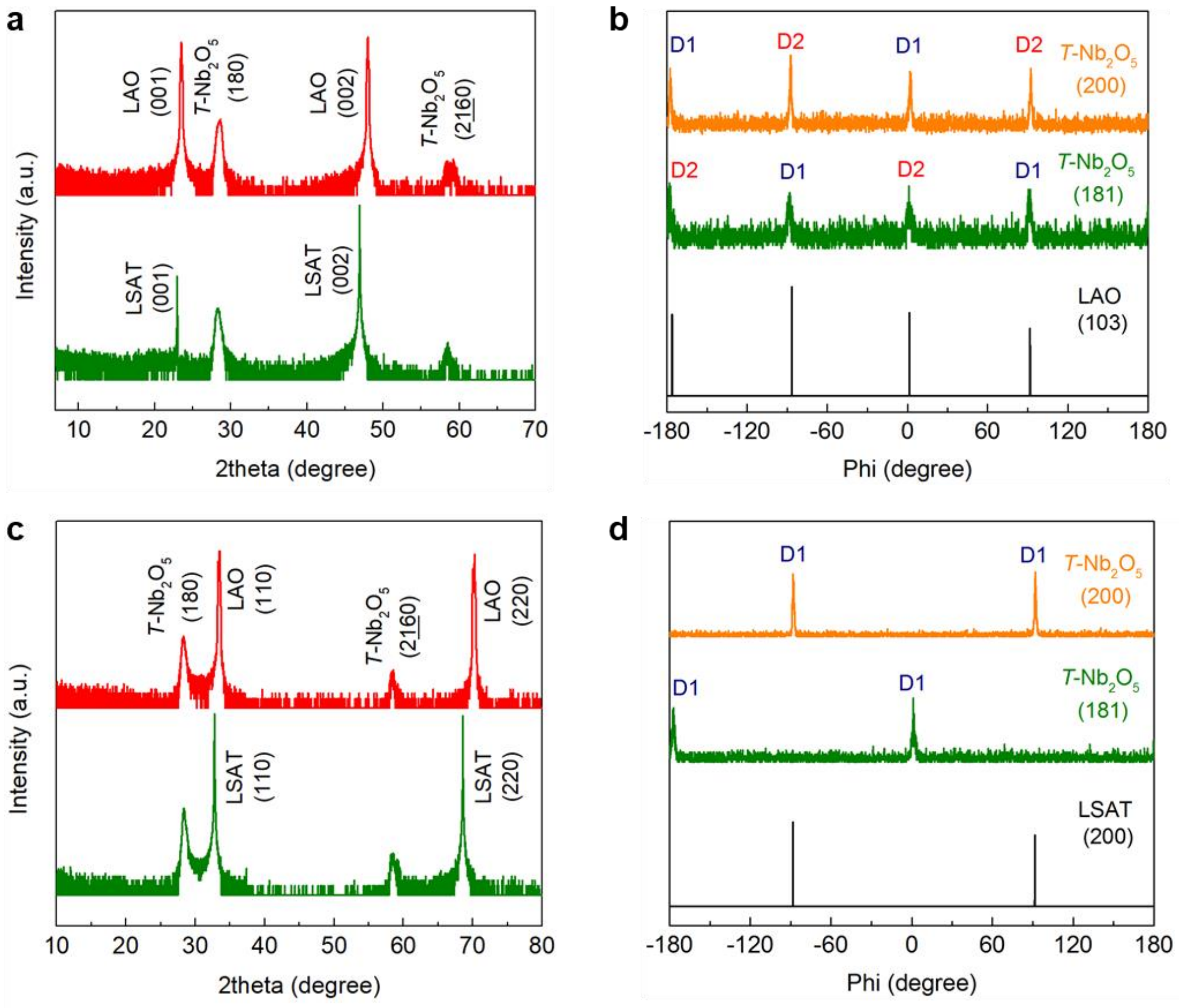

**Supplementary Fig. 4. XRD characterizations of *T*-$Nb_2O_5$ thin films and the schematic film structures.** **a**, Theta-2theta XRD scans of *T*-$Nb_2O_5$ thin films grown on (001)-oriented substrates such as LAO (001) and LSAT (001) substrates. **b**, Phi-scan of thin films grown on a LAO (001) substrate. The 4-fold symmetry of *T*-$Nb_2O_5$ (200) and (181) reflections represent the presence of multi-domains rotated 90° to each other. **c**, Theta-2theta XRD scans of *T*-$Nb_2O_5$ thin films grown on (110)-oriented substrates, including LAO (110) and LSAT (110) substrates. **d**, Phi-scan of thin films grown on a LSAT (110) substrate. Theta-2theta scans exhibit that all films are oriented along out-of-plane (180), and the 2-fold symmetry from phi scans of both (200) and (181) reflections are revealed.

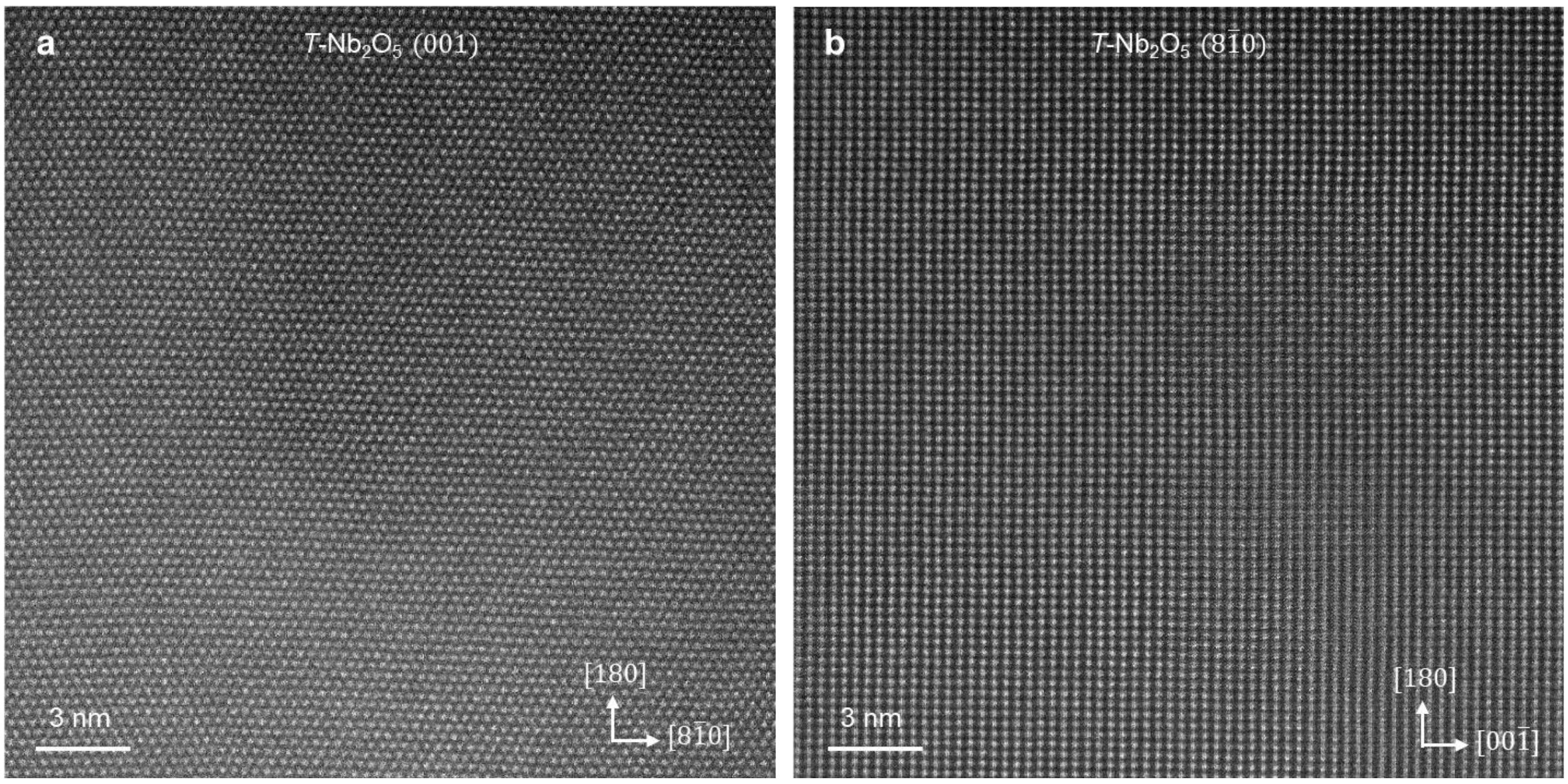

**Supplementary Fig. 5. Low magnification HAADF-STEM images of single-crystalline *T*-$Nb_2O_5$ thin films grown on a LSAT (110) substrate.** Cross-sectional STEM images of *T*-$Nb_2O_5$ thin film viewed along the **a**, *T*-$Nb_2O_5$ [001] and **b**, [8$\bar{1}$0] direction, respectively. The images reveal the single-crystalline structure.

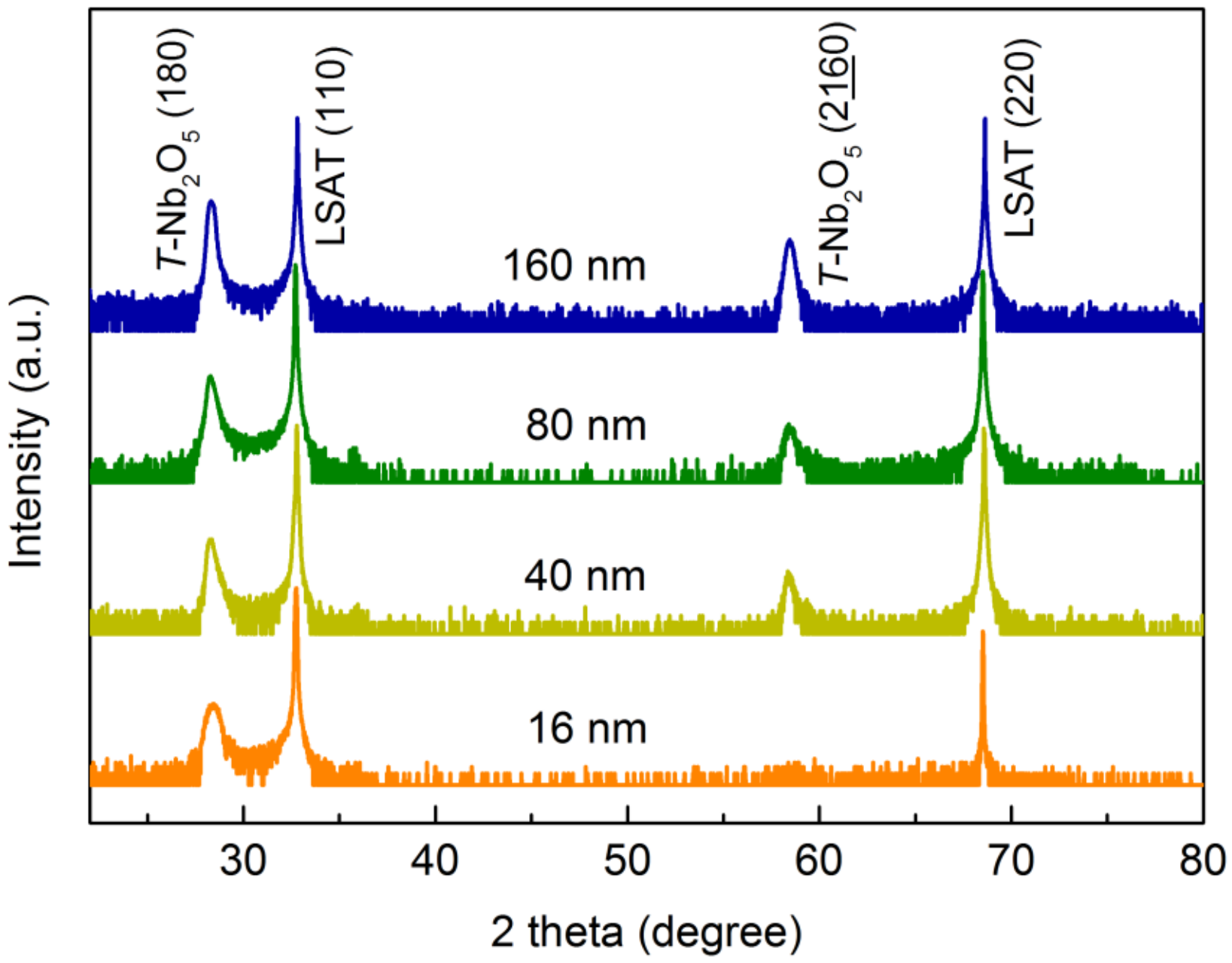

**Supplementary Fig. 6. Thickness-dependent theta-2theta XRD scans of *T*-$Nb_2O_5$ thin films grown on LSAT (110) substrates.** The XRD results show good crystallinity even at the high thickness of 160 nm.

## 2. *In situ* and *ex situ* structural analysis via lithiation

### 2.1. Monoclinic distortion of *T*-$Nb_2O_5$ thin films via lithiation

In Supplementary Fig. 8, *ex situ* XRD on the pristine and gated *T*-$Nb_2O_5$ thin films were carried out using a Gallium-metal jet x-ray source ($\lambda$=1.34 Å) and a six-circle diffractometer equipped with a Pilatus 100 K pixel detector. Reciprocal space maps (RSM) were collected under grazing incidence of the incoming beam using one set of lattice parameters for the pristine and gated sample in order to provide a relative measurement of the gating-induced modifications of the lattice of the film. We used the orthorhombic lattice characterized by a=6.12 Å, b=29.75 Å, and c= 3.89 Å and $\alpha=\beta=\gamma=90°$. This lattice was used throughout the analysis, i.e. before and after gating. Therefore, we are studying relative modifications of the lattice, most importantly the change of the angle $\gamma$ between the a- and the c-axis. Note that, in our setting of the film lattice this corresponds to the monoclinic angle $\beta$ used in the analysis of the bulk crystal.

The effect of gating can be directly seen positions, which indicates peak shifts in the range of up to several tens of reciprocal lattice units. The analysis was carried out by calculating the length of the scattering vectors and the angles between them. RSMs were collected in the vicinity of the (200) and the (0$\underline{16}$0) reflections for the pristine (p) and the gated (g) sample. Reflection indices are given with respect to the same lattice parameters. The inset in (a) sketches a schematic emphasizing the changes of the positions of the scattering vectors. For the (200) and the (0$\underline{16}$0) reflection we find rotations of the scattering vectors by +0.7 and -2.6 degrees which add up to 3.3 degrees by which the angle $\gamma^*$ between the reciprocal a* and b* axis decreases. In turn this corresponds to an increase of $\gamma$ by +3.3 degrees, i.e. the monoclinic angle is equal to 93.3 degrees, which is in reasonable agreement with the result derived for the bulk sample by powder x-ray diffraction.

In total we have analyzed several pairs of reflections where we find in general rotation of the scattering vector positions by up to 4 degrees, while simultaneously the magnitudes of the lattice vectors decrease between about 0.5 % and 3 % upon gating, which is related to the general lattice expansion.

### 2.2. Determination of the monoclinic structure for powder *T*-$Nb_2O_5$

In Supplementary Fig. 12 and Table 1, $Li_{1.2}Nb_2O_5$ and $Li_{1.6}Nb_2O_5$ (see methods for sample preparation) SXRD patterns could be indexed using two phases: a *Pbam* model similar to *T*- $Nb_2O_5$ together with a *P2/m* structural model built from a monoclinic (*m*-) distortion of *Pb*am with $\gamma \neq 90°$. Note, not all reflections could be indexed using a single unit cell. Rietveld refinement of the full SXRD patterns for $Li_{1.6}Nb_2O_5$ and $Li_{1.2}Nb_2O_5$ lead to a decent fit with some minor discrepancies on the relative intensities. Optimization of the atomic positions was not performed due to the large number Nb atoms, 8 and 16 Nb positions in the orthorhombic and monoclinic model respectively, together with overlapping reflections. Moreover, an anisotropic broadening fitted with a strain model is observed with narrow (00l) and large (hk0) reflections for both phases.

### 2.3. Determination of tetragonal structures for powder *T*-$Nb_2O_5$

In Supplementary Fig. 13 and Table 2, the structural model for the tetragonal (*t*-) phase was determined as follow. Unit cell was found using Dicvol. Space group was chosen to be *P4/mmm*, the highest symmetry space group. Nb and O positions were determined from simulated annealing as implemented in Fullprof, and Nb occupancies were refined leading to a Nb/O ratio of 0.42. Nb-O average distance is 2.1 Å, which is just slightly larger than the average Nb-O distance in $NbO_2$, 2.05 Å, ($Nb^{4+}$) suggesting a lower Nb oxidation state in this compound ($n<4$, with $Nb^{n+}$), consistant with its formation at low voltage. Due to the low scattering power of Li, the presence of Li in the structure is unknown. However, based on the Nb/O and the suggested oxidation state, we hypothesize the presence of Li. Completely filling the remaining octahedral sites with Li would lead to the composition $Li_{1.15}Nb_{0.85}O_2$ with a Nb average oxidation state of +3.35 which is possible, although further investigations are necessary to test this hypothesis.

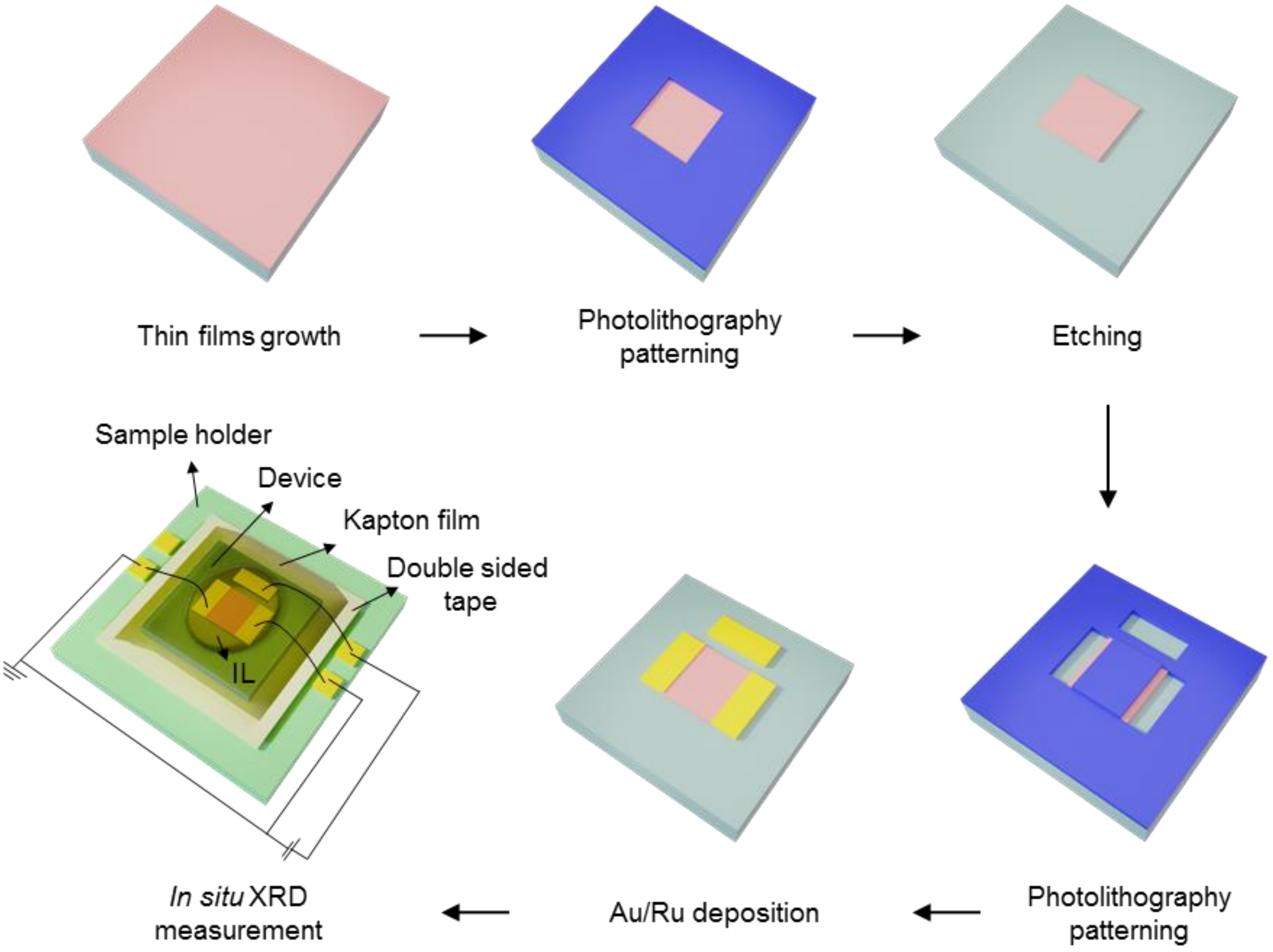

**Supplementary Fig. 7. Schematics of device fabrications for *in situ* thin film XRD.** The *T*-$Nb_2O_5$ channel with a size of 2 × 2 $mm^2$ was etched, then Au (70nm)/Ru (5 nm) layers were deposited for the gate and contact electrodes. The device was attached using double-side tape on a specially designed sample holder. The IL was placed on the device surface, and then Kapton film is attached to thin the IL thickness.

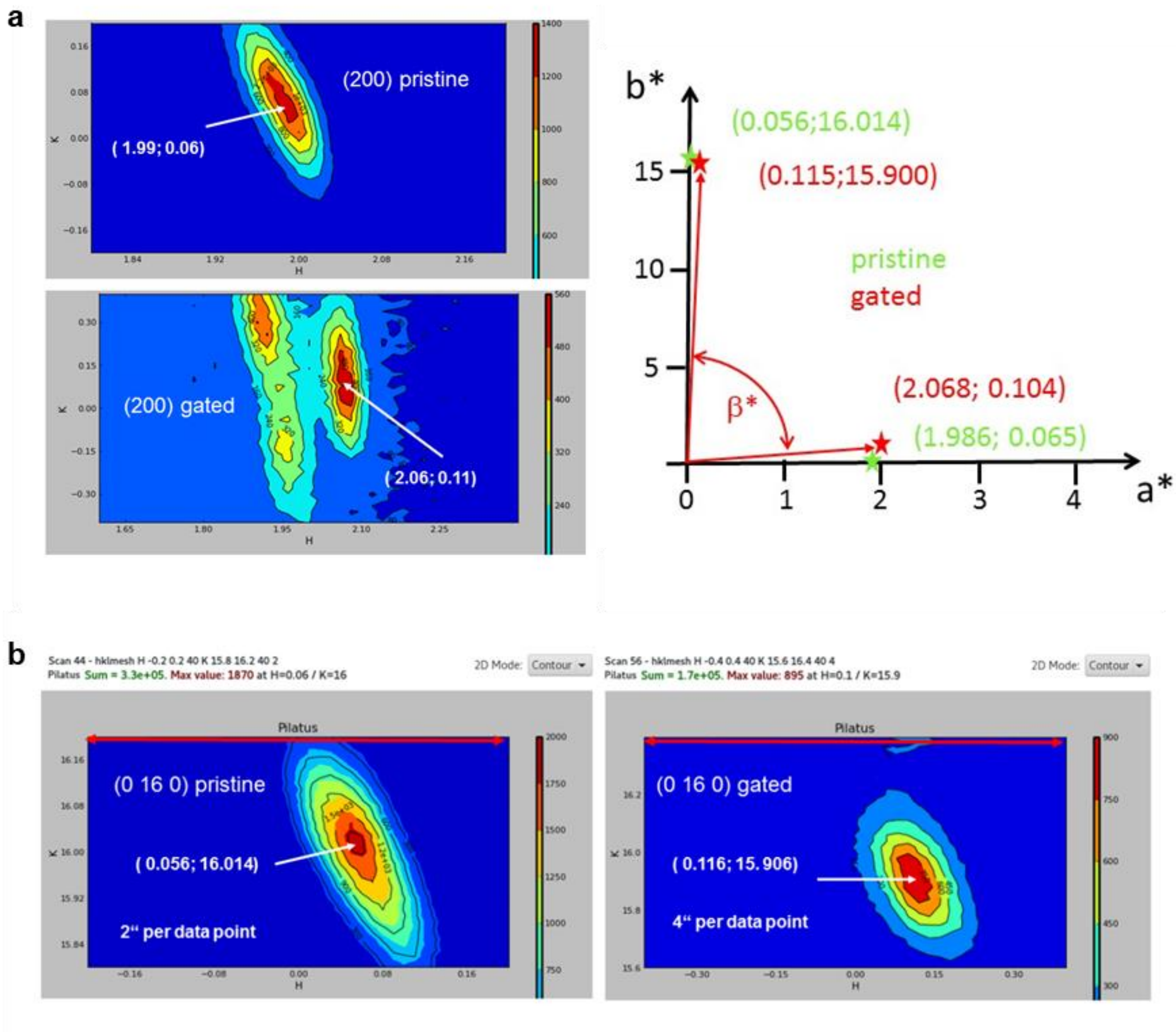

**Supplementary Fig. 8. Reciprocal space mappings (RSM) of 100 nm thick *T*-$Nb_2O_5$/LSAT(110) using GaJet X-Ray source (λ=1.3414 Å) XRD.** RSM of the pristine and gates samples for the **a,** (200) and **b,** $(0\underline{16}0)$ reflections. The positions of the peaks are indicated. The inset shows a schematic of the reflection positions for the pristine (p) and the gated (g) sample. After gating, the scattering vectors of the (200) and the $(0\underline{16}0)$ reflection are rotated by +0.7 and -2.6 degrees indicating a relative modification of the (reciprocal) angle γ* by 3.3 degrees.

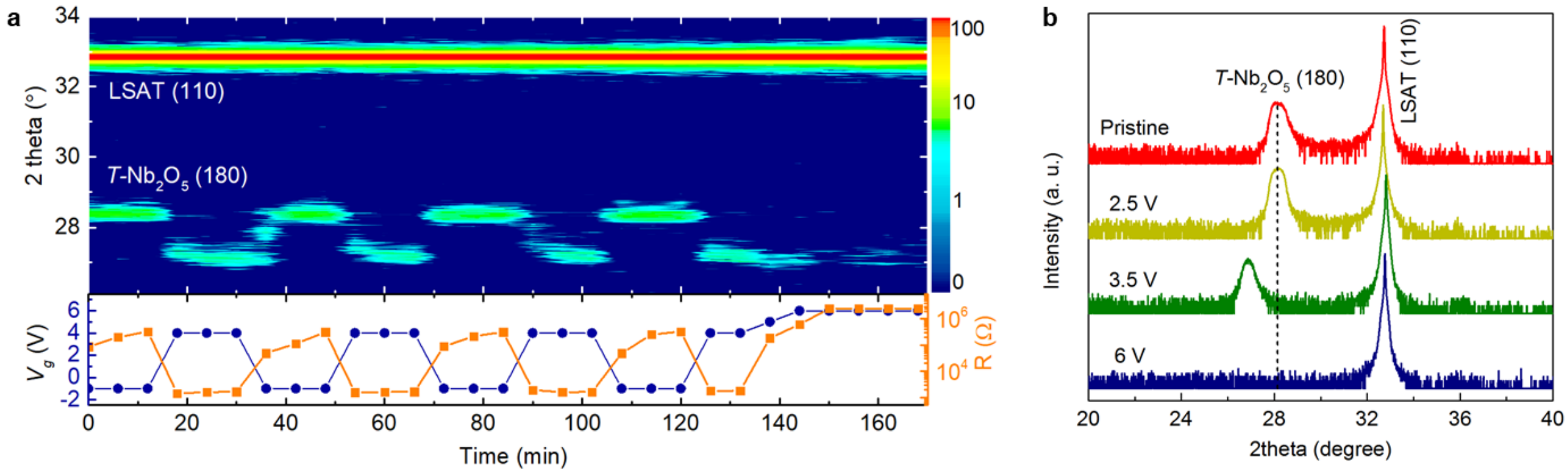

**Supplementary Fig. 9. *In situ* and *ex situ* XRD via Li-ionic liquid gating. a**, The *in situ* XRD measurements during Li-ionic liquid gating. The reversible structural and electronic property changes between the orthorhombic insulator and monoclinic metal are observed. The structure is collapsed after 6 V gating. 50 nm thick *T*-$Nb_2O_5$/LSAT (110) device is used. **B**, *ex situ* XRD with different gate voltage. 100 nm thick *T*-$Nb_2O_5$/LSAT (110) film is gated. The 2.5 V gated film does not show structural changes, while 3.5 V gated film shows the (180) peak shift, indicating a transition from orthorhombic to monoclinic phase. The 6 V gated film show the disappearance of the *T*-$Nb_2O_5$ (180) peak.

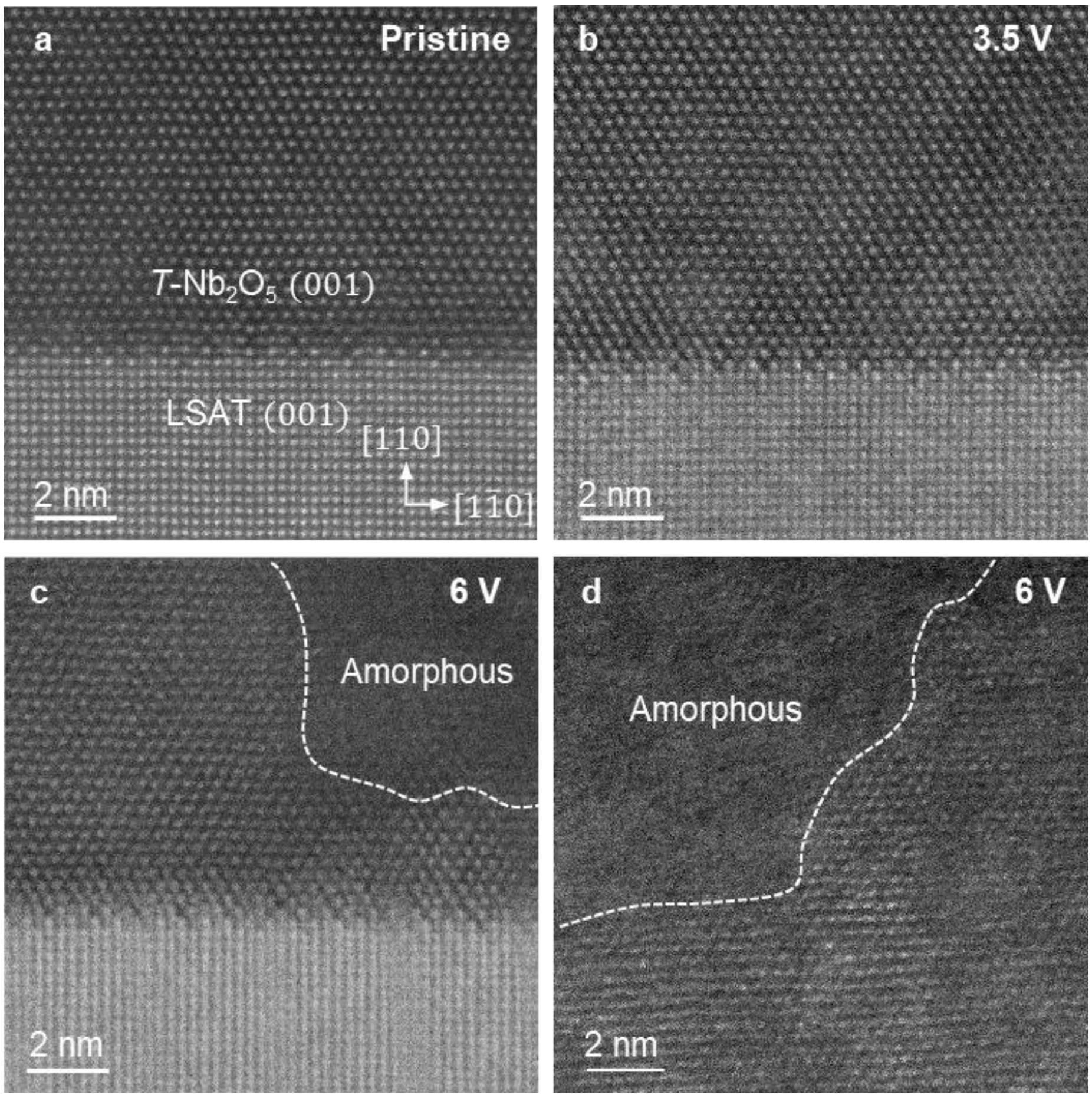

**Supplementary Fig. 10. Cross-sectional *ex situ* STEM-HAADF images of pristine and gated *T*-$Nb_2O_5$/LSAT(110). a**, pristine, **b**, 3.5 V, and **c, d** 6 V gated films. The 3.5 V gated film does not show noticeable changes of structures from the STEM image, while the 6 V gated film shows partially amorphous structures. The film thicknesses are 30 nm.

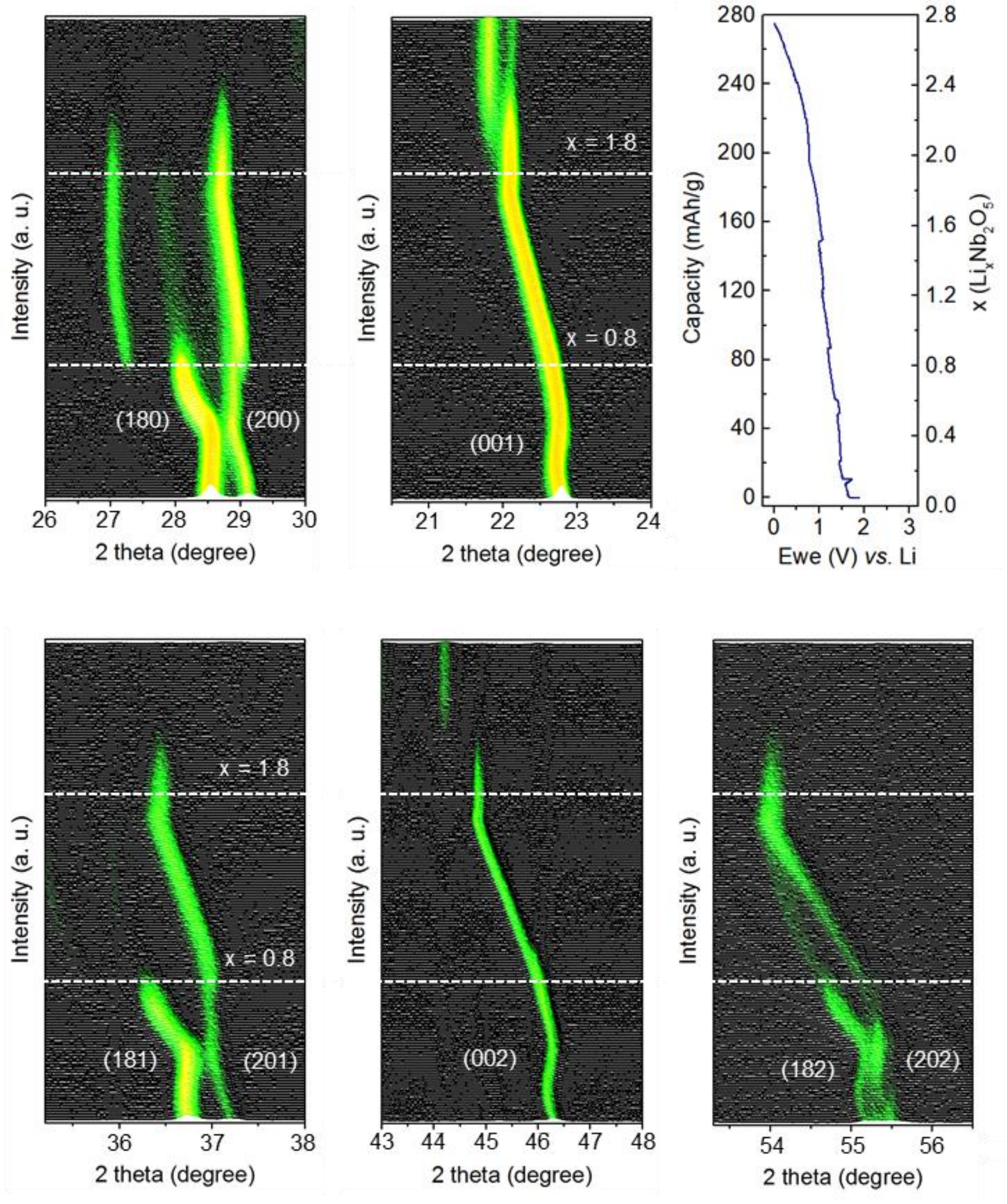

**Supplementary Fig. 11. *In situ* XRD data for bulk powder *T*-$Nb_2O_5$.** Monoclinic distortion starts at x = ≈0.8 (x in $Li_xNb_2O_5$), and the tetragonal structure shows from x = ≈1.8. The measurement was done until the deep discharge potential of 0.005 V vs Li with C/34 scan rate for electrochemical cycling. The upper right panel shows the corresponding discharge curve.

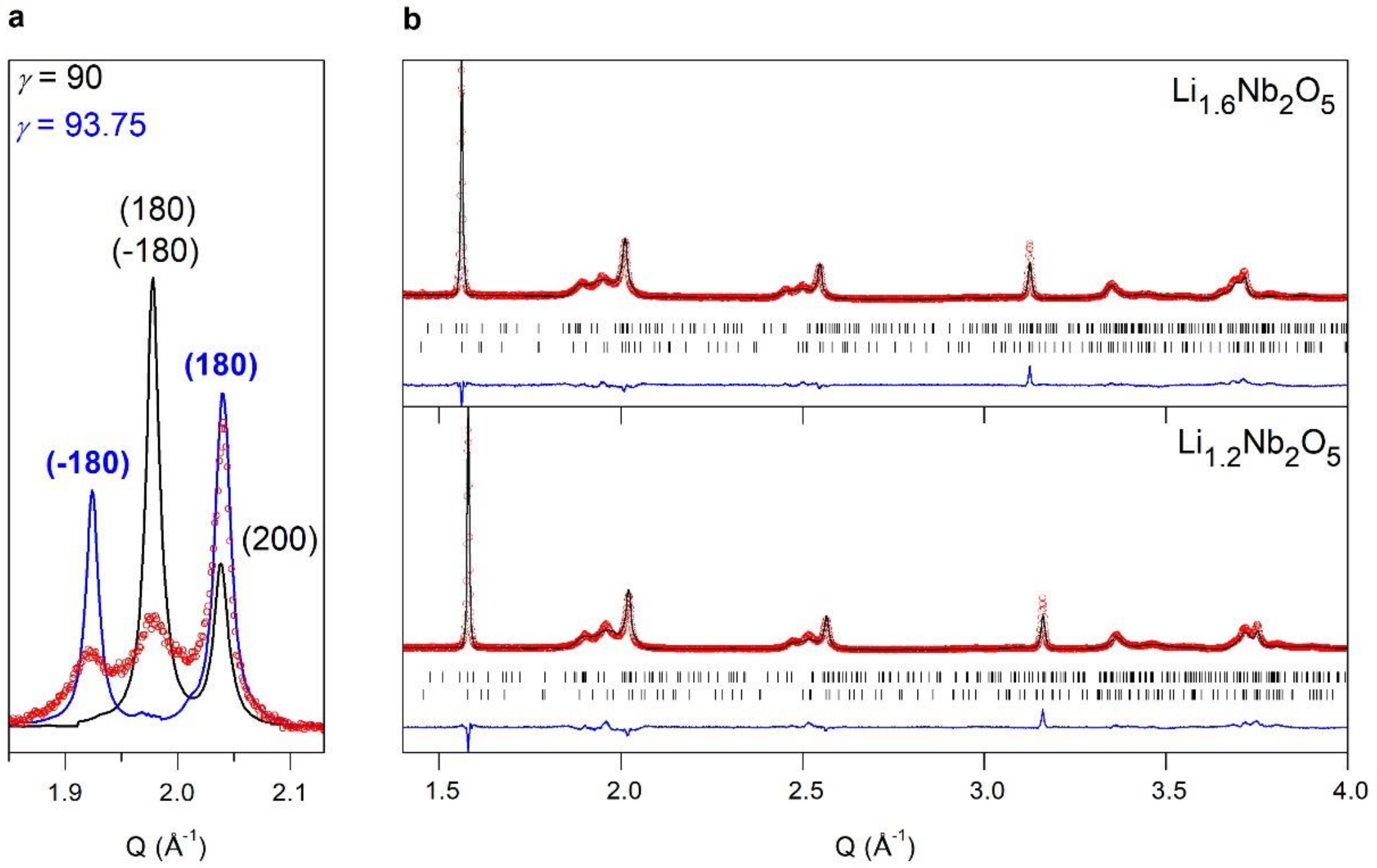

**Supplementary Fig. 12.** **a**, SXRD patterns of *ex situ* $Li_{1.6}Nb_2O_5$ together with simulated patterns of the *Pbam* and *P2/m*, in black and blue, respectively. **b**, SXRD patterns and Rietveld refinements of $Li_{1.2}Nb_2O_5$ and $Li_{1.6}Nb_2O_5$. Red circles, black line and blue line represent the observed, calculated and difference patterns, respectively.

**Supplementary Table 1.** Crystallographic table for the monoclinic phase in the SXRD pattern of $Li_{1.6}Nb_2O_5$. Li positions are omitted because unknown due to the low scattering power of Li compared to Nb and O.

***m*-$Li_{1.6}Nb_2O_5$ (*P2/m*): *a* = 6.262(4) Å, *b* = 30.074(12) Å, *c* = 4.0180(4) Å, *γ* = 94.13(2) °**

| | x | y | z | Occ | Biso (Å$^2$) | multiplicity |
|---|---|---|---|---|---|---|
| **O1_1** | 0.2823 | 0.0388 | 0 | 1 | 1 | 2 |
| **O1_2** | 0.7823 | 0.4612 | 0 | 1 | 1 | 2 |
| **Nb2_1** | 0.1857 | 0.1537 | 0.5477 | 1 | 0.5 | 4 |
| **Nb2_2** | 0.6857 | 0.3463 | 0.4523 | 1 | 0.5 | 4 |
| **Nb3_1** | 0.2522 | 0.27586 | 0.547 | 1 | 0.5 | 4 |
| **Nb3_2** | 0.7522 | 0.22414 | 0.453 | 1 | 0.5 | 4 |
| **Nb4_1** | 0.2376 | 0.40702 | 0.5543 | 1 | 0.5 | 4 |
| **Nb4_2** | 0.7376 | 0.09298 | 0.4457 | 1 | 0.5 | 4 |
| **Nb5_1** | 0.054 | 0.338 | 0 | 1 | 0.08 | 2 |
| **Nb5_2** | 0.554 | 0.162 | 0 | 1 | 0.08 | 2 |
| **Nb6_1** | 0.434 | 0.4776 | 0 | 1 | 0.08 | 2 |
| **Nb6_2** | -0.066 | 0.0224 | 0 | 1 | 0.08 | 2 |
| **Nb7_1** | 0.435 | 0.213 | 0 | 1 | 0.04 | 2 |
| **Nb7_2** | -0.065 | 0.287 | 0 | 1 | 0.04 | 2 |
| **Nb1_1** | 0.2807 | 0.03551 | 0.5445 | 1 | 0.5 | 4 |
| **Nb1_2** | 0.7807 | 0.46449 | 0.4555 | 1 | 0.5 | 4 |
| **O1_1** | 0.2823 | 0.0388 | 0 | 1 | 1 | 2 |
| **O1_2** | 0.7823 | 0.4612 | 0 | 1 | 1 | 2 |
| **O2_1** | 0.19 | 0.1562 | 0 | 1 | 1 | 2 |
| **O2_2** | 0.69 | 0.3438 | 0 | 1 | 1 | 2 |
| **O3_1** | 0.246 | 0.2782 | 0 | 1 | 1 | 2 |
| **O3_2** | 0.746 | 0.2218 | 0 | 1 | 1 | 2 |
| **O4_1** | 0.262 | 0.414 | 0 | 1 | 1 | 2 |
| **O4_2** | 0.762 | 0.086 | 0 | 1 | 1 | 2 |
| **O5_1** | 0.063 | 0.0879 | 0.5 | 1 | 1 | 2 |
| **O5_2** | 0.563 | 0.4121 | 0.5 | 1 | 1 | 2 |
| **O6_1** | 0.07 | 0.2199 | 0.5 | 1 | 1 | 2 |
| **O6_2** | 0.57 | 0.2801 | 0.5 | 1 | 1 | 2 |
| **O7_1** | 0.3423 | 0.3441 | 0.5 | 1 | 1 | 2 |
| **O7_2** | 0.8423 | 0.1559 | 0.5 | 1 | 1 | 2 |
| **O8_1** | 0.1166 | 0.4688 | 0.5 | 1 | 1 | 2 |
| **O8_2** | 0.6166 | 0.0312 | 0.5 | 1 | 1 | 2 |
| **O9_1** | 0.436 | 0.106 | 0.5 | 1 | 1 | 2 |
| **O9_2** | -0.064 | 0.394 | 0.5 | 1 | 1 | 2 |
| **O10_1** | 0.456 | 0.1964 | 0.5 | 1 | 1 | 2 |
| **O10_2** | -0.044 | 0.3036 | 0.5 | 1 | 1 | 2 |
| **O11_1** | 0 | 0 | 0.5 | 1 | 0.25 | 1 |
| **O11_2** | 0.5 | 0.5 | 0.5 | 1 | 0.25 | 1 |

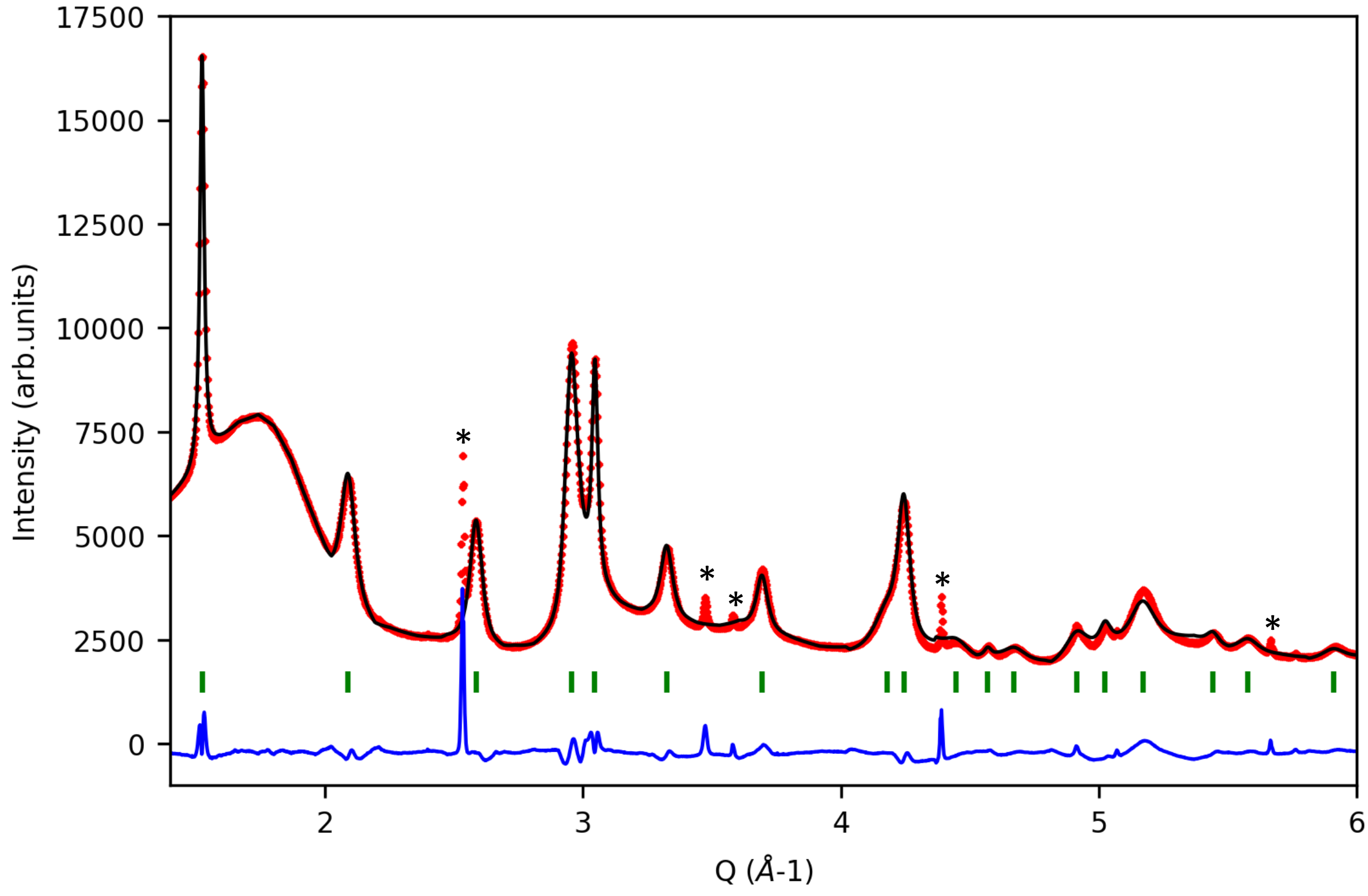

**Supplementary Fig. 13.** Rietveld refinement of an *operando* SXRD pattern done at 5 mV with the tetragonal phase described in Supplementary Table 1. The *operando* cell produces a large background together with peaks marked *. Red circles, black line and blue line represent the observed, calculated and difference patterns, respectively. Green vertical bars represent the Bragg position

**Supplementary Table 2.** Crystallographic table for the *t*-$Li_{2.5}Nb_2O_5$. Possible Li positions are omitted because unknown due to the low scattering power of Li compared to Nb and O.

***t*-$Li_{2.5}Nb_2O_5$ (*P4/mmm*): *a* = 3.0073(2) Å, *c* = 4.1249(3) Å**

| | x | y | z | Occ | Biso (Å²) | multiplicity |
|---|---|---|---|---|---|---|
| **O1** | 0 | 0 | ½ | 1 | 1 | 4 |
| **O2** | ½ | ½ | 0 | 1 | 1 | 4 |
| **Nb1** | 0 | 0 | 0 | 0.69(2) | 1 | 4 |
| **Nb2** | ½ | ½ | ½ | 0.151(2) | 1 | 4 |

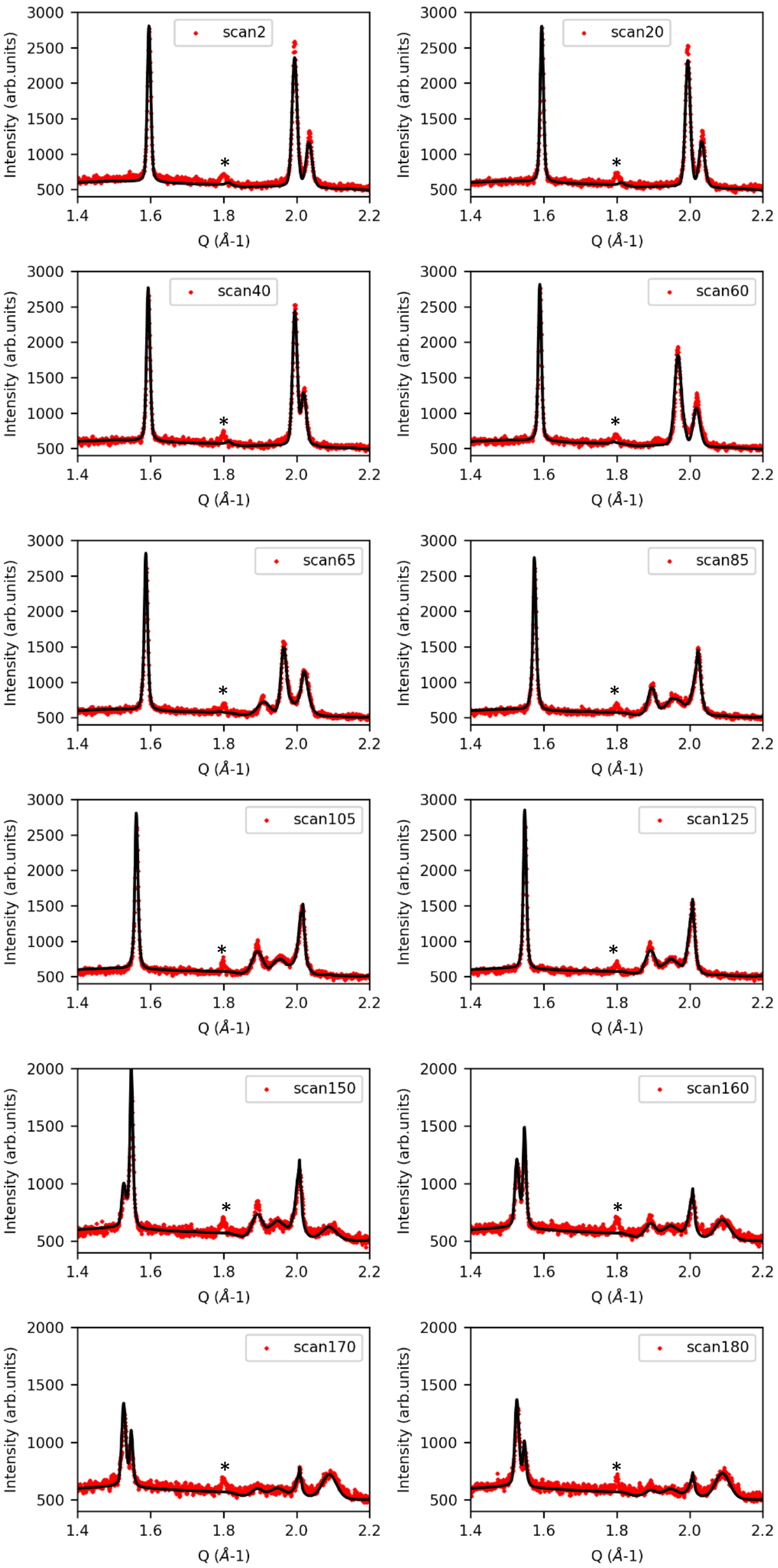

**Supplementary Fig. 14. Rietveld refinements of the *operando* XRD on powder *T*-$Nb_2O_5$.** Scans 2-60 are refined with the orthorhombic model only. Scans 65-125 are refined with the orthorhombic and monoclinic models. Scans 150-180 are refined all models (orthorhombic, monoclinic and tetragonal). * comes from the sample holder. Red crosses and black line

represent the observed and calculated patterns corrected form the shift due to the sample height, respectively.

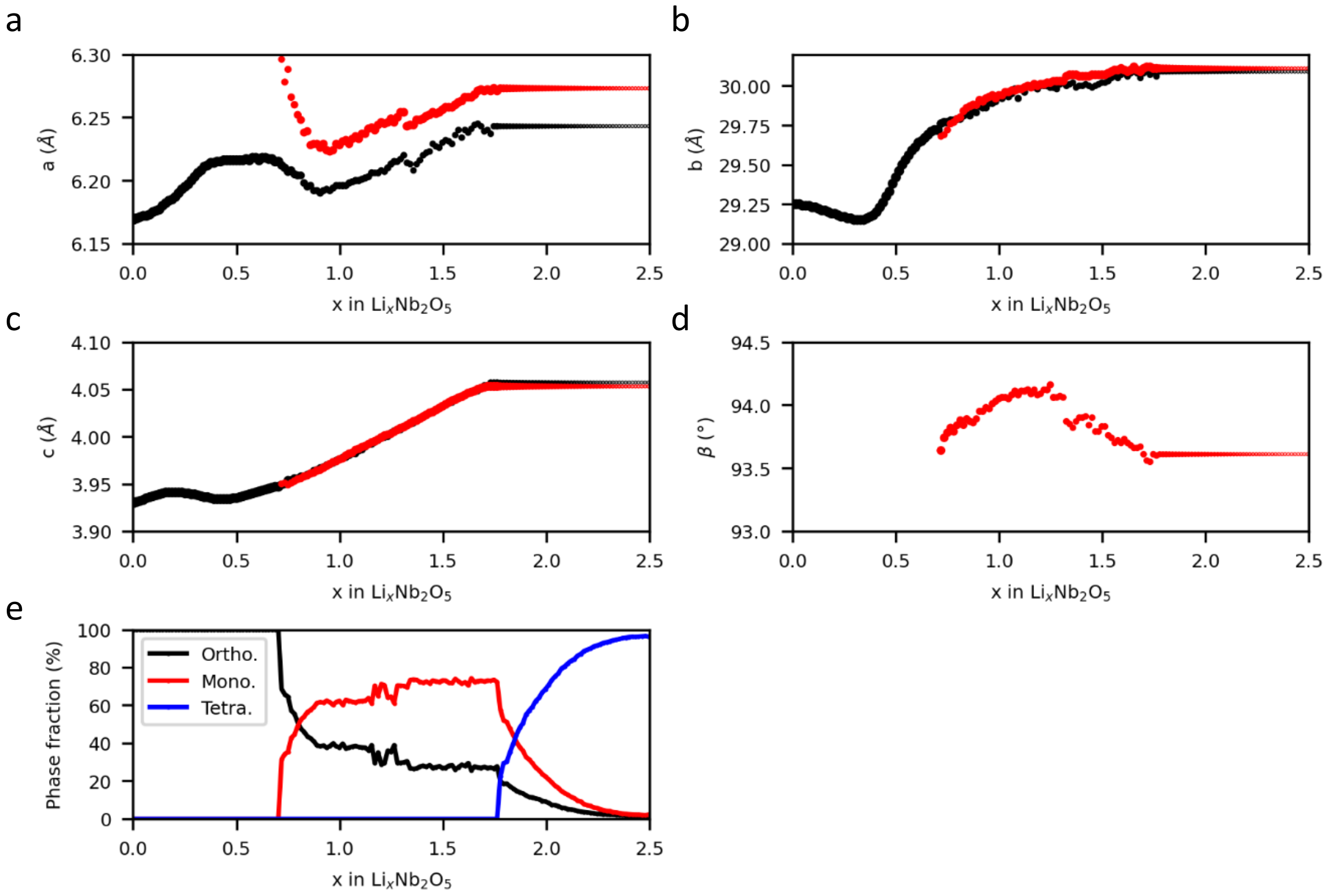

**Supplementary Fig. 15. a-d**, Cell parameters obtained from the Rietveld refinement for the orthorhombic and monoclinic phase in black and red, respectively. The marker size is proportional to the phase fraction. **e**, phase fraction obtained from the Rietveld refinement. X in $Li_xNb_2O_5$ is determined from the applied current and active material mass used during the *operando* XRD experiment.

## 3. Computational methods and results

### 3.1. Computational methods

First-principles calculations use density functional theory with a plane-wave basis set, as implemented in the QUANTUM ESPRESSO (QE) package[2]. Ultrasoft pseudopotentials[3] along with Perdew, Burke, Ernzerhof (PBE) exchange correlation (XC)[4] functional were used. Dispersion interactions are accounted for using the Grimme DFT-D3 method[5]. Phonon calculations are performed using the DFT-D2 scheme due to technical computational considerations in performing this type of calculations using QE. The DFT+$U$ method[6-8], with $U$=6.0 eV, is applied to account for d-electron localization on the Nb atoms[9]. The plane wave kinetic-energy cutoff for wave functions is set at 40 Ry, with the cutoff for charge density being ten times larger. The Brillouin zone is sampled using a 6×1×8 Monkhorst-Pack mesh[10] for the unit cell of un-lithiated [$T$-$Nb_2O_5$] ($Li_4Nb_{16}O_{42}$ in our model, see the introduction in Supplementary Section 3.2). The $2a \times b \times 2c$ and $a \times b \times 3c$ super cells of $T$-$Nb_2O_5$ are sampled with a 3×1×4 k-point mesh. We modeled the diffusion of one Li-ion in the unit cell of $T$-$Nb_2O_5$ by the nudged elastic band method with climbing image (CI-NEB)[11,12]. The free-energy profile of the most favorable diffusion pathway has been shown in Fig. 3g.

### 3.2. Introduction of our *T*-$Nb_2O_5$ simulation model

The conventional unit cell of $T$-$Nb_2O_5$ material (with a space group of *Pbam*, orthorhombic crystal system, shown in Supplementary Fig. 16a), is composed of 16.8 Nb and 42 O atoms in two layers. All the Nb sites are fractionally occupied, with 16 Nb atoms in 4h layer located at four 8i Wyckoff positions with half occupancy-corresponding to pairs of sites just above and below the 4h plane (Supplementary Fig. 16c). The remaining 0.8 Nb atoms per unit cell are randomly scattered into four sets of three different Wyckoff positions in the 4g layer with occupancies of 0.08, 0.08, and 0.04 (Supplementary Fig. 16b). The Nb atoms in the 4h layer coordinate into $NbO_6$ units (distorted octahedral) or $NbO_7$ units (pentagonal bipyramidal). Because of the disordered positions due to fractional occupancy, the computational treatment on the conventional unit cell of $T$-$Nb_2O_5$ structure is performed as follows: the pairs of sites in the 4h layer were consolidated into 16 Nb atoms in the 4h plane and allowed to relax, while the 0.8 Nb in the 4g layer were replaced by 4 Li atoms distributed in this layer for charge compensation. Thus, we provide a new simulation model of $Li_4Nb_{16}O_{64}$ (Fig. 3b) which we denote as [$T$-$Nb_2O_5$] model with the relaxed lattice parameters as shown. Through the density of states (DOS) analysis (Supplementary Fig. 17b), our simulation model (Fig. 3c) shows a larger band gap (2.3 eV) than that (1.8 eV) of the previous theoretical model (eliminating 0.8 Nb in the 4g layer and charge compensating with $2V_O$ in the 4h layer; $Nb_{16}O_{40}$[13,14] calculated with the same method. In comparison, our simulation model shows a better gap estimate, closer to the experimental band gap of 3.6 eV.

### 3.3. Determination of most favorable O sites for the external Li intercalation

Here, we investigated the activity of different O sites to adsorb one extra Li on their top sites in [*T*-$Nb_2O_5$]. As shown in Supplementary Fig. 18, *T*-$Nb_2O_5$ consists of two distinct layers (4g layer and 4h layer, Supplementary Fig. 18a. As for the 4h layer, it is a hybrid Nb-O layer, in which oxygen atoms, which may have two distinct coordinations ($O_{2c}$ and $O_{3c}$, Supplementary Fig. 18b), may be active sites that can adsorb more Li on top of it. $O_{3c}$ atoms are coordinated by three neighboring Nb atoms, which belong to two $NbO_7$ and one $NbO_6$ units, while $O_{2c}$ atoms are coordinated by two neighboring Nb atoms, which belong to one $NbO_7$ and one $NbO_6$. The bridge and hollow sites of $O_{2c}$ and $O_{3c}$ are found unfavourable, because Li will spontaneously transfer to the top site of either $O_{2c}$ or $O_{3c}$. In the case of these two candidates, Li intercalates into the 4g layer, on top of $O_{2c}$ and $O_{3c}$ sites (in 4h layer). The top views of these two structures (after relaxation of ion positions and lattice parameters) are shown in Supplementary Fig. 18c and 18e, respectively. Furthermore, we also considered another Li intercalation configuration, where Li is located in the 4h layer together with Nb cations (Supplementary Fig. 18d).

The O candidates in the 4g layer (Fig. 3b) are inactive based on the calculated $\Delta E_b$, because they are vertically coordinated by Nb atoms in 4h layer. Moreover, we found it can be spontaneously relaxed back to the ground-state configuration (describe later) without any barrier. As for O candidates in the 4h layer, they can be categorized into two different kinds: $O_{3c}$ and $O_{2c}$ as discussed above (Supplementary Fig. 18). The first extra Li prefers to be adsorbed on the top site of $O_{2c}$ with a larger binding energy ($\Delta E_b$ = -2.10 eV, defined in Methods) than that of $O_{3c}$ ($\Delta E_b$ = -2.03 eV) and other less favorable O sites. Therefore, we found the top sites of $O_{2c}$ provides the ground-state location (with strongest binding energy) for Li intercalation. Once $O_{2c}$ sites are fully occupied, more Li will start to be adsorbed on the top of less favorable $O_{3c}$ sites. Based on this, we determined the active sites (coordination environment) by computing the differential binding energy for each Li intercalation of $Li_x$-[T-$Nb_2O_5$] ($x$ = 0.12 ~ 1.08) as shown in Supplementary Fig. 19. We note that in this step, many trial sites have nearly identical binding energies. This, when coupled with the low barriers for Li diffusion suggests that there are actually an ensemble of structures with slightly different Li distributions, and our calculations identify one reasonable representative. This is consistent with previous findings from NMR showing that Li occupies a distribution of similar, but not identical, sites[15]. The configuration with the lowest energy is selected for the further analysis and is used as precursor for the next Li intercalation. The most favorable site identified by DFT for each Li interstitial (in the [$T$-$Nb_2O_5$] unit cell) is summarized in Supplementary Fig. 20. Moreover, it has been well-established (including with the XRD data presented in this manuscript) that intercalation of small amounts of Li causes an expansion in the cell volume without substantially changing the overall $Nb_2O_5$ structure[16]. This finding and the observation of fast, reversible Li movement suggests that Li is located in the relatively open 4g layer, which is supported both by Raman experiments[13] and our DFT calculations.

### 3.4. The onset Li concentration for conductivity in two kinds of super cells and unit cell

To model low concentrations of Li intercalation and avoid spurious interactions between periodic images in the smallest c direction in the unit cell, we created two kinds of super cells (2a × b × 2c and a × b × 3c) of $T$-$Nb_2O_5$ with 248 and 186 atoms in the model, respectively. The (a × b × 3c) super cell is the smallest model to observe the new valence band (induced by one extra Li interstitial) wholly below the Fermi energy as shown in Fig. 3d and Supplementary Fig. 17b, while unit cell and (2a × b × 2c) super cell are directly tuned to metal once even one extra Li atom inserted into the models. The above findings in three models indicate that the accurate DFT-estimation of the onset of conductivity requires a very large simulation model, especially in c direction. Therefore, our findings in the (a × b × 3c) super cell have already been computationally meaningful to unveil the electronic evolution properties of $T$-$Nb_2O_5$ affected by extra Li interstitial (see Figure 3 and related discussions in the main texts).

### 3.5. Appendix structural files corresponding to Figure 3

All fully relaxed configurations of $Li_x$-[$T$-$Nb_2O_5$] we study in this paper are provided as Supporting Data. Detailed description of the structures we used for the results in Figure 3 (shown in main text):

- Figure 3(b) and (c), the structure can be found in the folder named “Unit Cell” .
- Figure 3(d) and (e), the structures can be found in the folder named “Super Cell 33_Li_x” with the file names “Li0.04_T_Nb2O5.xsf” and “Li0.08_T_Nb2O5.xsf”.

- Figure 3(f), the structure can be found in the folder named "Unit Cell_Li_x" with the file name "Li1.08_T_Nb2O5.xsf"
- Figure 3(g), the structure can be found in the folder named "Unit Cell_Li_x" with the file name "Li0.12_T_Nb2O5.xsf"
- Figure 3(h), the structures can be found in the folders named "Unit Cell_Li_x" and "Super Cell 22_Li_x", all files are related.
- Figure 3(i), the structures can be found in the folder named "Monoclinic", all files are related.

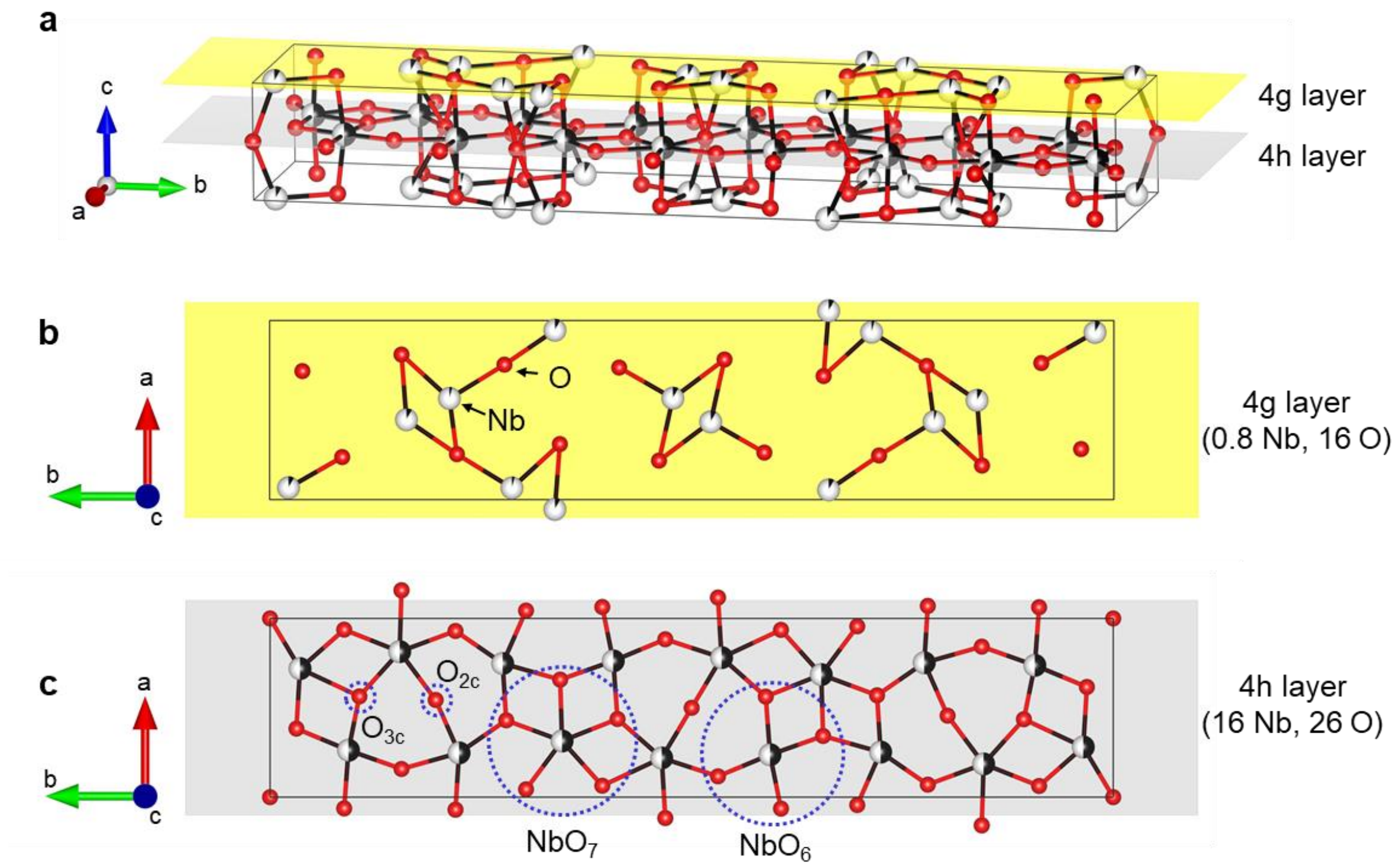

**Supplementary Fig. 16**. **a,** The front view of the conventional unit cell of *T*-$Nb_2O_5$ ($Nb_{16.8}O_{42}$), while top views of 4g layer and 4h layer are shown in **b** and **c**, respectively.

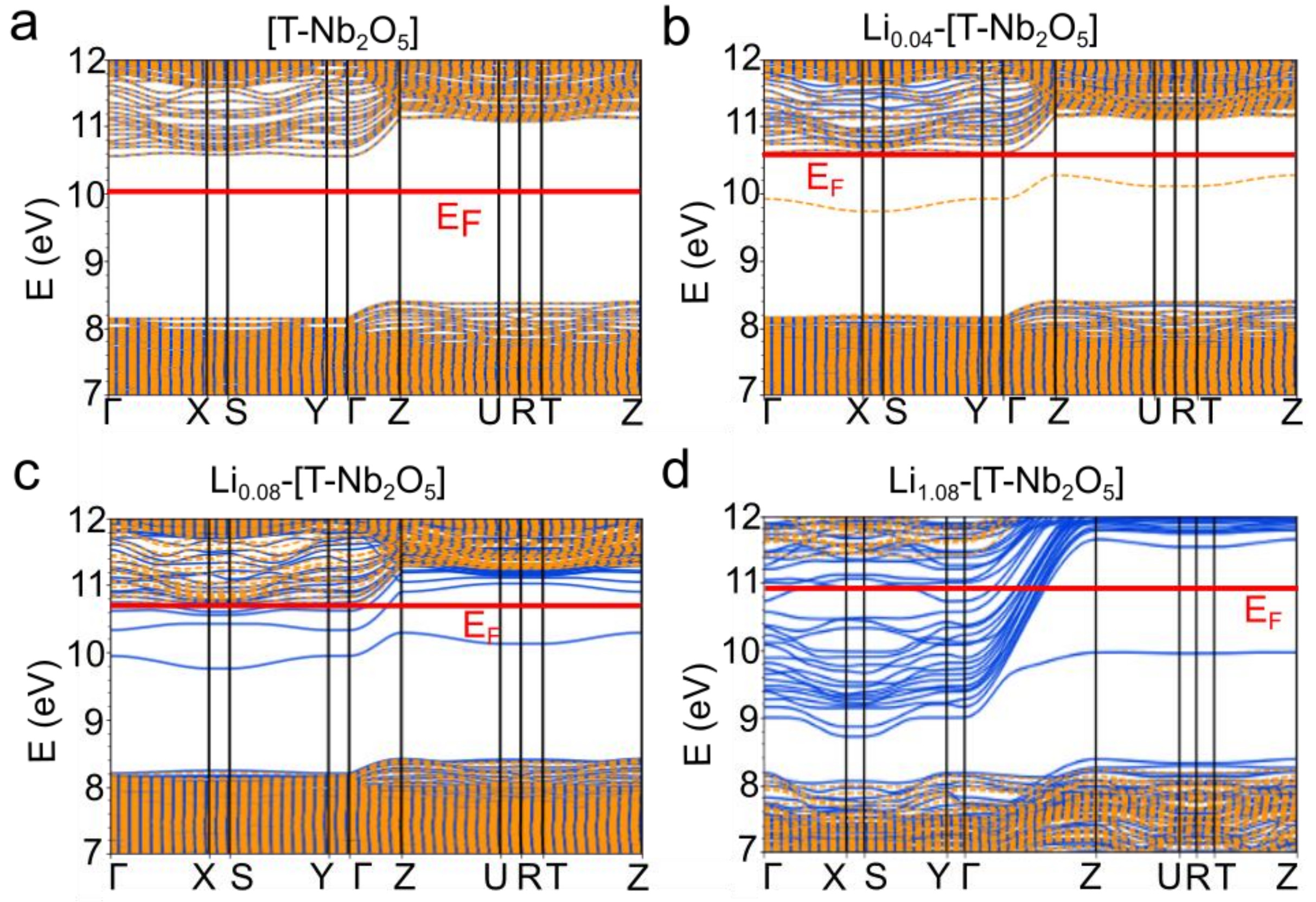

**Supplementary Fig. 17**. **a-c**, Band structures of the (a × b × 3c) [*T*-$Nb_2O_5$] super cell with 0, 1, and 2 extra Li interstitial. **d,** Band structure of the $Li_9Nb_{16.8}O_{42}$ model in monoclinic phase.

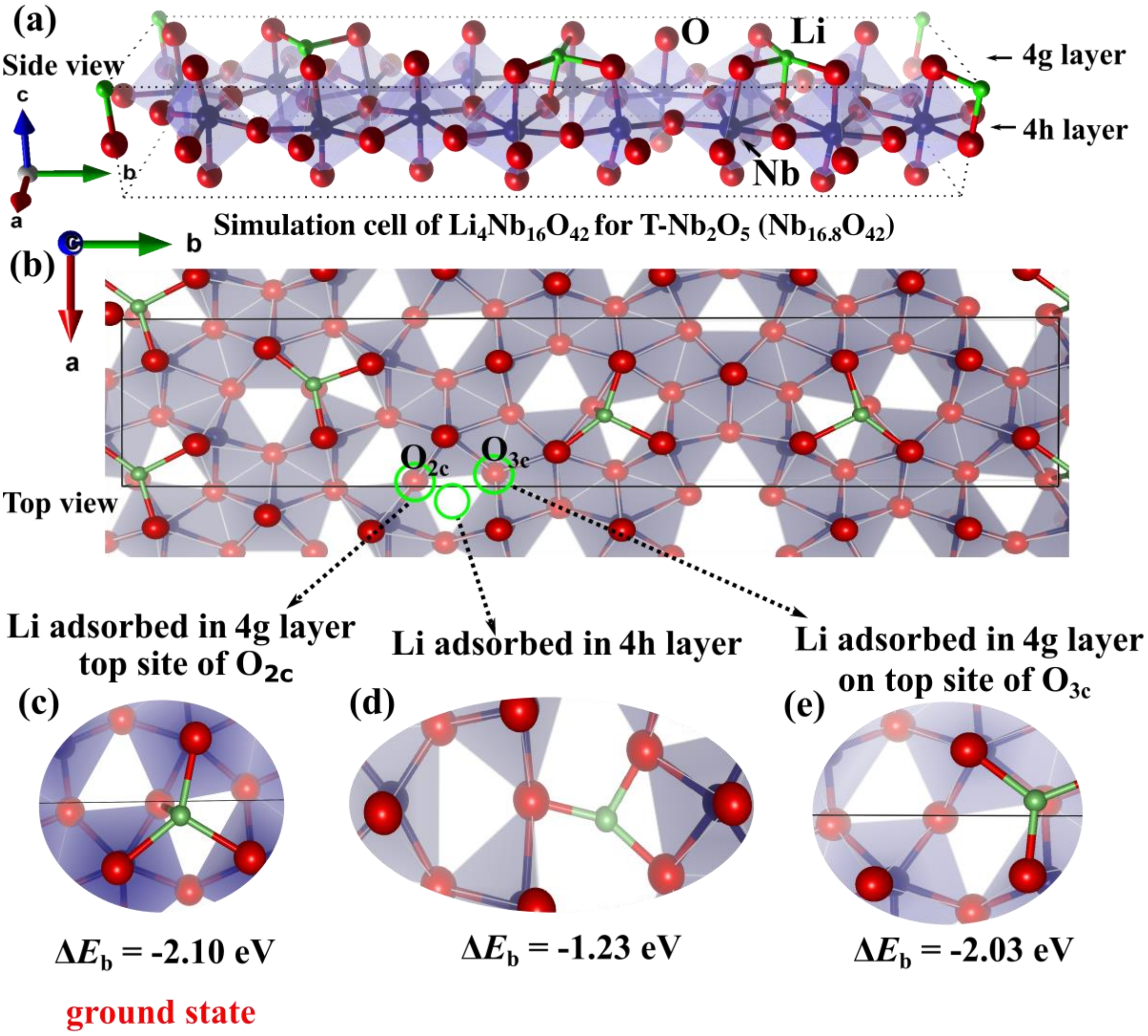

**Supplementary Fig. 18**. **(a)** and **(b)** show the side and top views, respectively, of the orthorhombic model system ($Li_4Nb_{16}O_{42}$) used to simulate the conventional unlithiated unit cell of *T*-$Nb_2O_5$ ($Nb_{16.8}O_{42}$). O2c and O3c represent the O ions in the 4h layer coordinated by two and three adjacent Nb ions in the same plane (4h layer). **(c)-(e)** Three possible configurations of an extra $Li^+/e^-$ intercalation in *T*-$Nb_2O_5$, where $Li^+/e^-$ intercalates the 4g layer in (c) and (e) (on top of $O_{2c}$ and $O_{3c}$ sites, respectively) and $Li^+/e^-$ is in the 4h layer (in panel (d)).

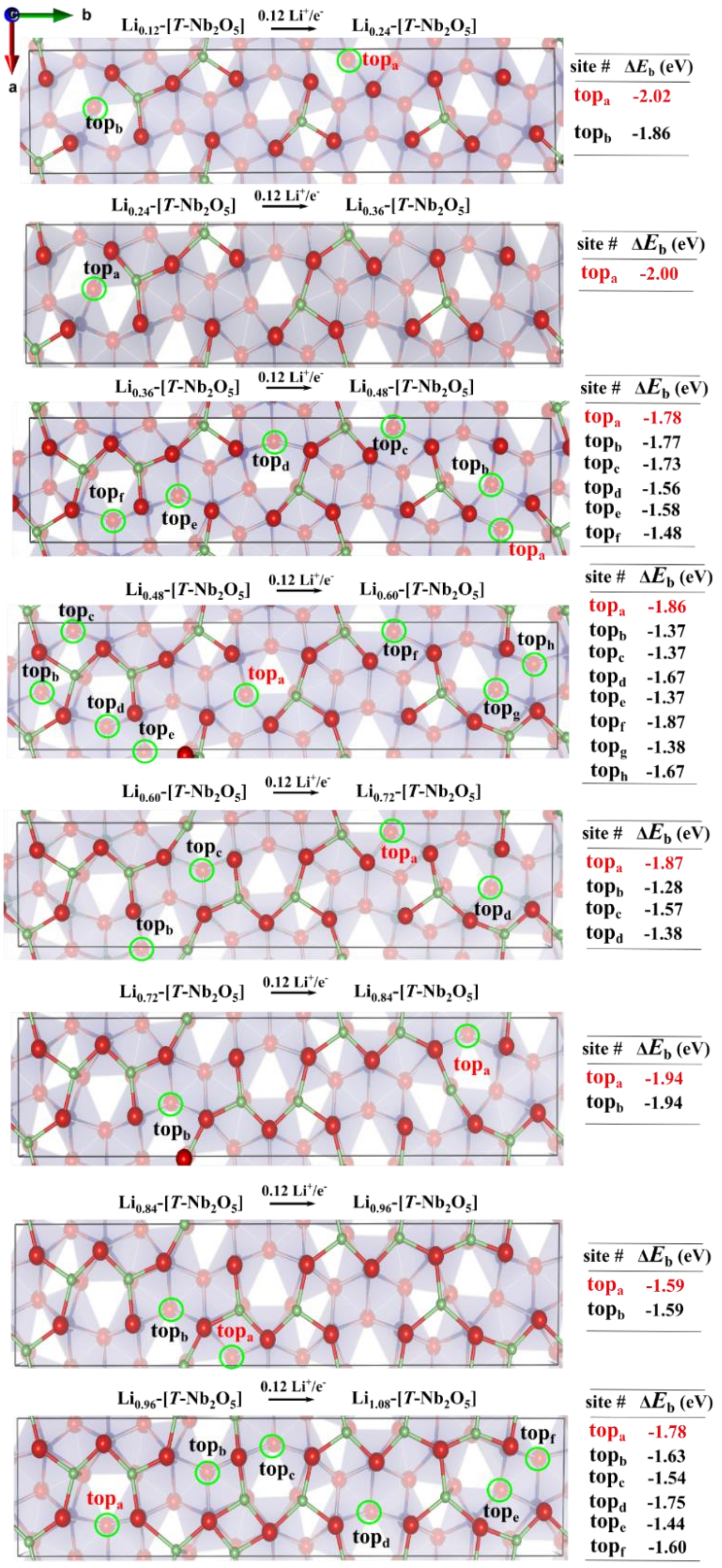

**Supplementary Fig. 19**. Top view of active sites associated with differential binding energies for each $Li^+/e^-$ intercalation of $Li_x$-[T-$Nb_2O_5$] (x = 0.12 ~ 1.08), where 0.12 $Li^+/e^-$ indicates a pair of $Li^+/e^-$ adsorbed in *T*-$Nb_2O_5$ (because we convert $Nb_{16.8}O_{42}$ to $Nb_2O_5$). $Li^+/e^-$ pair intercalate in 4g layer and on top of the candidate sites as we shown in each configuration of $Li_x$-[T-$Nb_2O_5$], while the most favorable site (providing the strongest differential binding

energy) for each $Li^+/e^-$ is labeled by $top_a$ highlighted in red. In the configuration of $Li_{0.72}$-[$T$-$Nb_2O_5$] and $Li_{0.84}$-[$T$-$Nb_2O_5$], both of two active-site candidates are favorable.

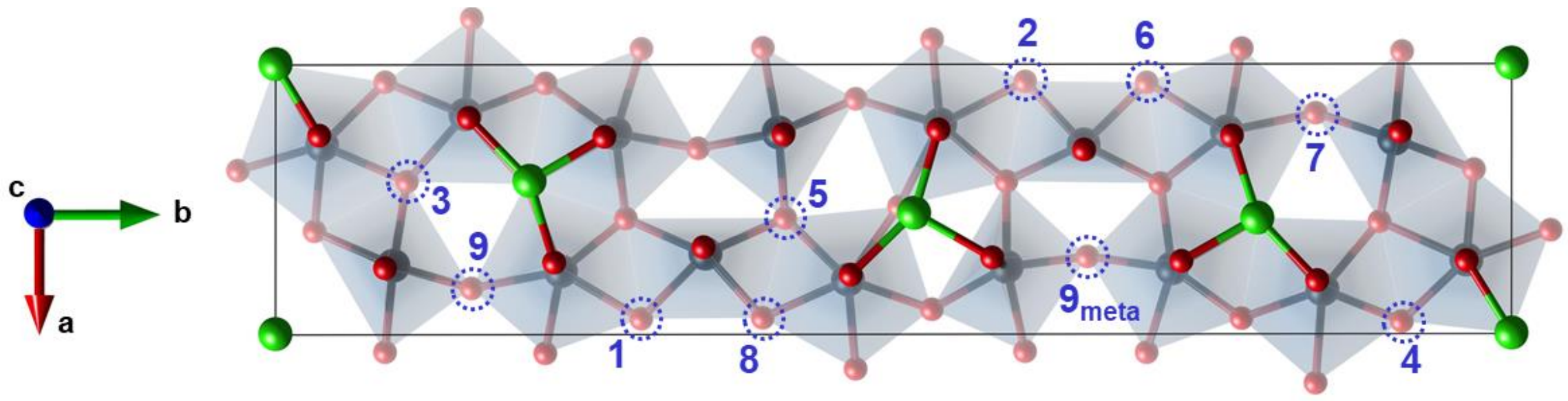

**Supplementary Fig. 20**. The most favorable sites for each Li interstitial (from 1st to 9th) in our $T$-$Nb_2O_5$ model (unit cell). The determination is based on our $\Delta E_b$ values when Li adsorbs atop the available O atoms in 4h layer. The site tagged with $9_{meta}$ is the metastable structure for $Li_{1.08}Nb_2O_5$, which is also favorable to be tilted into monoclinic phase (see Supplementary Fig. 21).

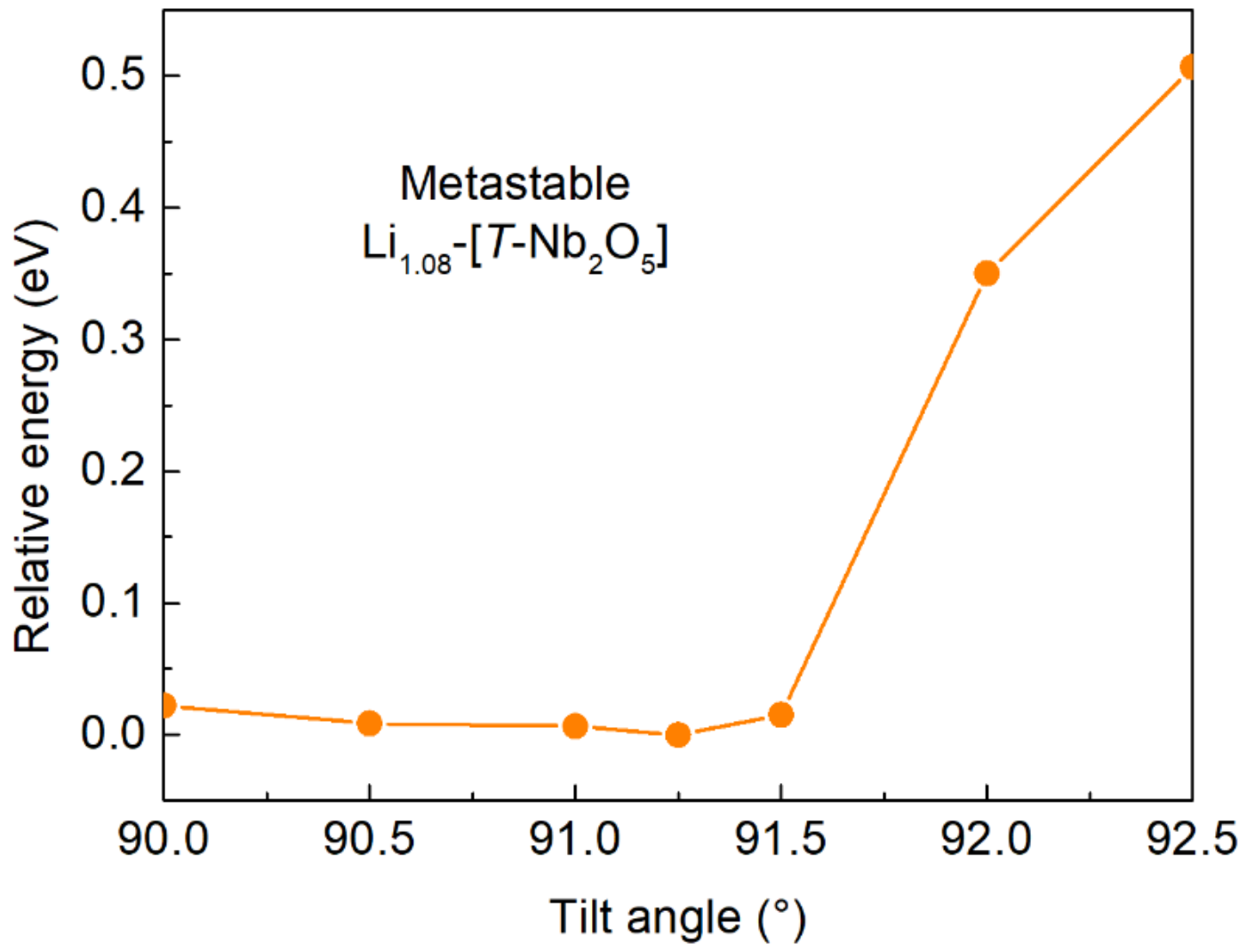

**Supplementary Fig. 21**. The relative energy evolution of the metastable structure $Li_{1.08}$-[*T*-$Nb_2O_5$] as a function of tilt angle, from 90° to 92.5°, where the Li from 1st to 8th are inserted at the most favorable sites shown in Supplementary Fig. 20, while the 9th Li is inserted at the $9_{meta}$ (less favorable O sites) site.

## 4. Electrochemical tests of *T*-$Nb_2O_5$

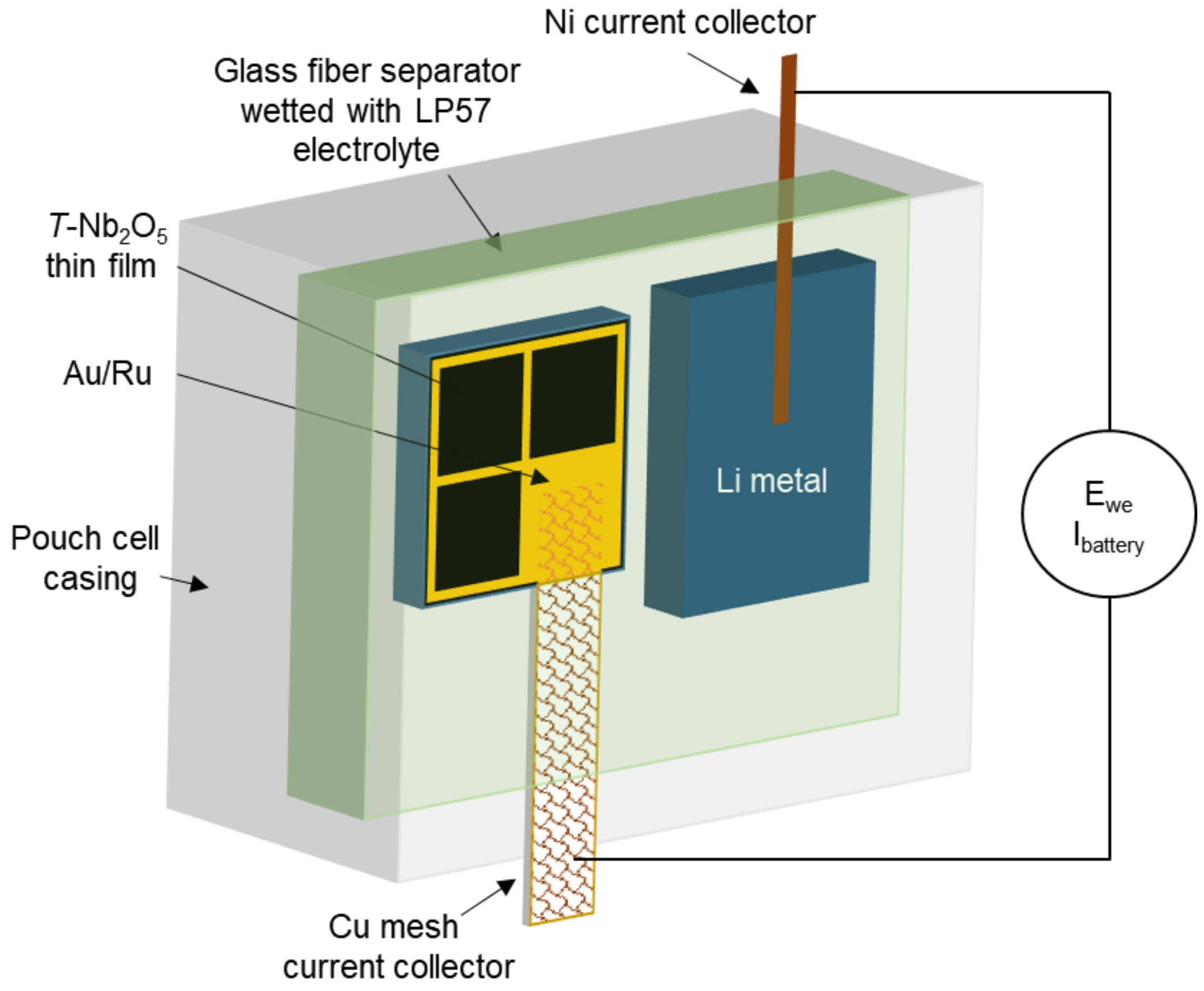

**Supplementary Fig. 22. A schematic Pouch cell of *T*-$Nb_2O_5$ thin film for battery testing.** The current, $I_{battery}$ and $E_{we}$ is measured between *T*-$Nb_2O_5$ thin film and Li anode. By analogy with the gating experiment, *T*-$Nb_2O_5$ is the channel and Li metal is the gate electrode, hence the $I_{battery}$ and $E_{we}$ correspond to the leakage current, $I_g$, and the gate voltage respectively. Note that the conventions are different: $I_{battery} = - I_g$.

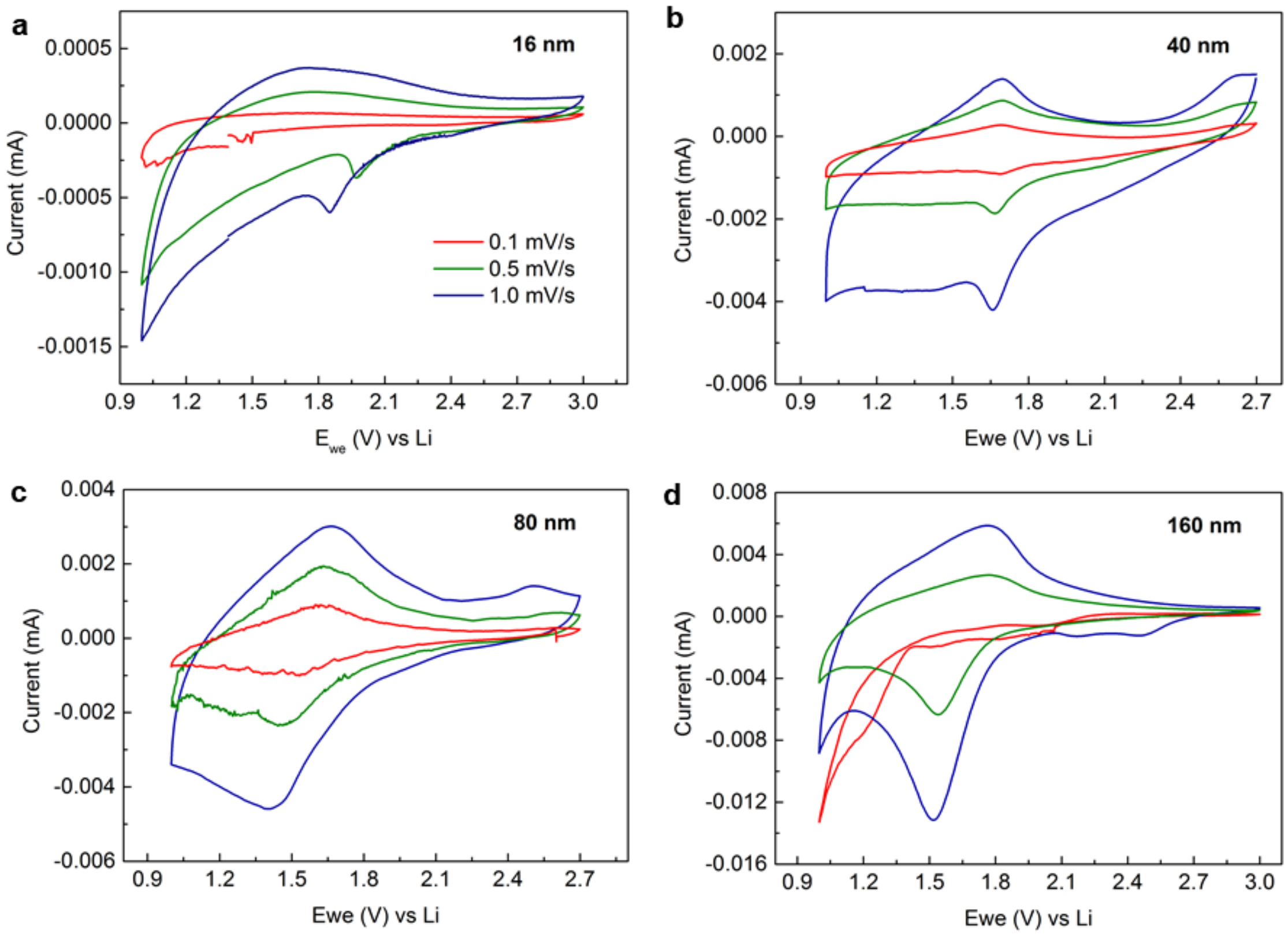

**Supplementary Fig. 23.** Cyclic voltammogram for *T*-$Nb_2O_5$/LSAT (110) thin films at different thicknesses such as **a**, 16 nm, **b**, 40 nm, **c**, 80 nm, and **d**, 160 nm. Cyclic voltammogram was recorded with different scan rate (0.1, 0.5 and 1.0 mV/s), and second cycle of each cycling rate is shown here. Irreversible reactions seem occurs at low voltages < 1 V as observed by the rise of the current.

## 5. Transport properties of *T*-$Nb_2O_5$ thin films via lithiation

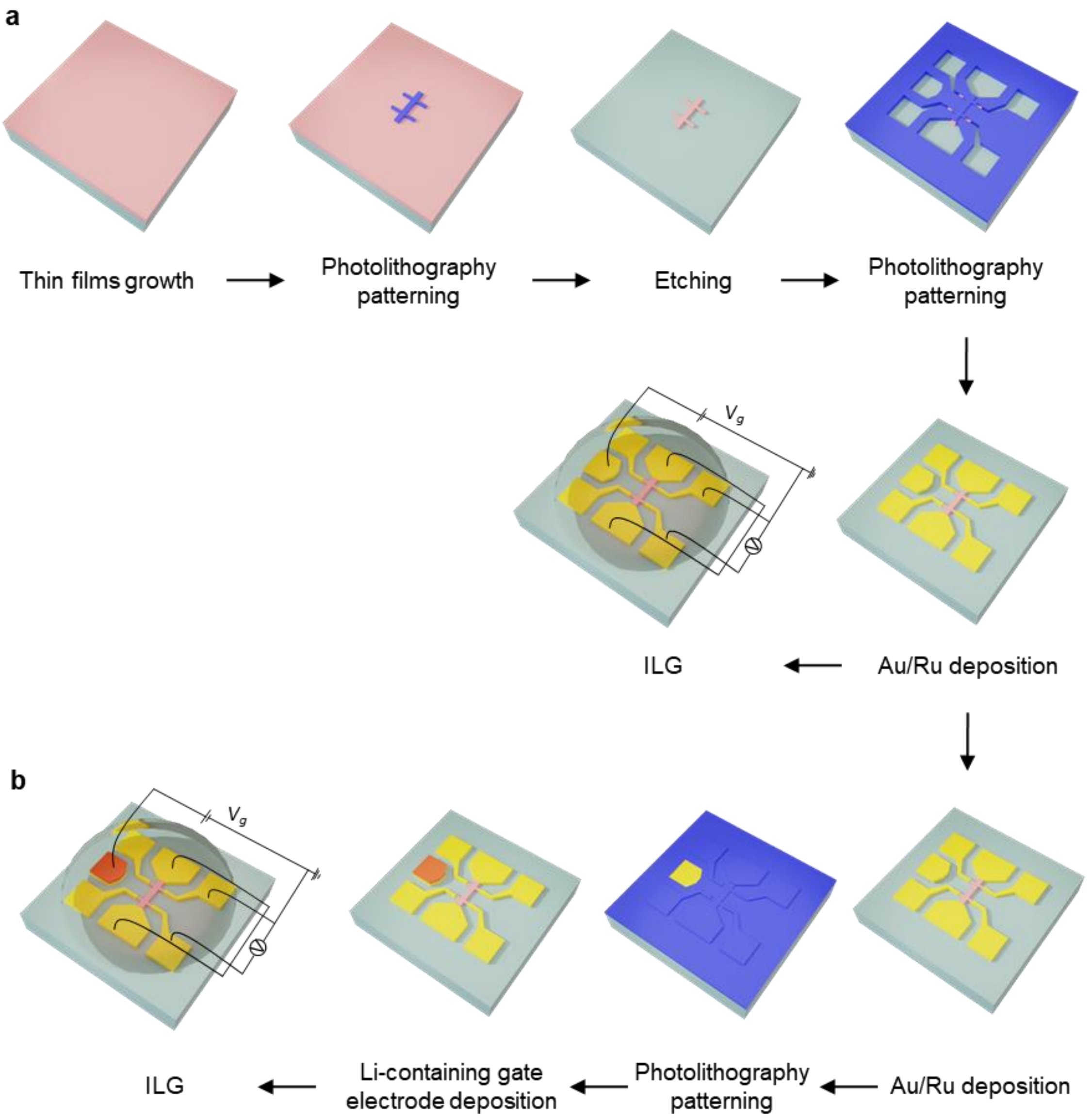

**Supplementary Fig. 24. Schematic diagrams of device fabrications. a**, Schematics of ionic liquid gating device fabrications. The standard photo-lithographic techniques are used. The channel was etched, and then Ru (5 nm) and Au (70 nm) are deposited successively for the gate electrode and channel contacts. The Li-ionic liquid was placed on the device for the gating experiment. **b**, Schematics of the device fabrications having Li-containing gate electrode. The Li-containing oxides are deposited on the Au/Ru gate electrode by pulsed laser deposition. The channel sizes are 65 × 30 $\mu m^2$.

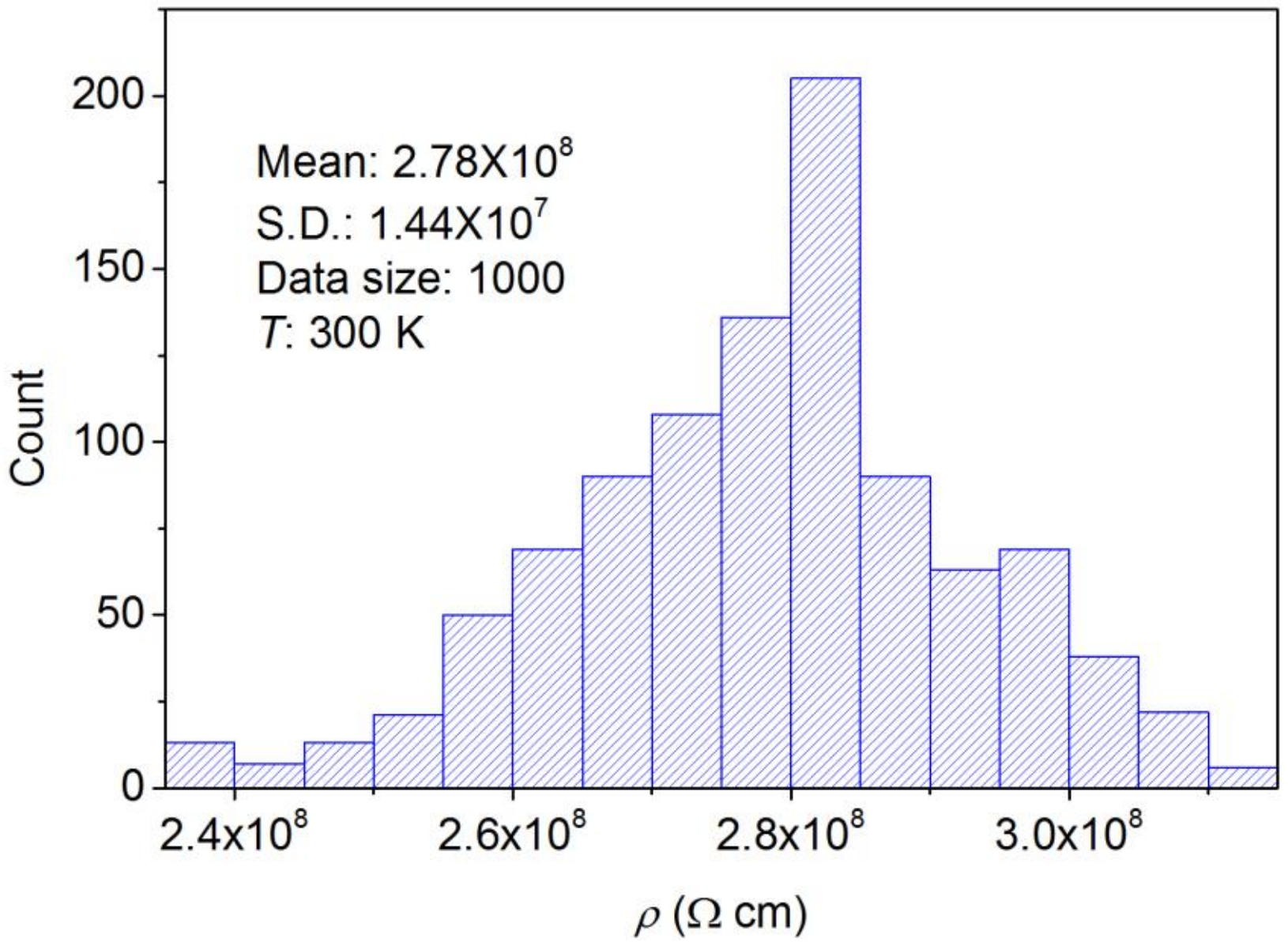

**Supplementary Fig. 25. Resistivity measurements of the pristine *T*-$Nb_2O_5$/LSAT (110).** The 1000 data points were averaged using a high resistance meter (B2985A, KEYSIGHT). During the measurement, the temperature of the sample was kept at 300 K in a vacuum environment.

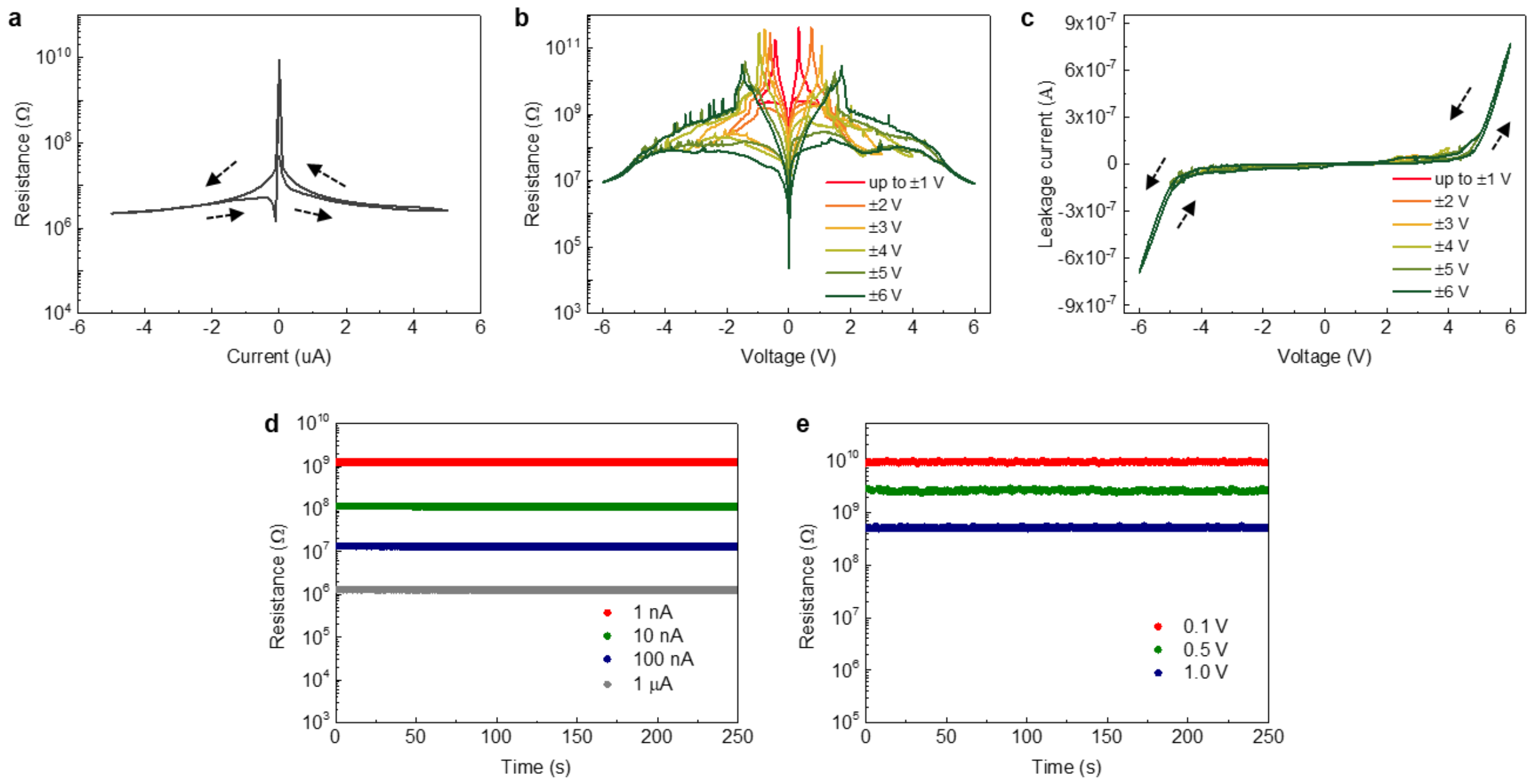

**Supplementary Fig. 26. Source-drain current ($I_{sd}$)- or voltage ($V_{sd}$)-dependent Li-IL resistance. a,** Resistance of Li-IL as a function of $I_{sd}$. **b**, Resistance and **c**, current of IL as a function of $V_{sd}$. Time-dependent changes in resistance of Li-IL at different **d**, $I_{sd}$- and **e**, $V_{sd}$. In order to measure the electrical resistance of ILs, we have placed the IL on a device made by LSAT (110) substrate, which does not interact via gating. The device has the same electrode structure (Fig. 2d) for ILG of *T*-$Nb_2O_5$ thin films for the comparison. The resistance of IL decreases from ≈$10^{10}$ to ≈$10^6$ Ω when the applied current or voltage is increased and saturates at ≈$10^6$ Ω even upon application of high current (5 μA) and voltage (6 V). The decrease in resistance along with the increase in leakage at high $I_{sd}/V_{sd}$ are likely due to the decomposition of IL. The time-dependent resistance is constant and strongly depends on the applied current or voltage.

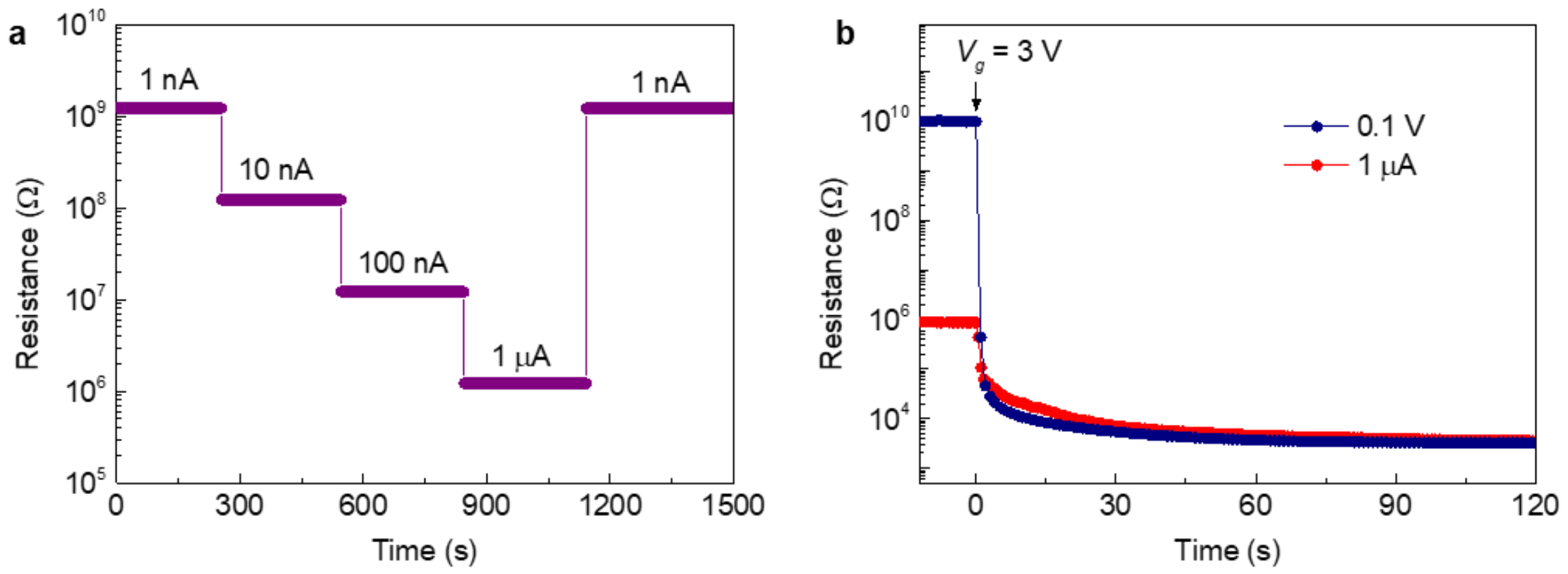

**Supplementary Fig. 27. Effect of source-drain current ($I_{sd}$) and voltage ($V_{sd}$) on the resistance change of the *T*-$Nb_2O_5$ device. a,** Time-dependent change in resistance of the *T*-$Nb_2O_5$ device at different $I_{sd}$ without applying a gate voltage. $I_{sd}$ strongly affects the resistance of device, and the resistance value is the same with the $I_{sd}$-dependent resistance of IL on the substrate (Supplementary Fig. 26), indicating that the observed resistance change is due to the IL, not the *T*-$Nb_2O_5$. Noteworthy is that the resistance change of IL is reversible even though the high current can lead to the decomposition of IL. **b**, Time-dependent resistance change of the *T*-$Nb_2O_5$ device at different $I_{sd}$/$V_{sd}$. The resistance decreases when gate voltage ($V_g$) of 3 V is applied. The initial resistance depends on $I_{sd}$ and $V_{sd}$. Thus, applying low $V_{sd}$ (0.1 V) shows larger change in resistance via metallization of *T*-$Nb_2O_5$.

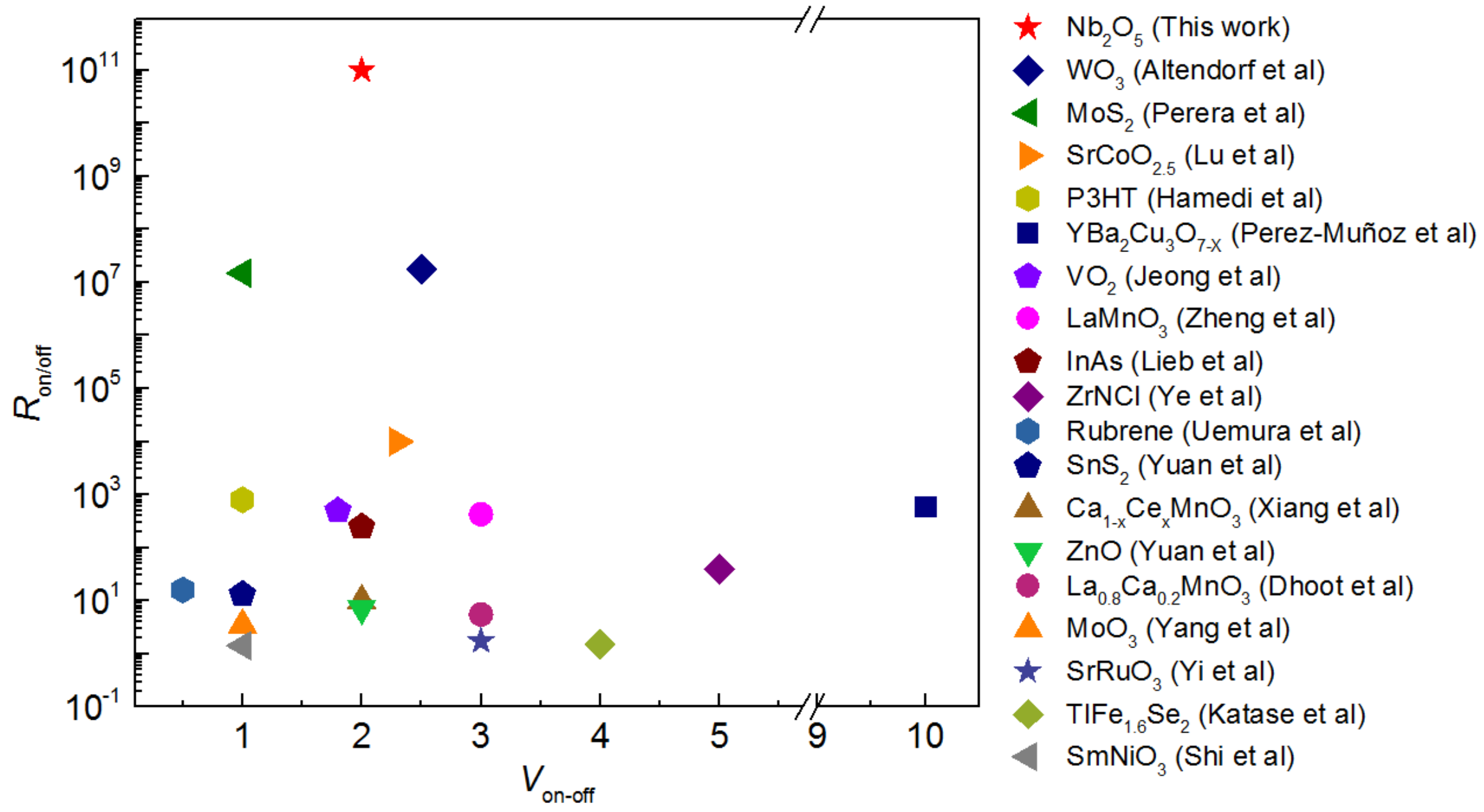

**Supplementary Fig. 28. Resistance changes vs. gate voltage plot of ionic liquid gating devices.** The insulating $SrTiO_3$ has shown emergent superconductivity at the low temperature after ILG, however the resistivity before gating is not presented in the reported paper[17]. Considering the high resistivity of $SrTiO_3$[18], we expect that the resistance changes of $SrTiO_3$ via ILG at ≈3.5 V will be $10^{10}$–$10^{11}$ orders of magnitude at room temperature. The detailed list are shown in Supplementary Table 1. The electrolyte gating studies using solid electrolytes are not shown here, which mostly shows slower responses and relatively small changes in resistance.

**Supplementary Table 3. Resistance changes via ionic liquid gating**

| Material | Von-off | Ron/off | Mechanism | Migrated ion | Ionic liquid |
|---|---|---|---|---|---|
| $Nb_2O_5$ (This work) | 2 | $\approx 1\times10^{11}$ | electrochemical | Lithium | Li containing DEME-TFSI |
| P3HT (Hamedi et al)[19] | 1 | $\approx 8\times10^{2}$ | electrochemical | Anions | BMIM-TFSI |
| $MoO_3$ (Yang et al)[20] | 1 | ≈3.5 | electrochemical | Hydrogen | EMIM-TFSI |
| $WO_3$ (Altendorf et al)[21] | 2.5 | $\approx 1.8\times10^{7}$ | electrochemical | Oxygen | HMIM-TFSI |
| $SrCoO_{2.5}$ (Lu et al)[22] | 2.3 | $\approx 1\times10^{4}$ | electrochemical | Oxygen | DEME-TFSI |
| $Yba_2Cu_3O_{7-X}$ (Perez-Muñoz et al)[23] | 10 | $\approx 6\times10^{2}$ | electrochemical | Oxygen | DEME-TFSI |
| $VO_2$ (Jeong et al)[24] | 1.8 | $\approx 5\times10^{2}$ | electrochemical | Oxygen | HMIM-TFSI |
| $SrRuO_3$ (Yi et al)[25] | 3 | ≈1.7 | electrochemical | Oxygen | EMIM-TFSI |
| $SmNiO_3$ (Shi et al)[26] | 1 | ≈1.4 | electrochemical | Oxygen | PEMIM-TFSI |
| $MoS_2$ (Perera et al)[27] | 1 | $\approx 1.5\times10^{7}$ | electrostatic | | DEME-TFSI |
| $LaMnO_3$ (Zheng et al)[28] | 3 | $\approx 4.3\times10^{2}$ | electrostatic | | DEME-TFSI |
| InAs (Lieb et al)[29] | 2 | $\approx 2.2\times10^{2}$ | electrostatic | | EMIM-TFSI |
| ZrNCl (Ye et al)[30] | 5 | ≈40 | electrostatic | | DEME-TFSI |
| Rubrene (Uemura et al)[31] | 0.5 | ≈16 | electrostatic | | EMIM-TFSI |
| $SnS_2$ (Yuan et al)[32] | 1 | ≈13 | electrostatic | | DEME-TFSI |
| $Ca_{1-x}CexMnO_3$ (Xiang et al)[33] | 2 | ≈10 | electrostatic | | DEME-TFSI |
| ZnO (Yuan et al)[34] | 2 | ≈7.2 | electrostatic | | DEME-TFSI |
| $La_{0.8}Ca_{0.2}MnO_3$ (Dhoot et al)[35] | 3 | ≈5.5 | electrostatic | | EMIM-TFSI |
| $TlFe_{1.6}Se_2$ (Katase et al)[36] | 4 | ≈1.5 | electrostatic | | EMIM-TFSI |

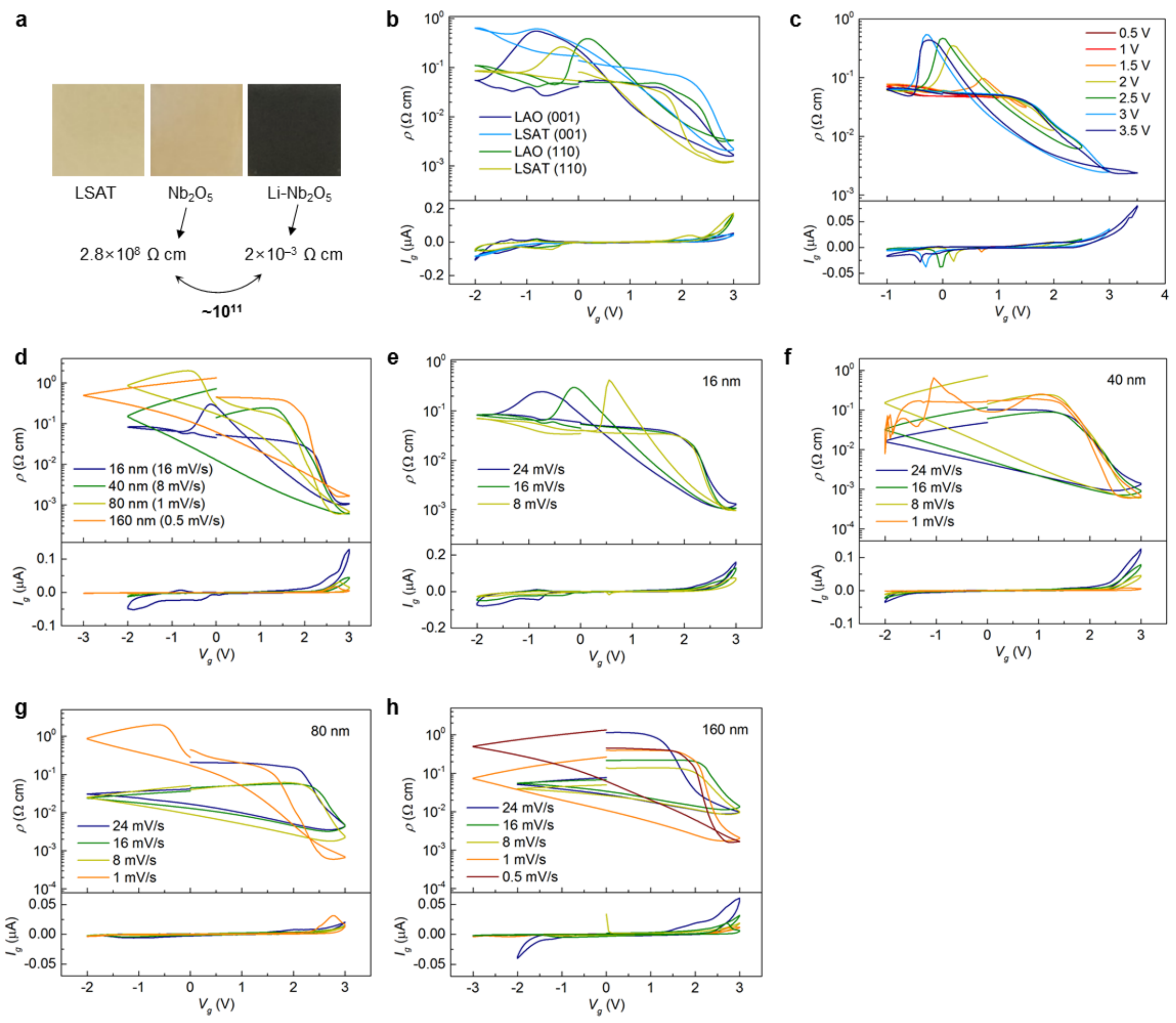

**Supplementary Fig. 29. Resistance changes via ionic liquid gating for different substrate and thicknesses. a**, Optical photographs of a LSAT (110) substrate, *T*-$Nb_2O_5$ thin films, and Li-ionic liquid gated *T*-$Nb_2O_5$ (Li-$Nb_2O_5$) thin films. The gated film shows a black color with the decrease of resistivity. The resistivity measurement was done by the Van der Pauw method at room temperature. **b**, $V_g$ dependent resistivity ($\rho$) and leakage current ($I_g$) curves of 16 nm thick *T*-$Nb_2O_5$ thin film devices grown on different substrates, including LAO(001), LSAT (001), LAO (110), and LSAT (110). All devices show similar responses from gating because all films have vertically formed ionic transport channels. **c**, $V_g$ dependent $\rho$ and $I_g$ curves with different maximum positive sweeping voltage for the 16 nm thick *T*-$Nb_2O_5$/LSAT(110) device. The resistivity decreases as increasing the maximum positive $V_g$. **d**, $V_g$ dependent $\rho$ and $I_g$ curves with different thicknesses. $V_g$ dependent $\rho$ and $I_g$ curves with different gate voltage sweeping rate for **e**, 16 nm, **f**, 40 nm, **g**, 80 nm, and **h**, 160 nm thick *T*-$Nb_2O_5$/LSAT(110) devices.

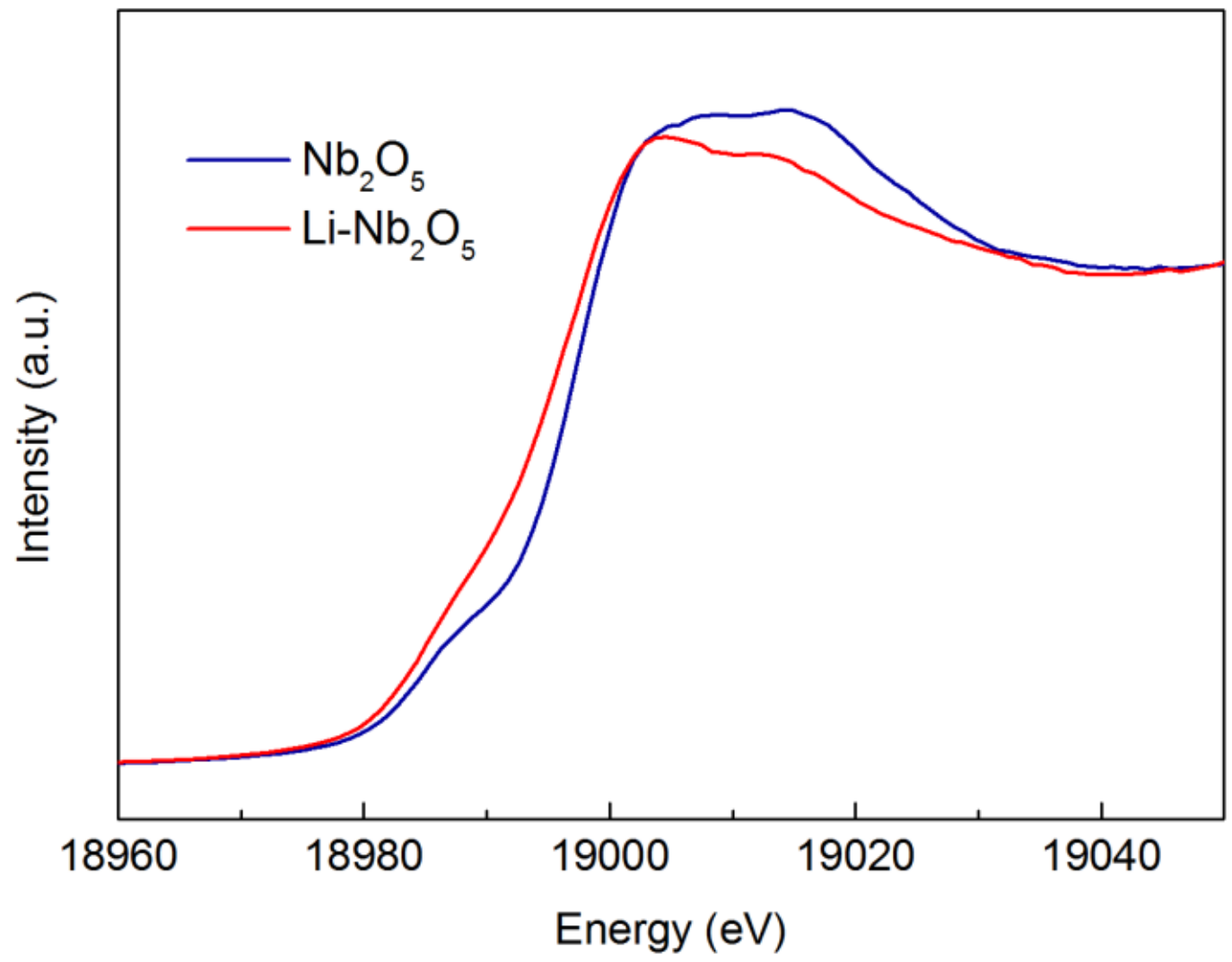

**Supplementary Fig. 30. Nb K-edge XANES spectra of pristine and gated 60 nm thick *T*-$Nb_2O_5$ thin films.** The absorption edge showed a chemical shift to a low-energy side, indicating a reduction of Nb state by Li intercalation into the lattice. The gating was done at 3 V for 20 min.

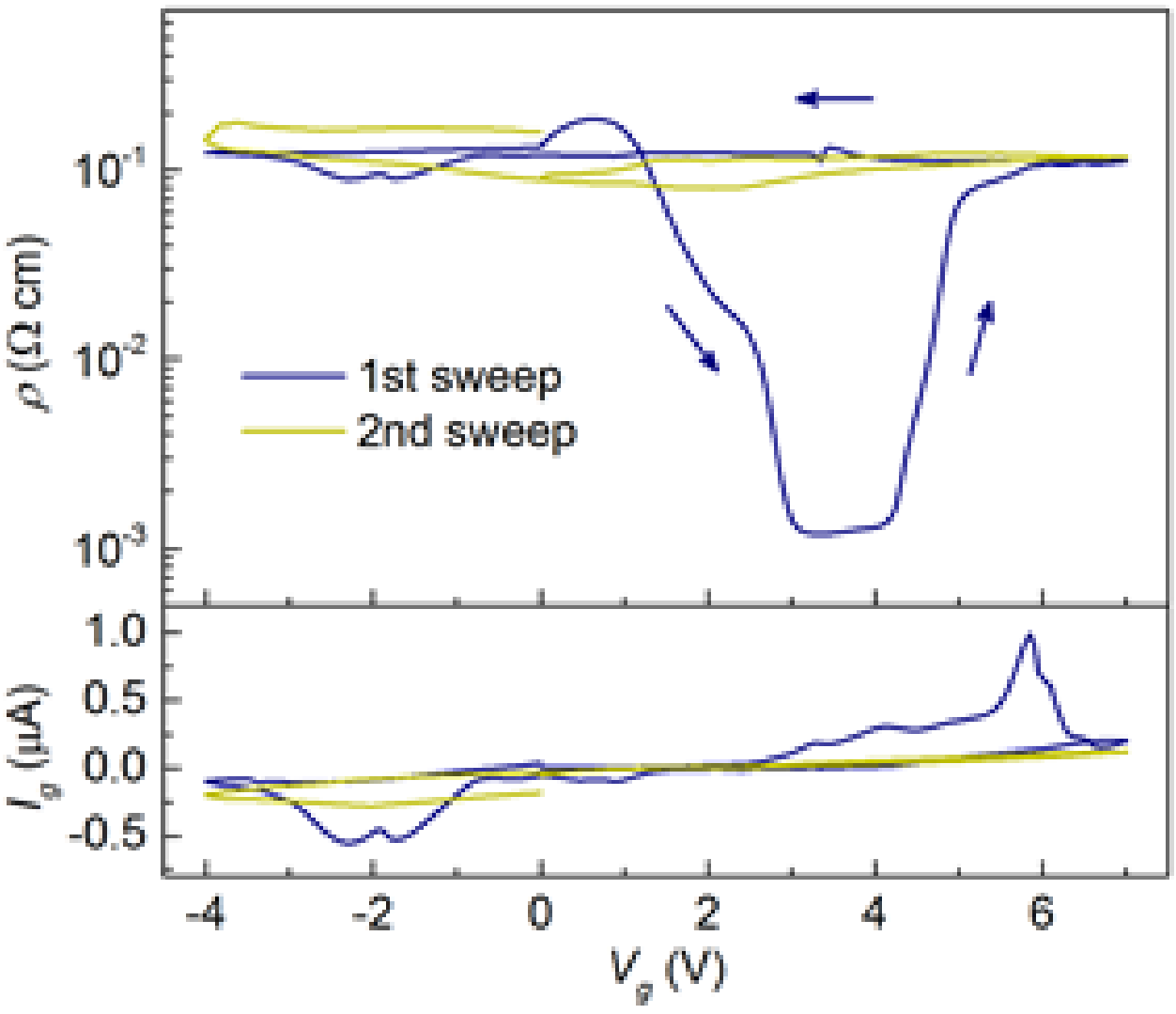

**Supplementary Fig. 31. $V_g$ dependent $\rho$ and $I_g$ curves with high voltages for the 16 nm thick film.** The sweeping rate was 16 mV/s. The high voltage gating above ≈4.5 V lead to the insulting behavior, and the second sweeping does not show metallization, indicating the degradation of the films from the high voltage gating.

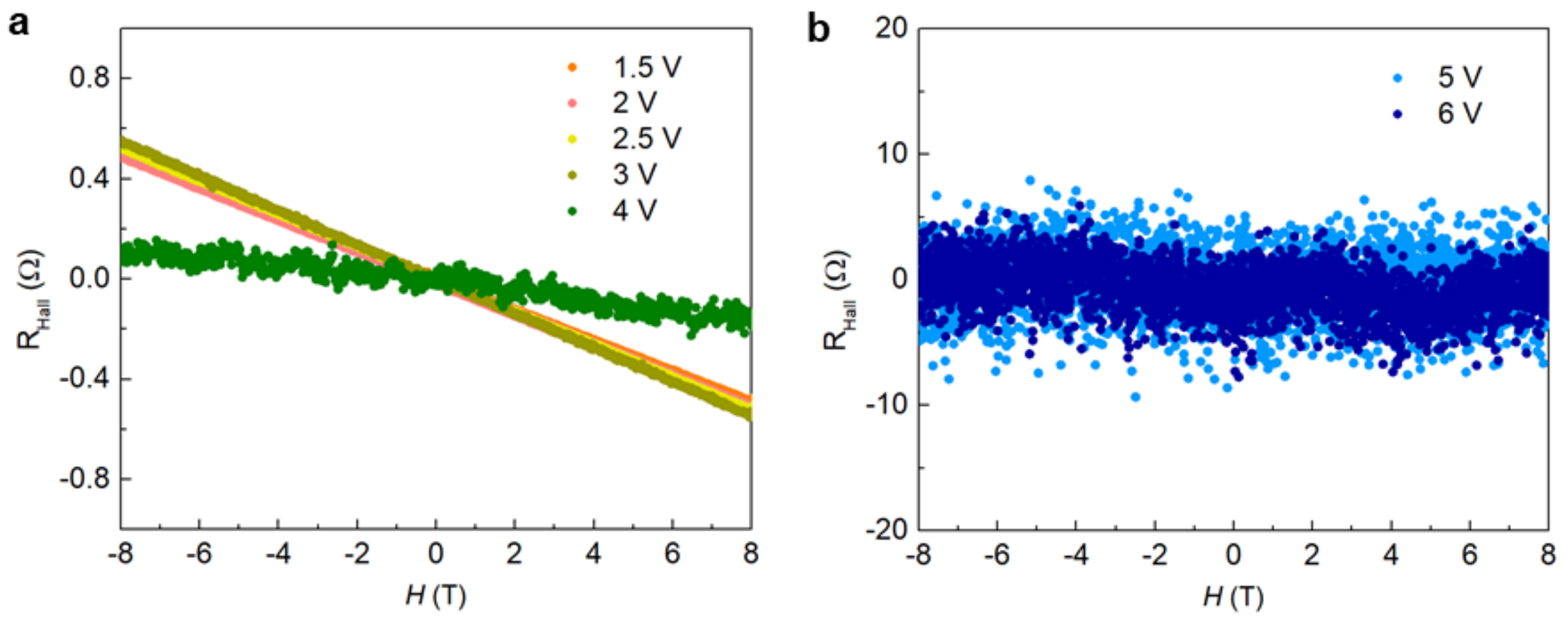

**Supplementary Fig. 32. Hall measurements of the *T*-$Nb_2O_5$/LSAT(110) thin film device. a**, Hall measurements with different gate voltages ranging from 1.5~4 V measured at 200 K. The negative signs represent that the mobile charge carriers are electrons. **b**, Hall measurements at high voltage (5 V and 6 V) lead to noisy signals due to the amorphous phase at the high voltage gating. All measurements were done for the 40 nm thick *T*-$Nb_2O_5$/LSAT(110) device.

## 6. Gate voltage tuning by using Li-containing gate electrodes

To explore the effect of the chemical potential of the gate electrode on the critical metallization voltage ($V_c$), the Au/Ru gate electrode is replaced by Li-containing electrode materials such as $LiFePO_4$ (LFP), $LiCoO_2$ (LCO), and $Li_xNb_2O_5$ (LNbO). The growth conditions of LFP and LCO are first optimized by PLD, and the chemical states were characterized by X-ray photoelectron spectroscopy (XPS) as shown in Supplementary Fig. 33. Then, the materials were deposited at room temperature on the Au/Ru gate electrode of the *T*-$Nb_2O_5$ devices (Supplementary Fig. 24b). The deposited LCO and LFP films do not show any crystalline peak from laboratory XRD measurements. In the case of the LNbO gate electrode, two *T*-$Nb_2O_5$ windows were first etched, and the Au/Ru electrodes and the LFPO film were deposited successively, as depicted in Supplementary Fig. 34. Then, Li ions were transferred from the LFP electrode to *T*-$Nb_2O_5$ gate electrode by L-ILG to form the LNbO gate electrode.

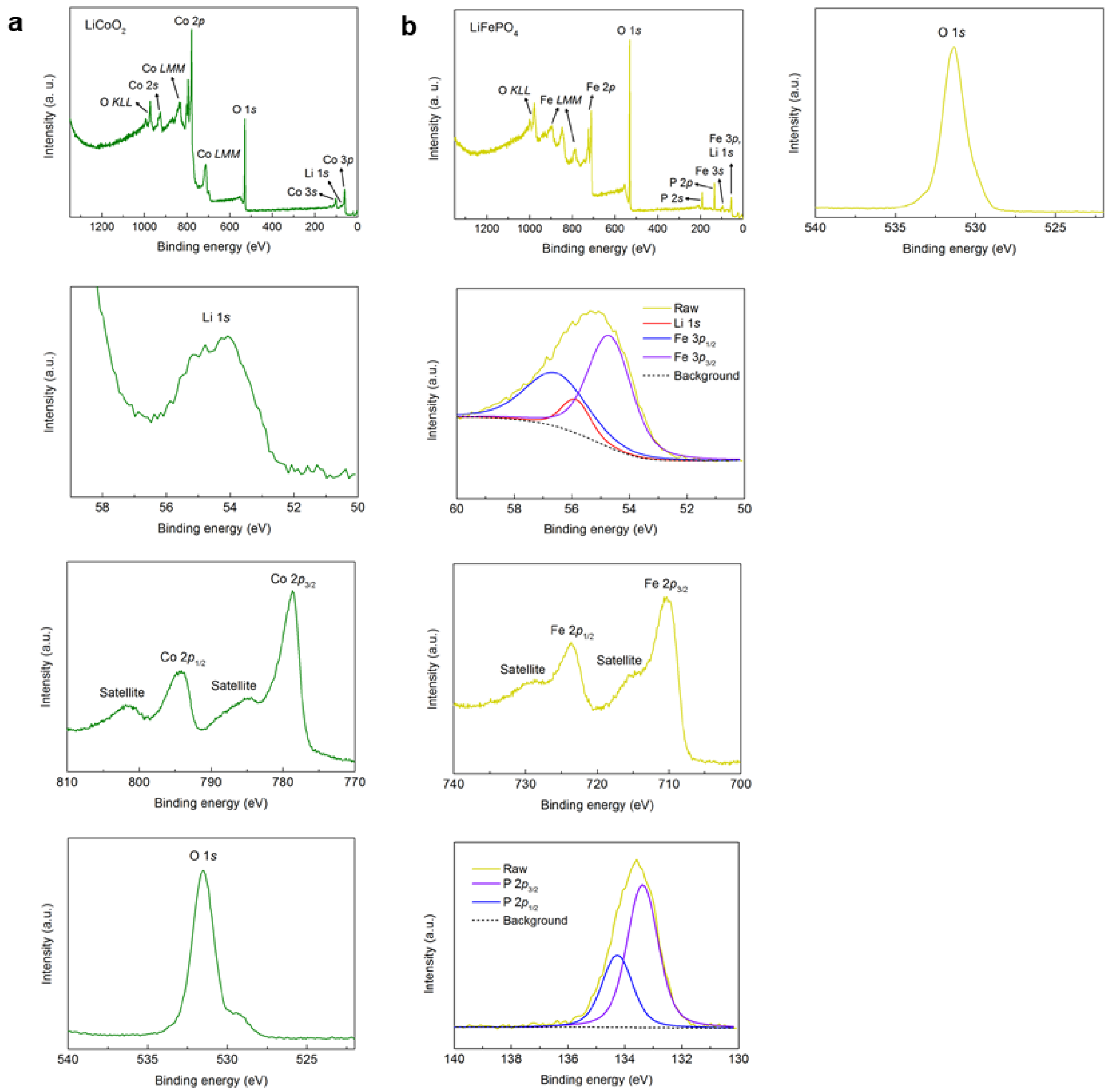

**Supplementary Fig. 33. XPS results of Li-containing oxides.** XPS spectra of **a**, $Li_{1+x}CoO_2$, and **b**, $Li_{1+x}FePO_4$ after growth on LSAT (110) substrates using pulsed laser deposition.

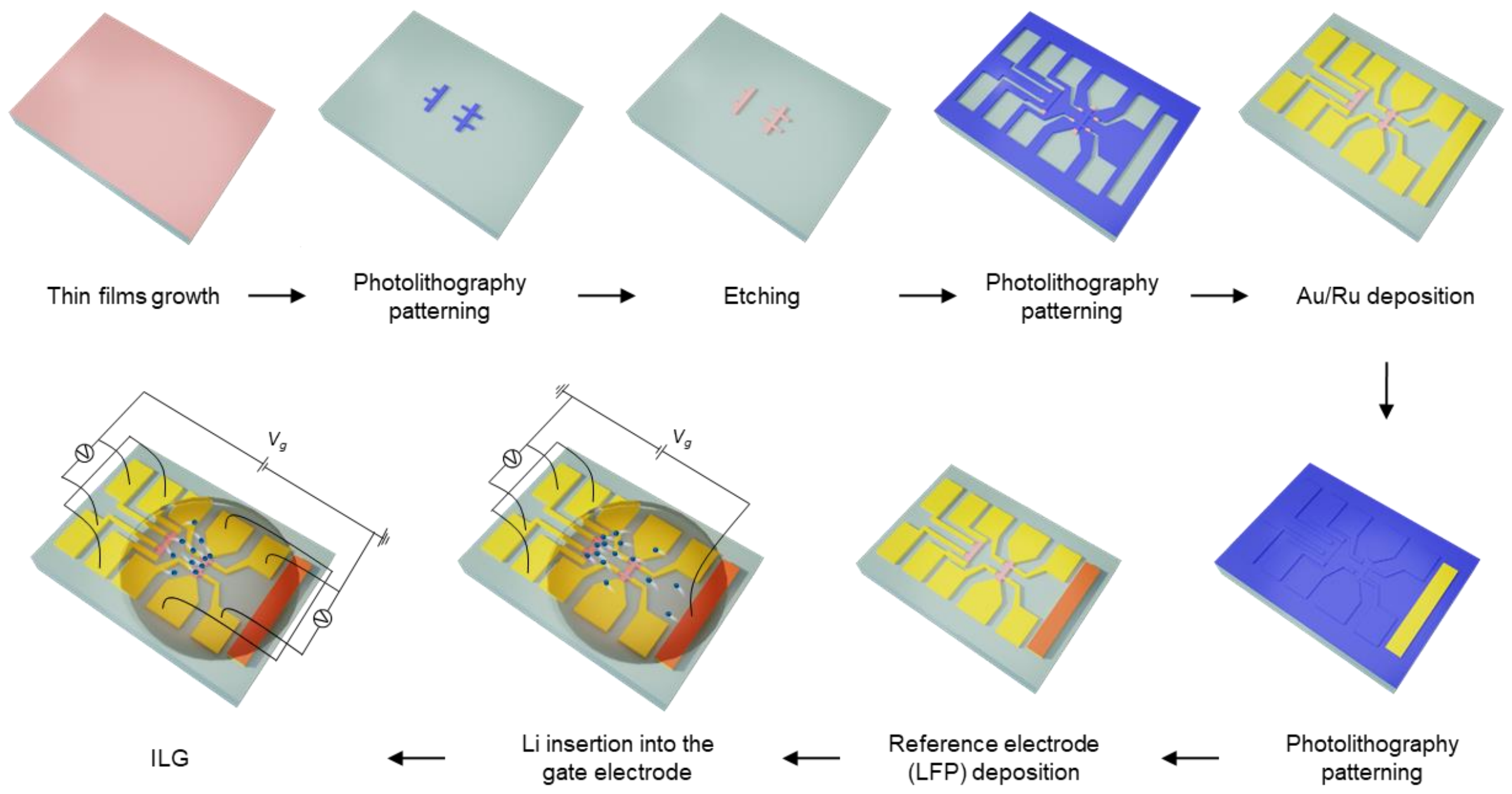

**Supplementary Fig. 34. Schematic diagrams of Li-$Nb_2O_5$ gate electrode device fabrications.** The standard photo-lithographic techniques are used. The channel and gate were etched, and then Au (70 nm)/Ru (5 nm) are deposited for the gate and channel contacts. The reference electrode (LFP) is deposited using pulsed laser deposition. After the device fabrication, Li ions are moved from the LFP to the gate electrode by ILG to make Li-$Nb_2O_5$ gate electrode. Then, the gating was done to the $Nb_2O_5$ channel using the Li-$Nb_2O_5$ gate electrode.

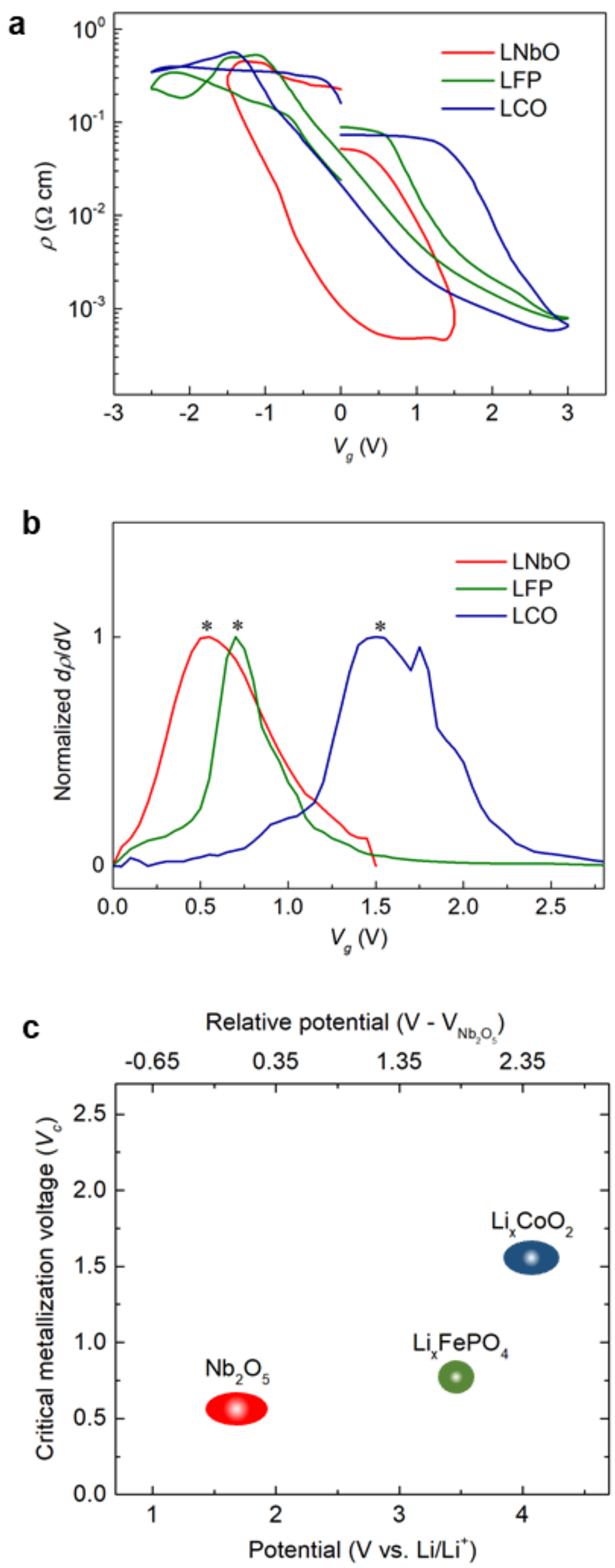

**Supplementary Fig. 35. Tunable critical metallization voltage ($V_c$) of *T*-$Nb_2O_5$ thin film devices via gate electrode potential control. a**, $V_g$ dependent resistivity curves with different gate electrodes. The sweeping rate was 16 mV/s. **b**, Normalized first derivative resistive-Vg curves to determine the $V_c$. **c**, The chemical potential dependent $V_c$ for different gate electrodes.

## 7. Electronic device performance

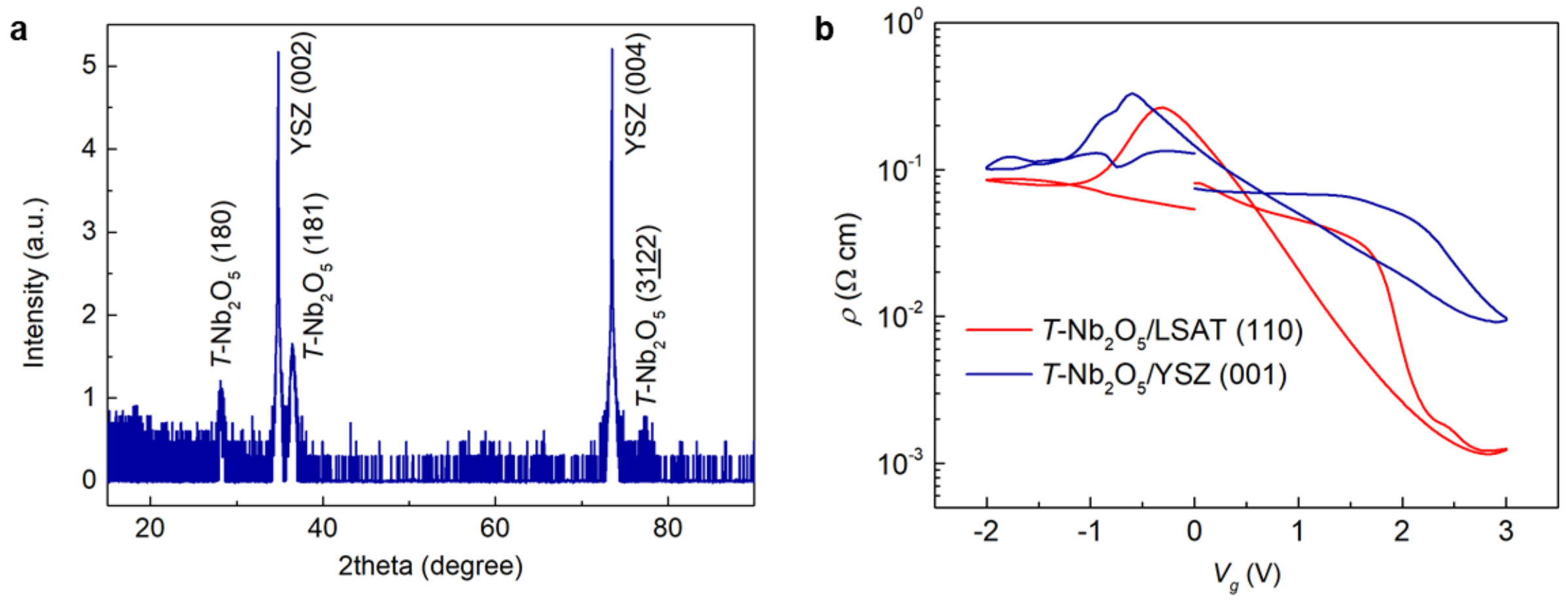

**Supplementary Fig. 36. Structure and electronic properties of poly-crystalline *T*-$Nb_2O_5$ thin film device**. **a**, Theta-2theta XRD scans of *T*-$Nb_2O_5$ thin films grown on a YSZ (001) substrate. The film shows multiple facets, indicating that the ionic channels are oriented in various directions. **b**, Gate voltage (Vg)-dependent resistivity curves of *T*-$Nb_2O_5$ thin films. The poly-crystalline *T*-$Nb_2O_5$/YSZ (001) shows a smaller resistance change compared to the single-crystalline *T*-$Nb_2O_5$/LSAT (110) having unidirectional ionic channel, for the same sweeping rate of 16 mV/sec. This represents that the vertical ionic channel orientation leads to the faster ionic migration from gating.

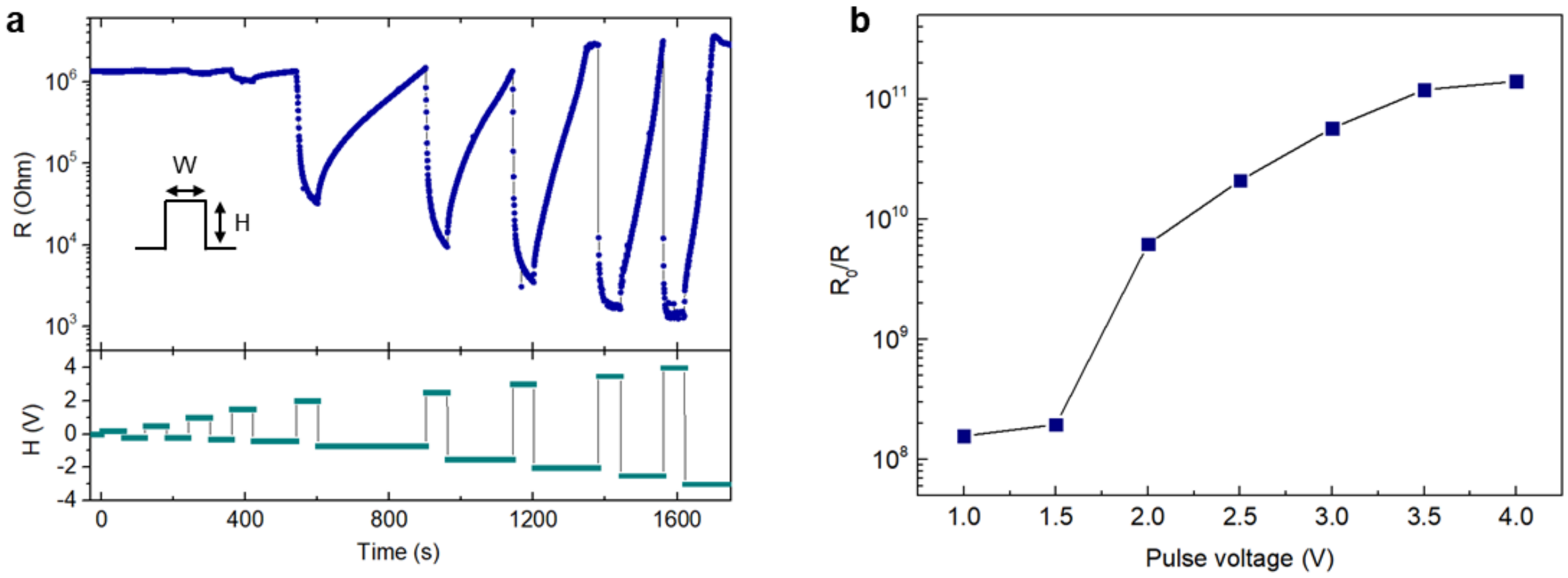

**Supplementary Fig. 37. Pulsed voltage-dependent resistance change in the single crystaline *T*-$Nb_2O_5$ thin film device. a**, Time-dependent resistance data with change of H at the fixed W = 0.5 sec. **b**, The ratio between the resistance of pristine *T*-$Nb_2O_5$ ($R_0$) and the gated resistance (R), i.e., $R_0/R$ as a function of H, obtained from (a).

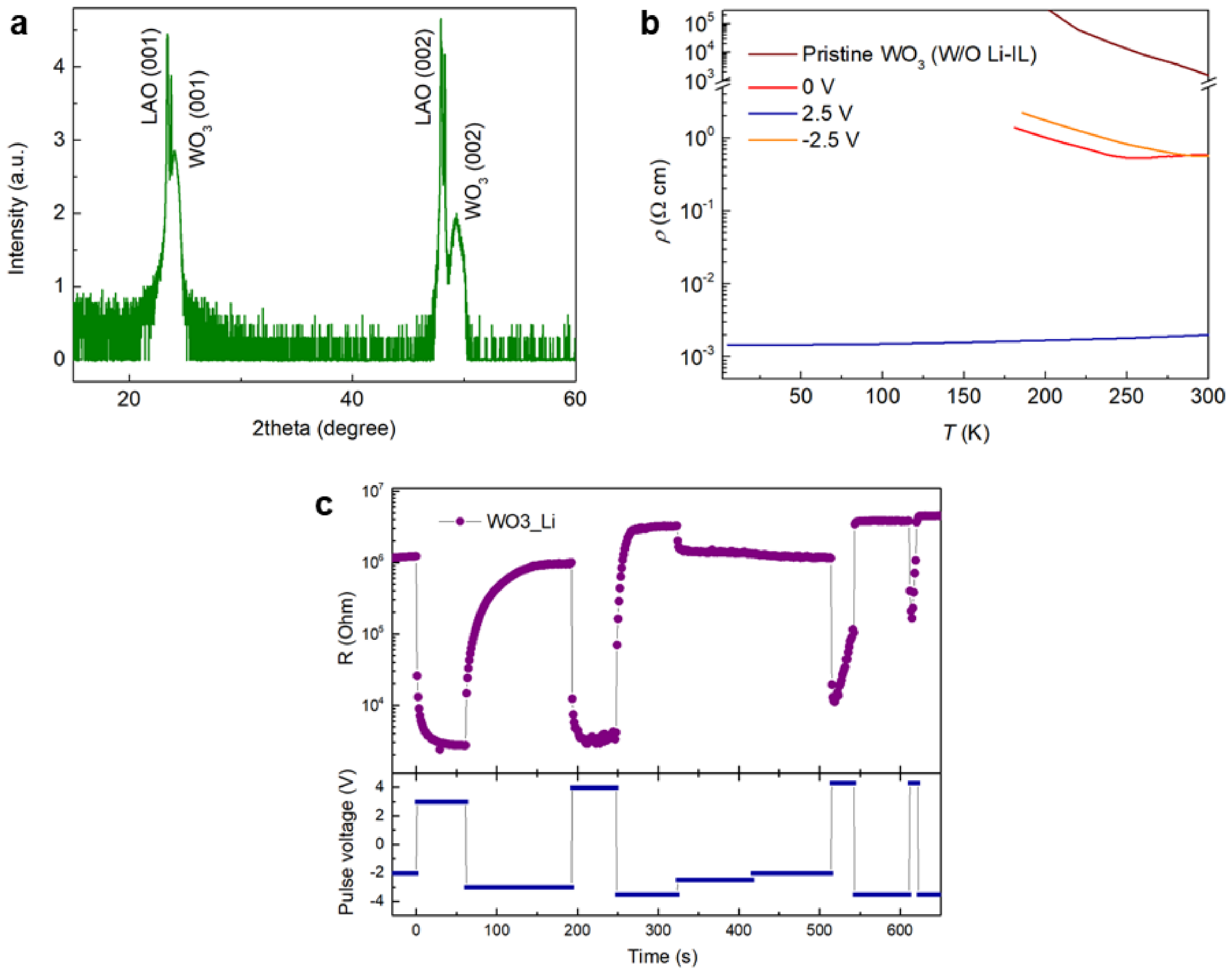

**Supplementary Fig. 38. Structure and electronic properties of $WO_3$ thin film device**. **a**, Theta-2theta XRD scans of epitaxial $WO_3$ thin films grown on a LAO (001) substrate. **b**, Temperature-dependent resistivity curves via Li-ILG. The resistivity of the pristine $WO_3$ film was measured by a high resistance meter (B2985A, KEYSIGHT). The resistivity at 0 V originates from the lower resistance of IL than the $WO_3$ film, as also observed for *T*-$Nb_2O_5$ (Fig. 2g). The film becomes metallic after gating at 2.5 V for 30 min, while it changes back to an insulating state after gating at -2.5 V for 30 min. The resistivity change from the initial film to the metallic film is ≈6 orders of magnitude. **c**, Pulse voltage gating using Li-IL. The pulse width (W) was 0.8 sec. The film becomes metallic at 3 V and 3.5 V, but the resistance rather increases when using 4~4.3 V. This could be due to the conversion reaction of $WO_3$ via Li intercalation.

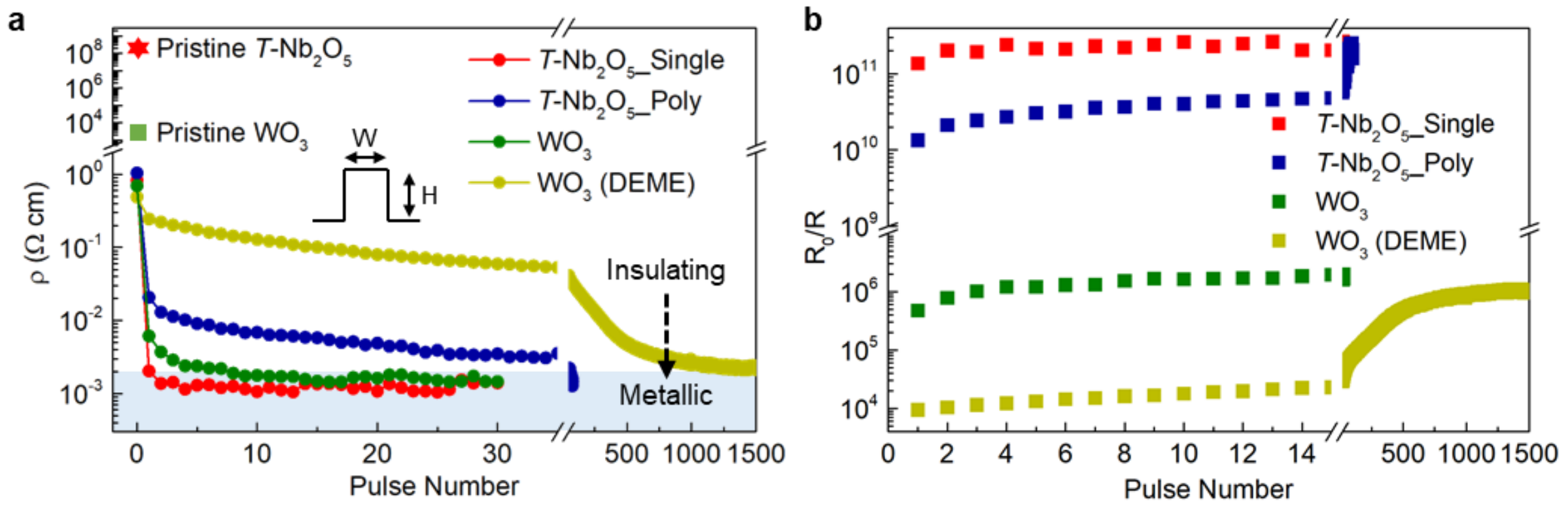

**Supplementary Fig. 39. Pulsed voltage gating of single-crystalline *T*-$Nb_2O_5$, poly-crystalline *T*-$Nb_2O_5$, and $WO_3$ thin films. a**. Resistivity changes of thin film devices as a function of number of pulses. The red, blue, and green lines denote the single-crystalline *T*-$Nb_2O_5$, poly-crystalline *T*-$Nb_2O_5$, and $WO_3$ thin films, respectively, gated by Li-IL. The yellow line represents the $WO_3$ thin film gated using DEME-TFSI. The pulse voltages (H) of 4.3, 4.3, 3.5, 4.3 V were applied for the single-crystalline *T*-$Nb_2O_5$, poly-crystalline *T*-$Nb_2O_5$, $WO_3$ with Li-IL, and $WO_3$ with DEME-TFSI, respectively. The $WO_3$ thin film shows a limited voltage range as it shows an increase in resistance when applying a voltage above 4 V possibly due to a conversion reaction with Li ions (See supplementary Fig. 38c). A pulse width (W) of 0.8 sec was used for all devices. The resistivities of pristine *T*-$Nb_2O_5$ and $WO_3$ are marked by a black pentacle and a square, respectively. The critical resistivity for metallization was defined as $2 \times 10^{-3}$ Ω cm at room temperature, which shows a decrease in resistivity with decreasing temperature (Fig. 4f and supplementary Fig. 38b). **b**, The corresponding resistance change ($R_0/R$) of each film as a function of pulse numbers. $R_0$ represents the resistance of pristine film, while R denotes the changed resistance during gating. All film thicknesses and channel sizes were 30 nm and 60×30 $\mu m^2$, respectively.

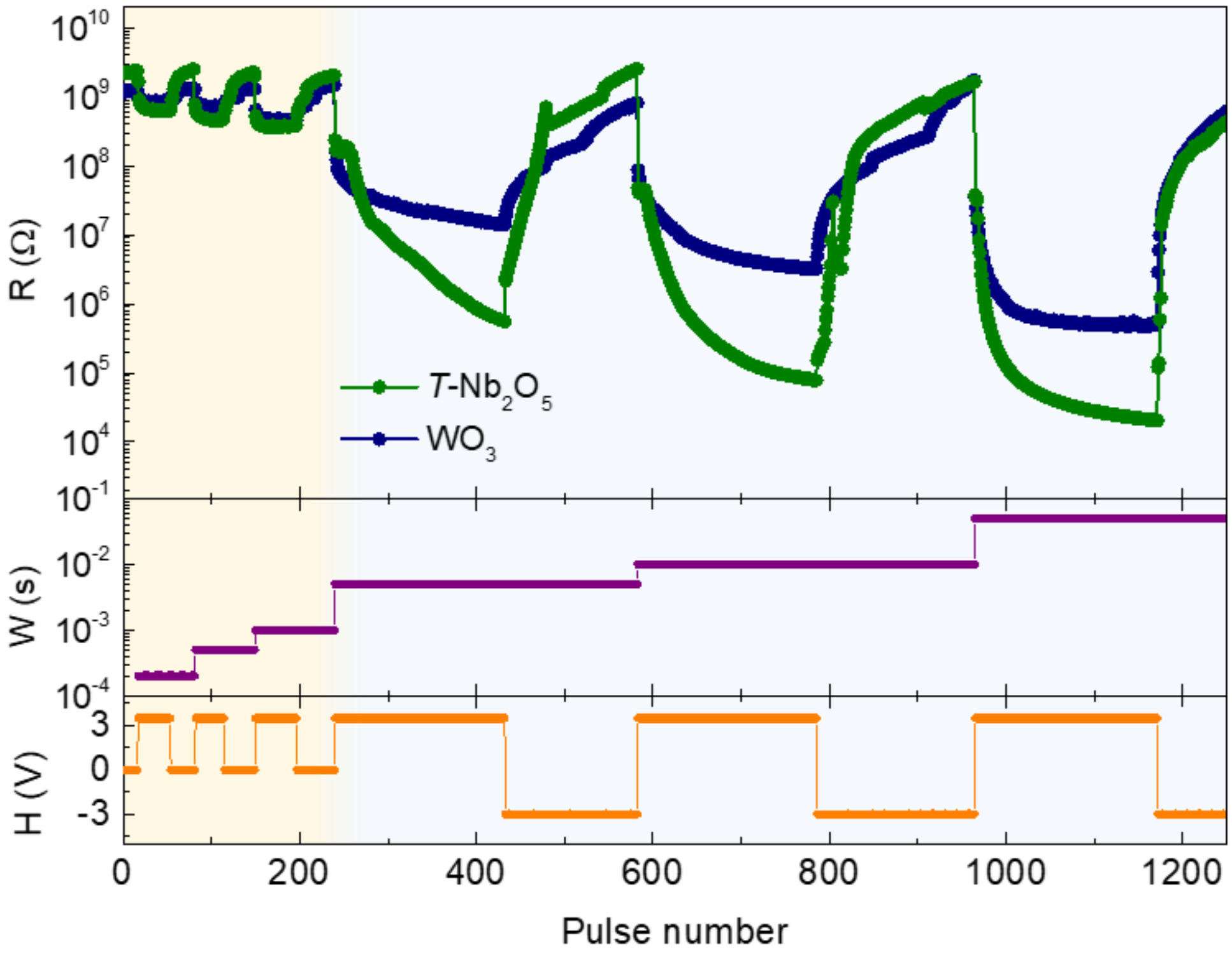

**Supplementary Fig. 40. Pulse width (W)-dependent resistance changes of *T*-$Nb_2O_5$ and $WO_3$ devices**. Resistance changes at different pulsed widths (W) are measured for *T*-$Nb_2O_5$ and $WO_3$ devices at $V_{sd}$ of 0.1 V. The film thicknesses are 30 nm. The resistance of both devices do not show noticeable changes up to W of 1 ms at the gate (pulse) voltage (H) of 3.5 V, and the resistance increases back even at 0 V. This indicates that electrostatic effects are likely to be dominant at short values of W, while electrochemical intercalation into the material dominates when W is longer than 5 ms. As the electrochemical reaction via electrolyte gating occurs first via electric double layer (EDL) formation on the surface, and then subsequent ion migration into the material[37], the small change of resistance at short W (≤ 1 ms) seems to be because applied W is smaller than the EDL formation time (≈ms)[38], while the large resistance changes are observed from electrochemical intercalation into the film at longer W (5 ms) on a timescale that is longer than the EDL formation time. Notably, in the electrochemical reaction regions (above 5 ms), the resistance change of the *T*-$Nb_2O_5$ device is larger and faster than that of $WO_3$.

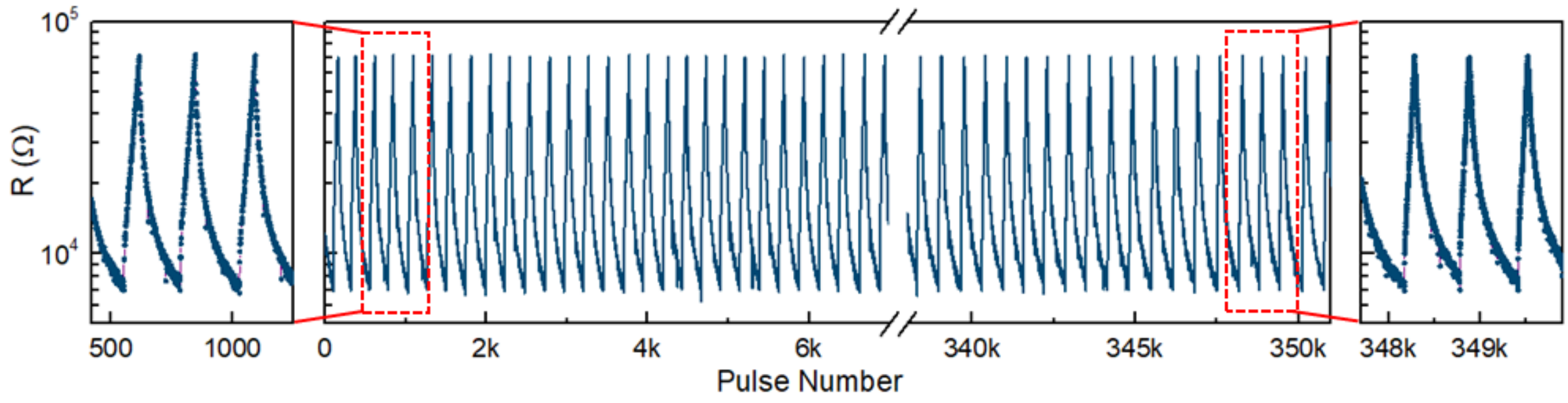

**Supplementary Fig. 41. Repeatable gating of single-crystalline *T*-$Nb_2O_5$ thin films.** Repeatable gating responses of the single-crystalline *T*-$Nb_2O_5$ thin film device. Pulse voltages of 3.8 V/-2 V were applied with a pulse with of 50 msec. The film thickness and channel size were 30 nm and 60×30 μm$^2$, respectively.

## 8. References in Supplementary Information.